\documentclass[fleqn,10pt]{wlscirep}
\usepackage[utf8]{inputenc}
\usepackage[T1]{fontenc}
\usepackage{amsmath,amssymb}
\usepackage{bm}
\usepackage{subcaption}
\usepackage[justification=justified,singlelinecheck=false]{caption}
\usepackage{lmodern}
\usepackage[version=4]{mhchem}
\usepackage{float} 
\usepackage[capitalise]{cleveref}
\usepackage[
  backend=biber,
  style=numeric,
  sorting=none
]{biblatex}
\Crefname{equation}{Eq.}{Eqs.}
\crefname{equation}{Eq.}{Eqs.}
\Crefname{figure}{Fig.}{Figs.}
\crefname{figure}{Fig.}{Figs.}

\usepackage{graphicx}
\setkeys{Gin}{trim=0.20in 0.26in 0.20in 0.26in, clip}

\let\hbar\relax
\DeclareRobustCommand{\hbar}{\mathchar'26\mkern-9mu h}

\newcommand{\nrg}{{\mathcal{E}}}

\newlength{\figheight}
\begin{document}

\title{Efficient Quantum-Mechanical Modeling of Nonradiative Charge Transfer Processes}

\author[1]{Alexander Karl}
\author[1]{Dominic Waldhoer}
\author[1]{Theresia Knobloch}
\author[1]{Christoph Wilhelmer}
\author[1, *]{Tibor Grasser}
\affil[1]{Institute for Microelectronics, Technical University Vienna, Austria}
\affil[*]{e-mail: grasser@iue.tuwien.ac.at}

\begin{abstract}
Nonradiative charge transfer processes play a central role in a wide range of physical phenomena, including reliability phenomena in semiconductor devices such as bias temperature instability, hysteresis, random telegraph noise, and trap-assisted tunneling. nonradiative multiphonon (NMP) theory provides a physically rigorous framework for describing such charge transitions, but its full quantum-mechanical formulation is computationally too demanding for large-scale simulations. In this work, we present a systematic and implementation-oriented treatment of NMP-based models for practical large-scale simulations. Starting from the quantum-mechanical foundations of coupled electron--phonon dynamics, we derive computationally efficient approximations for charge capture and emission rates and clearly identify the underlying assumptions and validity regimes. In particular, we introduce an effective crossing-preserving approximation that yields fully analytic, numerically stable, and computationally inexpensive transition rates while retaining the essential quantum-mechanical physics. The resulting expressions are therefore well suited for large-scale device simulations, where capture coefficients must be evaluated repeatedly over broad multidimensional parameter spaces. Furthermore, we derive continuum formulations for transitions between localized defect states and extended electronic bands, enabling direct incorporation into semiconductor-device simulations. The resulting framework bridges microscopic defect physics and practical
large-scale simulations of charge transfer processes in complex semiconductor
devices. At the same time this work serves as a practical guide for implementing
physically grounded NMP-based models, providing both a
systematic derivation of the underlying theory and a clear guidance on the validity limits.

\end{abstract}

\flushbottom
\maketitle

\thispagestyle{empty}

 \newpage
 
\section*{Introduction}

Nonradiative charge transfer processes at defects in semiconductor devices are a key physical mechanism governing the behavior, reliability, and long-term stability of modern semiconductor devices. Phenomena such as bias temperature instability (BTI), hysteresis, random telegraph noise (RTN) and gate leakage due to trap assisted tunneling (TAT) are all governed by the dynamics of underlying charge capture and emission processes at defects. As device dimensions continue to scale, an accurate physical description of defect-mediated charge capture and emission processes has become essential for the correct description of device behavior. \\

In most Technology Computer-Aided Design (TCAD) tools, defect-mediated charge transfer processes are commonly described using the Shockley–Read–Hall (SRH) model \cite{Shockley1952, Hall1952, Sze2006} due to its simplicity and computational efficiency. However, the SRH formalism assumes that charge transitions occur without significant relaxation of the atoms surrounding the defect. This assumption breaks down for virtually all technologically relevant defects, where a change in the defect's charge state is accompanied by substantial structural rearrangement of the neighboring atoms. Consequently, the SRH description is inherently incomplete and generally fails to reproduce the experimentally observed temperature and field dependence of charge trapping and detrapping. To reconcile SRH-based simulations with experimental observations, the missing relaxation physics is typically absorbed into empirical fitting parameters, most notably capture cross sections that are allowed to vary with temperature and electric field \cite{MinimosNT,SentaurusDevice}. While such empirical extensions can be calibrated to reproduce measured trapping and detrapping kinetics, the resulting capture cross sections often span many orders of magnitude and generally lack a direct connection to the microscopic properties of the underlying defects  \cite{Kirton1989, Ryan2015, Ruch2021}. \\

A physically rigorous description of defect-mediated charge transfer is provided by Nonradiative multiphonon (NMP) theory \cite{Huang1950, Kubo1955, Freed1970, Stoneham1981, Alkauskas2014, Turiansky2021, Shi2015}. Conceptually, electron capture by a defect may be viewed as a two-step process within this framework: First, the electron tunnels to the defect, changing its charge state and destabilizing the initial atomic configuration. Subsequently, the atoms at the defect site relax toward a different equilibrium configuration corresponding to the new charge state. The NMP framework therefore naturally captures the interplay between electronic transitions and atomic relaxation. This framework has been successfully used to quantitatively reproduce a wide range of experimentally observed reliability phenomena in MOSFETs, including BTI \cite{Vandelli2014, Rzepa2018, Putcha2020}, hysteresis \cite{Knobloch2017, Karl2025, Lv2026}, RTN \cite{Grasser2012, Mao2016, Guo2017}, and TAT \cite{Larcher2003, Liu2020, Waldhoer2023}. In parallel, substantial progress has been made in determining the underlying model parameters from first-principles calculations, particularly using density functional theory (DFT) \cite{Alkauskas2014, Turiansky2021}. The parameters obtained from DFT can be directly compared with those extracted from experimental data, enabling the identification or exclusion of microscopic defect candidates responsible for the observed reliability phenomena \cite{Goes2018, Waldhoer2021, WaldhoerDISS}. Owing to this capability to link macroscopic reliability effects to their atomistic origins, the NMP framework is widely regarded as the state-of-the-art approach for modeling defect dynamics in modern semiconductor devices, providing a unified and physically well-founded description of diverse degradation mechanisms. \\

Despite these advances, the full quantum-mechanical NMP formulation remains too computationally demanding for direct use in TCAD simulations. Practical device-level modeling therefore relies on simplified rate expressions. However, the assumptions behind these expressions are often not systematically documented, and their derivation and range of validity are not always clearly defined. This is particularly problematic for the widely used classical approximation, which is formally valid only in the high-temperature limit. Depending on the specific defect parameters the classical approximation can already break down at room temperature \cite{Michl2021_1, Michl2021_2}. Applying such approximations outside their validity range can therefore lead to misleading interpretations of simulation results. In this work, we address this gap by systematically deriving practical,
computationally efficient approximations for nonradiative charge-transition
rates starting from the full quantum-mechanical NMP formalism, while explicitly
discussing their underlying assumptions and validity regimes. \\

Specifically, we address the additional complexity arising in the general case where the potential-energy surfaces of the defect have different curvatures. In this case, the full quantum-mechanical NMP problem becomes significantly more challenging, since the transition rates involve a double sum over vibrational states that is generally not analytically tractable. To overcome this limitation, we develop a crossing-preserving approximation (CPA) for general NMP systems. This approach maps the original unequal-curvature problem onto an effective equal-curvature model while preserving the dominant crossing point that governs the transition dynamics. Combined with a continuum approximation for the vibronic spectrum, the mapping yields a fully analytic closed-form approximation to the quantum-mechanical capture and emission rates. The resulting expression retains the essential quantum-mechanical physics while remaining computationally inexpensive and numerically stable, making it well suited for integration into large-scale TCAD workflows, where capture coefficients must be computed repeatedly over broad multidimensional parameter spaces. \\

Building on this efficient approximation for transitions between two discrete electronic states, we extend the formalism to charge exchange between  defects and continua of electronic states, such as the conduction or valence band of semiconductor devices. Starting from the state-to-state transition rates, the total capture and emission
rates are obtained by summing over the electronic states in the continuum. The resulting expressions are then written in terms of the local density of states of the continuum and the occupation probabilities of the corresponding electronic states. Finally, we simplify the continuum rates using a band-edge approximation, thereby avoiding explicit energy integration and obtaining a compact, computationally efficient representation suitable for large-scale TCAD simulations. We then compare the resulting continuum rates with SRH theory and show that the conventional SRH framework is fundamentally phenomenological. In particular, the temperature-
and field-dependent capture and emission rates predicted by NMP theory can only be reproduced
within the SRH framework by introducing phenomenological capture cross sections with ad-hoc dependencies on temperature and field. \\

Beyond introducing these new approximations for the NMP transition rates, the present work is intended as a practical guide for modeling charge transitions in semiconductor devices. The manuscript aims to bridge the gap between defect physics and practical TCAD implementation by providing both a systematic derivation of the underlying theory and an intuitive interpretation of the relevant model parameters. In this way, the paper is designed not only as a theoretical treatment of NMP transitions, but also as a structured reference for implementing physically grounded charge-trapping models in semiconductor device simulations. \\

The paper is organized as follows: In Section~\ref{sec:qm-foundations}, we derive the quantum-mechanical foundations of the NMP model starting from the coupled electron--ion Hamiltonian and introduce the diabatic representation of charge transitions. Section~\ref{sec:configuration-coordinate} introduces the scalar configuration-coordinate picture, which transforms the multidimensional phonon problem into an effective one-dimensional model. In Section~\ref{sec:transition-rates}, we derive the quantum-mechanical transition-rate expression and discuss three complementary evaluation strategies: direct quantum-mechanical evaluation, the classical high-temperature approximation, and the newly developed effective crossing preserving approximation. Section~\ref{sec:continuum-transitions} extends the formalism to transitions between localized defect states and electronic bands, establishing the connection to TCAD-compatible capture and emission models. Here, we also derive a band-edge approximation for the resulting continuum rates and compare these rates with conventional SRH theory. Finally, Section~\ref{sec:examples} present representative BTI simulations that illustrate the differences between the transition-rate approximations developed throughout this work.

\section{Quantum-Mechanical Foundations of Nonradiative Charge Transfer}
\label{sec:qm-foundations}
We consider a nonradiative charge transition in a system consisting of a localized defect and an electronic reservoir that can exchange carriers with the defect. To describe the coupled electron--nuclear system, we denote the electronic coordinates and momenta by $\bm{x}$ and $\bm{p}$, and the mass-weighted nuclear coordinates and momenta by $\bm{Q}$ and $\bm{P}$. The full Hamiltonian is given by
\begin{equation}
H = T_\mathrm{N} + T_\mathrm{e} + V_\mathrm{ee} + V_\mathrm{eN} + V_\mathrm{NN},
\end{equation}
where $T_\mathrm{N}$ and $T_\mathrm{e}$ denote the nuclear and electronic kinetic-energy operators, respectively, while $V_\mathrm{ee}$, $V_\mathrm{eN}$, and $V_\mathrm{NN}$ describe the electron--electron, electron--ion, and ion--ion interactions. The Hamiltonian describes the complete system, including both the defect and the reservoir. A prototypical example is a defect at an insulator--semiconductor interface, where the defect can capture an electron from the semiconductor conduction band or emit its electron into the conduction band. More generally, the reservoir may represent any continuum of electronic states that can exchange charge with the defect. \\

To calculate nonradiative charge transition rates within perturbation theory, the Hamiltonian must be partitioned into an unperturbed part and a weak perturbation that induces transitions. Since perturbation theory yields transition rates between the eigenstates of the unperturbed Hamiltonian, these states define the basis of the perturbative treatment and must therefore be chosen such that they can be identified with the physical charge states of the defect. Only then can the calculated transition rates be interpreted as the physical rates for charge capture and emission. The choice of basis is therefore of central importance, as it determines the form of the Hamiltonian partitioning and consequently the coupling responsible for the transitions. As a natural starting point, we first introduce the adiabatic Born--Oppenheimer approximation, which provides a convenient representation of the coupled electron--nuclear system. This representation subsequently serves as the foundation for constructing the diabatic basis employed in the perturbative treatment in the rest of this work. \\

The Born--Oppenheimer approximation \cite{Shi2015} exploits the separation of electronic and nuclear time scales arising from the large mass difference between electrons and nuclei. As a result, electrons adjust quasi-instantaneously to the slow nuclear motion. Within this approximation, the total wave function can be factorized into a nuclear (vibrational) wavefunction $\chi_{a,n}(\bm{Q})$ and an electronic wave function $\psi_a(\bm{x};\bm{Q})$ at fixed nuclear configuration $\bm{Q}$:
\begin{equation}
\Psi_{a,n}(\bm{x},\bm{Q})
=
\psi_a(\bm{x};\bm{Q})\,
\chi_{a,n}(\bm{Q}),
\label{BOBASIS}
\end{equation}
where $a$ labels the electronic state and $n$ the nuclear (vibrational) state. The set of such product states defines a basis for the coupled electron--nuclear Hilbert space. For each $\bm{Q}$, the electronic wave function $\psi_a(\bm{x};\bm{Q})$ is obtained from the eigenvalue problem
\begin{equation}
H_\mathrm{e}(\bm{Q}) \,\psi_a(\bm{x};\bm{Q}) = E_a(\bm{Q}) \,\psi_a(\bm{x};\bm{Q}), \quad H_\mathrm{e}(\bm{Q}) = T_\mathrm{e} + V_\mathrm{ee} + V_\mathrm{eN}(\bm{Q}) + V_\mathrm{NN}(\bm{Q}),
\end{equation}
where the nuclear coordinates $\bm{Q}$ enter only as parameters. The eigenvalues $E_a(\bm{Q})$ define the adiabatic potential energy surfaces (PES). Note that the electronic Hamiltonian $H_\mathrm{e}$ includes the ion--ion interaction $V_\mathrm{NN}(\bm{Q})$, such that $E_a(\bm{Q})$ represents the total potential energy of the system. The nuclear wave functions $\chi_{a,n}(\bm{Q})$ are obtained from the effective Schrödinger equation
\begin{equation}
\left[ T_\mathrm{N} + E_a(\bm{Q}) \right] \, \chi_{a,n}(\bm{Q}) = E_{a,n} \,\chi_{a,n}(\bm{Q}),
\end{equation}
which describes the nuclei moving under the effect of the adiabatic PES $E_a(\bm{Q})$ defined by the electronic problem. In the following, we consider the adiabatic ground-state PES of the system. As the nuclear configuration $\bm{Q}$ is varied, the electronic charge distribution adjusts continuously to the nuclear motion and the character of the electronic ground state may change from a defect-localized state to one in which the electron is no longer localized at the defect. We assume that the relevant ground-state PES possesses two stable local minima, denoted by $\bm{Q}_i$ and $\bm{Q}_f$, corresponding to the relaxed atomic geometries with the carrier localized at the defect and after it has left the defect (e.g., by occupying a delocalized band state), respectively. In practice, these equilibrium geometries are typically obtained from independent DFT calculations. For example for isolated bulk defects the calculation is restricted to the host material containing the defect, and the relevant electronic configurations are realized by prescribing different total charges of the supercell. The relaxed structures of the corresponding charged supercells are then identified with the equilibrium geometries $\bm{Q}_i$ and $\bm{Q}_f$ \cite{WaldhoerDISS}. \\
 
In the adiabatic Born--Oppenheimer basis defined by \Cref{BOBASIS}, the electronic states $\psi_a(\bm{x};\bm{Q})$ depend parametrically on the nuclear configuration and therefore evolve continuously as $\bm{Q}$ changes. As a result, their physical character may change. In particular, near regions where different adiabatic potential-energy surfaces approach each other, the electronic states can become strong mixtures of states with different charge-state character, such as a defect-localized state and a conduction-band-like state. In such regions, the index $a$ no longer labels a well-defined charge state of the defect. To describe charge transitions between well-defined charge states, it is therefore necessary to employ a basis whose electronic character remains fixed throughout configuration space. For this purpose, we adopt the \textit{static approximation} \cite{Shi2015}, in which the adiabatic electronic eigenstates are frozen at a fixed reference geometry $\bm{Q}_0$ and used as a new electronic basis,
\begin{equation}
\phi_a(\bm{x};\bm{Q}_0)
\equiv
\psi_a(\bm{x};\bm{Q}_0).
\end{equation}
The corresponding static basis states for the coupled electron–nuclear system are
\begin{equation}
\Psi_{i,m}(\bm{x},\bm{Q})
=
\phi_i(\bm{x};\bm{Q}_0)\,
\chi_{i,m}(\bm{Q}),
\qquad
\Psi_{f,n}(\bm{x},\bm{Q})
=
\phi_f(\bm{x};\bm{Q}_0)\,
\chi_{f,n}(\bm{Q}).
\end{equation}
Here, $i$ and $f$ label the initial and final charge states of the defect. Unlike the adiabatic states $\psi_a(\bm{x};\bm{Q})$, the basis functions $\phi_i$ and $\phi_f$ no longer evolve with the nuclear configuration. Instead, they retain their electronic character throughout configuration space and therefore provide a diabatic-like basis in which the charge states remain well defined. Throughout this work, the equilibrium geometry of the initial charge state is chosen as the reference configuration for the diabatic basis, i.e. $\bm{Q}_0=\bm{Q}_i$. This, however, is merely a practical choice and does not alter the underlying formalism. In principle, the final equilibrium geometry $\bm{Q}_f$ could equally well be chosen as the reference. In first-principles calculations, the most convenient reference is often the geometry at which the electronic states entering the transition can be identified and evaluated most reliably within DFT. \\

Having fixed the basis states, we now derive an effective description of the coupled electron--ion system by projecting the full Hamiltonian onto the electronic subspace spanned by the initial and final diabatic electronic states. The resulting matrix elements
\begin{align}
H_{ab}(\bm{Q}) &=
T_\mathrm{N}\delta_{ab}
+
\langle\phi_a|H_\mathrm{e}(\bm{Q})|\phi_b\rangle.
\end{align}
define operators acting on the nuclear degrees of freedom. 
This projected Hamiltonian still retains the full dependence on the nuclear coordinates and is therefore generally too complicated for practical calculations. To obtain a model that captures the essential physics using only a small number of parameters, we now approximate the electronic matrix elements by their Taylor expansions truncated at leading order. \\

The diagonal electronic matrix elements
\begin{equation}
V_a(\bm{Q})
=
\langle\phi_a|H_\mathrm{e}(\bm{Q})|\phi_a\rangle,
\qquad
a\in\{i,f\},
\end{equation}
define the diabatic PESs, which describe the total potential energy of the system as a function of the nuclear coordinates while the defect retains a fixed electronic charge-state character. Provided the potential-energy surfaces are sufficiently smooth in the vicinity of their equilibrium geometries, they may be approximated by a second-order Taylor expansion,
\begin{equation}
V_a(\bm{Q})
=
E_a
+
\frac12
(\bm{Q}-\bm{Q}_a)^\mathsf{T}
\bm{K}_a
(\bm{Q}-\bm{Q}_a), \qquad
a\in\{i,f\},
\end{equation}
where $E_a$ denotes the minimum energy and $\bm{K}_a$ the Hessian matrix evaluated at the equilibrium geometry $\bm{Q}_a$. Since the expansion is performed about an equilibrium configuration the linear term in the Taylor expansion vanishes identically, leaving the quadratic (harmonic) contribution as the leading order. This harmonic approximation provides an accurate description of the dynamics associated with each charge state while remaining sufficiently simple for practical calculations. \\

The off-diagonal matrix elements $
V_{if}(\bm{Q})
=
\langle\phi_i|H_\mathrm{e}(\bm{Q})|\phi_f\rangle,
$
describe the electronic coupling between the two diabatic charge states. In principle, the off-diagonal matrix elements may be expanded about any arbitrary nuclear configuration. However, choosing the expansion point to coincide with the reference configuration $\bm{Q}_0 = \bm{Q}_i$, is particularly advantageous, since the electronic Hamiltonian is diagonal at this geometry. Consequently, the constant term in the Taylor expansion vanishes identically, yielding a linear electron--phonon coupling model to leading order:
\begin{equation}
V_{if}(\bm Q)
=
\bm W_{if}\cdot(\bm Q-\bm Q_i).
\end{equation}
As a result, the off-diagonal coupling is induced solely by the small nuclear displacements away from the initial equilibrium geometry $\bm Q_i$. This provides a well-controlled starting point for treating the electron--phonon interaction perturbatively with Fermi's golden rule. \\

Using these definitions, the Hamiltonian can be expanded to leading order in the nuclear displacements and written compactly as
\begin{equation}
H =
|\phi_i\rangle\langle \phi_i|
\left(
\frac{1}{2}\bm{P}^\mathsf{T} \bm{P} + V_i(\bm{Q})
\right)
+
|\phi_f\rangle\langle \phi_f|
\left(
\frac{1}{2}\bm{P}^\mathsf{T} \bm{P} + V_f(\bm{Q})
\right)
+
\left(|\phi_f\rangle\langle \phi_i| + |\phi_i\rangle\langle \phi_f|\right)
\, \bm{W}_{if} \cdot (\bm{Q} - \bm{Q}_i).
\label{eq:NMP_multidimensional_1}
\end{equation}
\begin{equation}
V_i(\bm{Q})
=
E_i
+
\frac{1}{2}
(\bm{Q} - \bm{Q}_i)^\mathsf{T}
\bm{K}_{i}
(\bm{Q} - \bm{Q}_i),
\qquad
V_f(\bm{Q})
=
E_f
+
\frac{1}{2}
(\bm{Q} - \bm{Q}_f)^\mathsf{T}
\bm{K}_{f}
(\bm{Q} - \bm{Q}_f).
\label{eq:NMP_multidimensional_2}
\end{equation}
Importantly, this Hamiltonian admits a simple physical interpretation. An electronic transition transfers the system from the initial diabatic PES $V_i(\bm{Q})$ to the final diabatic PES $V_f(\bm{Q})$. Before the transition, the nuclei fluctuate around the equilibrium geometry $\bm{Q}_i$ of the initial surface. Immediately after the transition, however, the nuclei remain at essentially the same atomic configuration while the governing PES changes to $V_f(\bm{Q})$, whose minimum is located at the different equilibrium geometry $\bm{Q}_f$. The nuclei therefore experience forces that drive a structural relaxation from the former equilibrium $\bm{Q}_i$ toward the new equilibrium $\bm{Q}_f$. This structural relaxation is central to charge trapping processes and determines the rate at which charge-state transitions occur.

\section{Effective Configuration-Coordinate Model}
\label{sec:configuration-coordinate}
The general NMP model defined in \Cref{eq:NMP_multidimensional_1,eq:NMP_multidimensional_2} is intrinsically multidimensional, since the nuclear motion generally involves all vibrational degrees of freedom of the system. For realistic systems, this multidimensional problem becomes difficult to evaluate explicitly due to the large number of vibrational degrees of freedom. A considerable simplification is obtained by describing the nuclear motion in terms of a single scalar configuration coordinate that parametrizes the structural relaxation along the straight line connecting the equilibrium configurations of the two charge states. This reduces the multidimensional problem to an effective one-dimensional model, which forms the basis for the remainder of this work. \\

The relaxation associated with the charge transition is characterized by the displacement
$\Delta\bm{Q}=\bm{Q}_f-\bm{Q}_i
$, between the equilibrium configurations of the two charge states, and by the energy difference between the minima of the corresponding diabatic potential-energy surfaces, $\Delta E = E_f-E_i$. To simplify the notation throughout the remainder of this work, we choose the equilibrium configuration and energy of the initial charge state as the origin of the coordinate system. Accordingly, we choose $\bm Q_i=\bm 0$ and $E_i=0$, such that the equilibrium configuration and energy of the final charge state are simply given by $\bm Q_f=\Delta\bm Q$ and $E_f=\Delta E$. This is merely a choice of coordinate system that simplifies the subsequent notation and does not affect the underlying physics or any of the resulting transition rates. \\

\begin{figure}[!hbt]
    \centering

    \begin{subfigure}{0.33\textwidth}
        \centering
        \includegraphics[width=\textwidth]{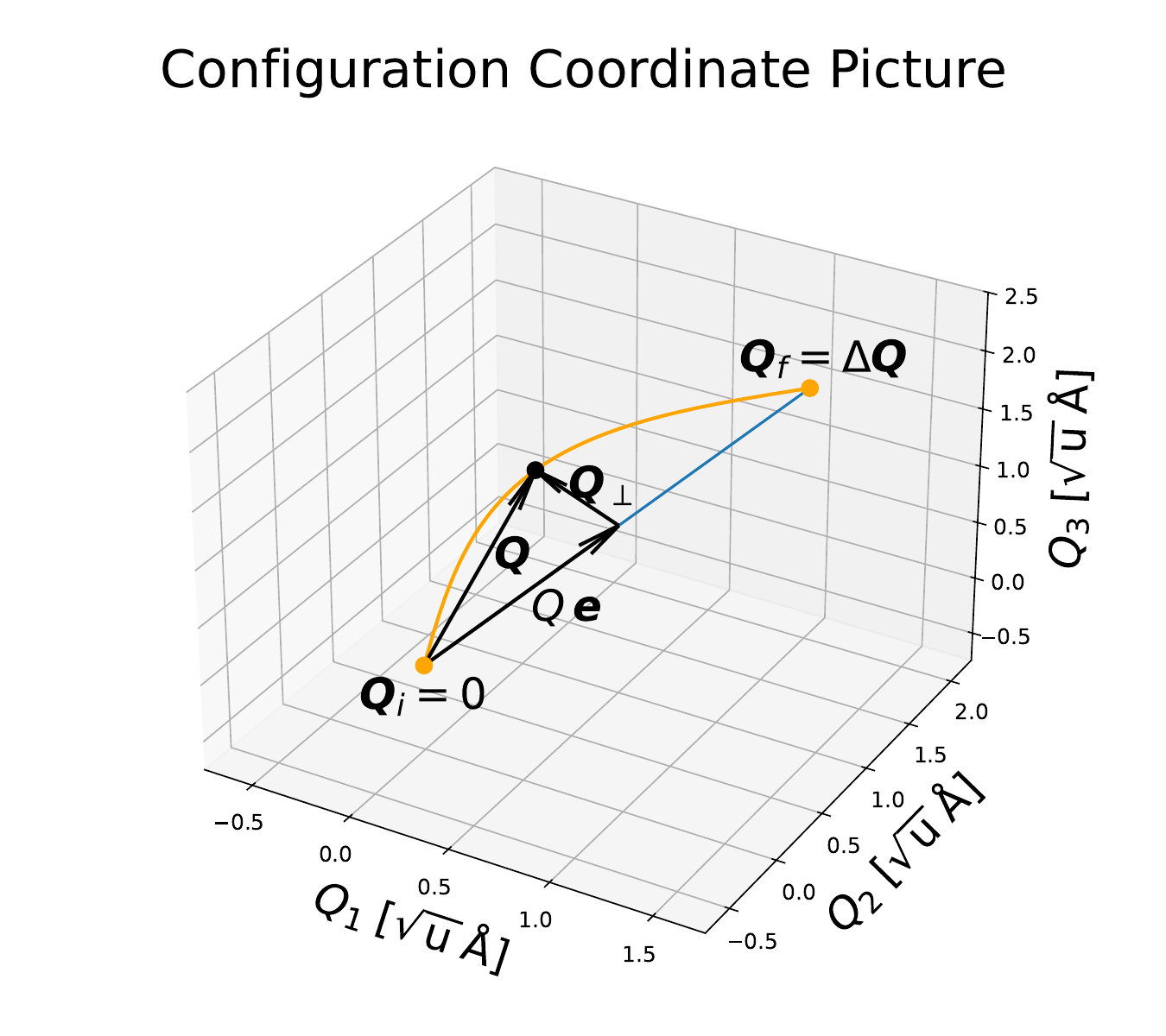}
        \caption{}
        \label{fig:config_coord_1}
    \end{subfigure}
    \begin{subfigure}{0.33\textwidth}
        \centering
        \includegraphics[width=\textwidth]{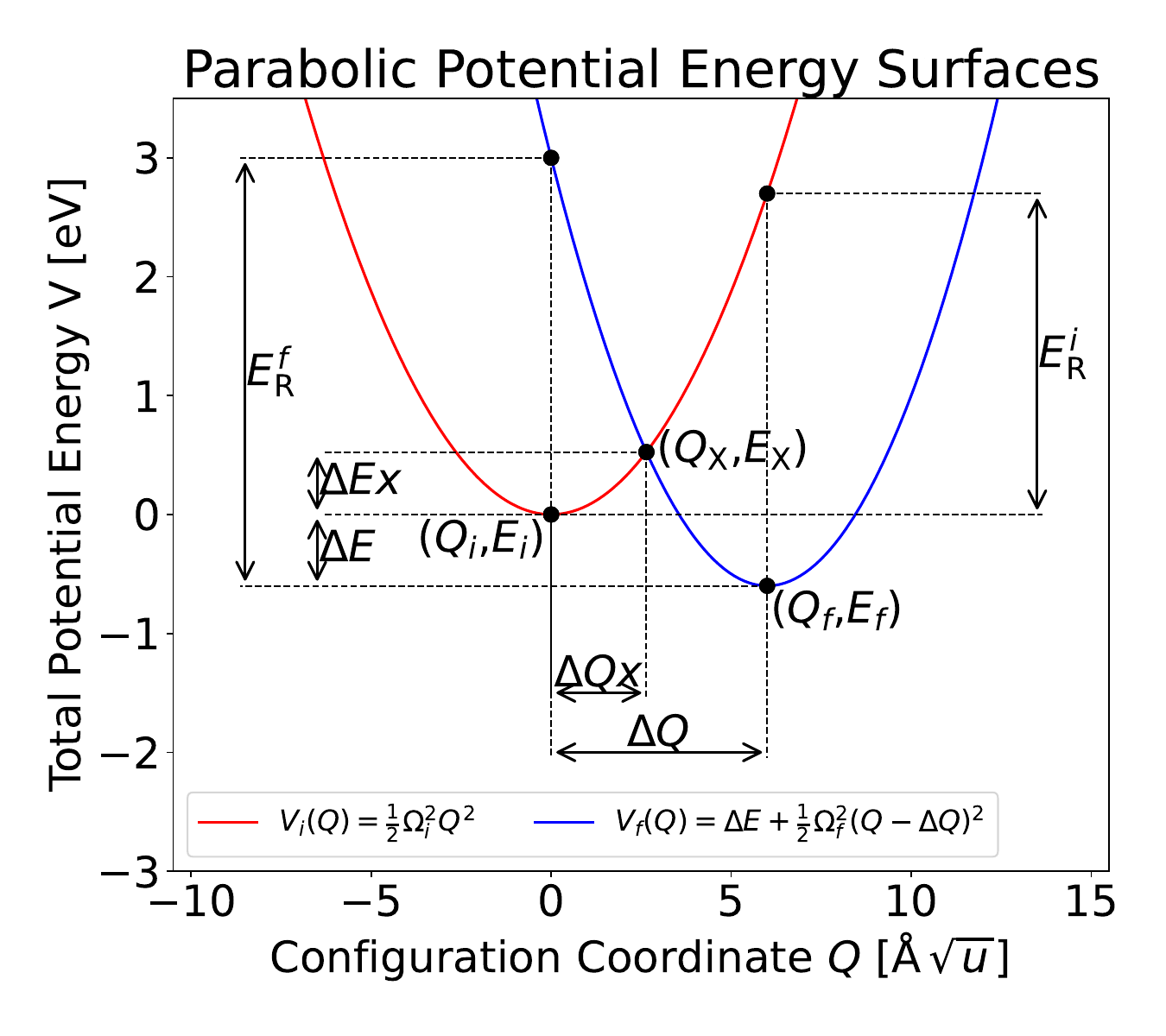}
        \caption{}
        \label{fig:config_coord_2}
    \end{subfigure}

    \caption{
    Schematic representation of the scalar configuration coordinate reduction.
    (a) Visualization of different paths in nuclear configuration space. The configuration coordinate path shown in blue connects the minima of the initial and final potential-energy surfaces, located at \(\bm{Q}_i=\bm{0}\) and \(\bm{Q}_f=\Delta\bm{Q}\), along a straight line. The orange curve illustrates a general path in configuration space. A general configuration $\bm{Q}$ can be decomposed into a component $\bm{Q}_{\parallel} = Q \, \bm{e} $ along the configuration coordinate path and an orthogonal component $\bm{Q}_{\perp}$. (b) Corresponding configuration coordinate diagram showing the effective potential energy surfaces $V_i(Q)$ and $V_f(Q)$ along the one-dimensional configuration coordinate $Q$. The minima of these surfaces are located at $(Q_i,E_i)=(0,0)$ and $(Q_f,E_f)=(\Delta Q,\Delta E)$, respectively. The two parabolic potentials are fully characterized by the energy difference $\Delta E$, the displacement between both equilibrium configurations $\Delta Q$, and the relaxation energies $E_\mathrm{R}^{f} = (1/2) \, \Omega_f^2 \, \Delta Q^2$ and $E_\mathrm{R}^{i} = (1/2) \, \Omega_i^2 \, \Delta Q^2$.
    }
    \label{fig:config_coord}
\end{figure}

To reduce the multidimensional problem to an effective one-dimensional description, we define the unit vector pointing from the initial equilibrium to the final equilibrium
\begin{equation}
\bm e=\frac{\Delta\bm Q}{\Delta Q} \qquad\mathrm{with} \qquad \Delta Q
=
\|\Delta\bm Q\|.
\end{equation}
As shown in \Cref{fig:config_coord}, every vector in configuration space can be decomposed into components parallel and orthogonal to the relaxation direction $\bm e$. For a general nuclear configuration $\bm Q$, this decomposition is expressed as
\begin{equation}
\bm Q
=
Q\,\bm e
+
\bm Q_\perp,
\qquad
Q=\bm e\cdot\bm Q,
\qquad
\bm Q_\perp
=
\bm Q-Q\,\bm e.
\end{equation}
Here, the scalar configuration coordinate $Q$ measures the displacement along the dominant relaxation direction, while $\bm Q_\perp$ describes displacements orthogonal to this path. Analogously, the nuclear momentum $\bm P$ is decomposed as
\begin{equation}
\bm P
=
P\,\bm e
+
\bm P_\perp,
\qquad
P=\bm e\cdot\bm P,
\qquad
\bm P_\perp
=
\bm P-P\,\bm e.
\end{equation}
Accordingly, the scalar momentum $P$ is the momentum conjugate to the configuration coordinate $Q$ and describes nuclear motion along the dominant relaxation direction, while $\bm P_\perp$ contains the momentum associated with motions orthogonal to this direction. \\

Substituting this decomposition into the Hamiltonian and neglecting terms
involving the orthogonal degrees of  freedom, we obtain an effective one-dimensional
model in which nuclear motion is confined to the line connecting the two
equilibrium configurations. The neglected orthogonal phonon modes are thereby treated
as weakly coupled environmental degrees of freedom, whose primary effect is to
provide dissipative corrections rather than to determine the
dominant transition dynamics. The resulting Hamiltonian reads
\begin{equation}
H =
|\phi_i\rangle\langle \phi_i|
\left(
\frac{P^2}{2} + V_i(Q)
\right)
+
|\phi_f\rangle\langle \phi_f|
\left(
\frac{P^2}{2} + V_f(Q)
\right)
+
\left(|\phi_f\rangle\langle \phi_i| + |\phi_i\rangle\langle \phi_f|\right)
\, W_{if} \, Q
+ \ldots .,
\label{eq:H_1D}
\end{equation}
where the effective one-dimensional diabatic potential-energy surfaces are given by
\begin{equation}
V_i(Q)
=\frac{1}{2}\Omega_i^2 Q^2,
\qquad
V_f(Q)
=
\Delta E
+
\frac{1}{2}\Omega_f^2 (Q - \Delta Q)^2.
\label{eq:1D_PES_absolute}
\end{equation}
The effective parameters of this one-dimensional model are obtained by projecting the corresponding multidimensional quantities onto the relaxation direction:
\begin{equation}
\Omega_i^2
=
\bm{e}^\mathsf{T} \bm{K}_{i} \bm{e},
\qquad
\Omega_f^2
=
\bm{e}^\mathsf{T} \bm{K}_{f} \bm{e},
\qquad
W_{if}
=
\bm{W}_{if}\cdot \bm{e}.
\label{eq:effective_parameters_1D}
\end{equation}

At first sight, neglecting the phonon modes orthogonal to the effective configuration coordinate may appear to be a rather drastic approximation. Within the effective one-dimensional model, the nuclei are constrained to move along the straight line connecting the equilibrium configurations of the two charge states. In general, however, this straight-line path does not coincide with the true minimum-energy path in the full multidimensional configuration space  \cite{Waldhoer2019}. One might therefore expect the neglected orthogonal degrees of freedom to significantly alter the transition rates. Nevertheless, the general validity of this one-dimensional reduction has been demonstrated in several first-principles studies of nonradiative carrier capture. In particular, Alkauskas \emph{et al.}~\cite{Alkauskas2014} developed a comprehensive framework based on a single effective configuration coordinate and showed that the resulting capture coefficients are in quantitative agreement with experimental measurements for defects in wide-band-gap semiconductors such as GaN and ZnO. Their work highlights that the dominant contribution to the transition rate arises from the collective distortion associated with the change in charge state, thereby justifying the reduction to an effective mode. Complementary to this, Shi \emph{et al.}~\cite{Shi2015} performed a systematic comparison of different theoretical approaches and demonstrated that formulations based on the static coupling picture and an effective scalar configuration coordinate yield capture rates in good agreement with experiment. 
Recent investigations by Waldhoer \emph{et al.}~\cite{WaldhoerDISS} further support this approximation while also clarifying its range of validity. In particular, the effective mode was found to capture the dominant contribution to the transition rate in most cases. While orthogonal phonon modes are not always negligible and may individually contribute with a magnitude comparable to that of the effective mode, their inclusion generally changes the overall transition rate by less than one order of magnitude. These findings indicate that, despite its simplicity, the effective one-dimensional model provides an accurate and computationally efficient description of nonradiative capture processes at defects. Consequently, this effective one-dimensional model forms the basis for the subsequent development. \\

The effective one-dimensional potentials \(V_i(Q)\) and \(V_f(Q)\) are most clearly visualized in a configuration-coordinate diagram, which represents the total potential energy as a function of the scalar configuration coordinate \(Q\). As shown in \Cref{fig:config_coord_2}, this diagram provides an intuitive one-dimensional picture of the structural relaxation process: Following the charge transition, the structural relaxation is described by the scalar configuration coordinate \(Q\), which serves as the reaction coordinate connecting the initial equilibrium geometry at \(Q_i = 0\) with the final equilibrium geometry at \(Q_f = \Delta Q\). Within this representation, the effective one-dimensional potentials \(V_i(Q)\) and \(V_f(Q)\) can be conveniently parametrized by the four parameters
\begin{equation}
\Delta E, \qquad
\Delta Q, \qquad
E_\mathrm{R}^{i} = \tfrac{1}{2}\Omega_i^2 \Delta Q^2, \qquad
E_\mathrm{R}^{f} = \tfrac{1}{2}\Omega_f^2 \Delta Q^2,
\label{eq:effective_NMP_parameters}
\end{equation}
where \(E_\mathrm{R}^{i}\) and \(E_\mathrm{R}^{f}\) denote the relaxation energies associated with structural relaxation on the respective potential-energy surfaces. This parametrization is physically transparent and useful for analytical treatments of nonradiative transitions. Furthermore, an additional convenient parameter is the dimensionless curvature ratio
\begin{equation}
R = \frac{\Omega_i}{\Omega_f}
= \sqrt{\frac{E_\mathrm{R}^{i}}{E_\mathrm{R}^{f}}}.
\end{equation}
The special case \(R=1\) corresponds to equal curvatures, for which the two harmonic potentials have identical shapes and differ only by displacement and energy offset. In contrast, \(R\neq 1\) defines the unequal-curvature case, for which the two harmonic potentials have different shapes and the nuclei therefore experience different restoring forces in the two charge states. \\

As shown in \Cref{fig:config_coord_2}, another important geometric feature of the two potential-energy surfaces is
given by their crossing points
\((Q_\mathrm{X}^\alpha, E_\mathrm{X}^\alpha)\) where \(\alpha\) labels the possible intersections of the two surfaces. The
crossing points are central to NMP transitions because they identify the
configurations at which a charge transfer can occur without changing the total potential
energy of the system. Since the probability of reaching a crossing point is thermally activated, the transition is generally dominated by the crossing with the lowest energy. In the following, we therefore omit the index \(\alpha\) and denote this dominant crossing point simply by \(( Q_\mathrm{X}, E_\mathrm{X})\). The crossing point may be specified either in absolute coordinates
$(Q_\mathrm{X},E_\mathrm{X})$ or in relative coordinates
$(\Delta Q_\mathrm{X},\Delta E_\mathrm{X})$ with respect to the initial equilibrium. Within the coordinate system adopted
throughout this work, these quantities are
numerically identical. Nevertheless, we retain the $\Delta$ notation whenever it is important to
emphasize that the corresponding quantity represents a coordinate or energy
difference, particularly in Arrhenius laws
$\exp(-\Delta E_\mathrm{X}/k_\mathrm{B}T)$, where only energy differences are
physically meaningful. \\

For unequal curvatures, \(E_\mathrm{R}^i  \neq E_\mathrm{R}^f \) the dominant crossing point is located at
\begin{equation}
\Delta Q_\mathrm{X}
=
\Delta Q\,
\frac{E_\mathrm{R}^f - D}
{E_\mathrm{R}^f - E_\mathrm{R}^i},
\qquad
\Delta E_\mathrm{X}
=
E_\mathrm{R}^i
\left(
\frac{E_\mathrm{R}^f - D}
{E_\mathrm{R}^f - E_\mathrm{R}^i}
\right)^2,
\qquad
D=
\sqrt{
E_\mathrm{R}^iE_\mathrm{R}^f
+
(E_\mathrm{R}^i-E_\mathrm{R}^f)\Delta E
}.
\label{eq:crossing_point_unequal_curvatures}
\end{equation}

For equal curvatures, \(E_\mathrm{R}^i=E_\mathrm{R}^f\equiv E_\mathrm{R}\), this simplifies to
\begin{equation}
\Delta Q_\mathrm{X}
=
\Delta Q\,\frac{\Delta E+E_\mathrm{R}}{2E_\mathrm{R}},
\qquad
\Delta E_\mathrm{X}
=
E_\mathrm{R}
\left(
\frac{\Delta E+E_\mathrm{R}}{2E_\mathrm{R}}
\right)^2 .
\label{eq:crossing_point_equal_curvatures}
\end{equation}

Note, that in practical first-principles calculations, the parameters of the effective one-dimensional model given by \cref{eq:effective_NMP_parameters} are usually not constructed by projecting the Hessian matrices onto the relaxation direction. Instead, the effective harmonic potentials are commonly constructed by fitting two parabolas to four total-energy calculations \cite{WaldhoerDISS}. Specifically, the energies of both charge states are evaluated at the relaxed geometries of the initial and final configurations, yielding
$V_i(0)$, $V_f(0)$, $V_i(\Delta Q)$, $V_f(\Delta Q)$. These four energies uniquely determine the harmonic PES. Besides avoiding the computationally demanding evaluation of the Hessian, this procedure is also often preferable from a physical point of view. Whereas a Taylor expansion reproduces the local curvature only in the immediate vicinity of the equilibrium configurations, the four-point construction fits the effective harmonic potentials over the entire interval between the two minima and therefore provides a more faithful effective description when the underlying multidimensional potential-energy surfaces are anharmonic. \\

\section{Nonradiative Charge Capture and Emission Rates}
\label{sec:transition-rates}
\begin{figure}[!hbt] 
\centering 
\begin{subfigure}{0.33\textwidth} 
\centering 
\includegraphics[width=\textwidth]{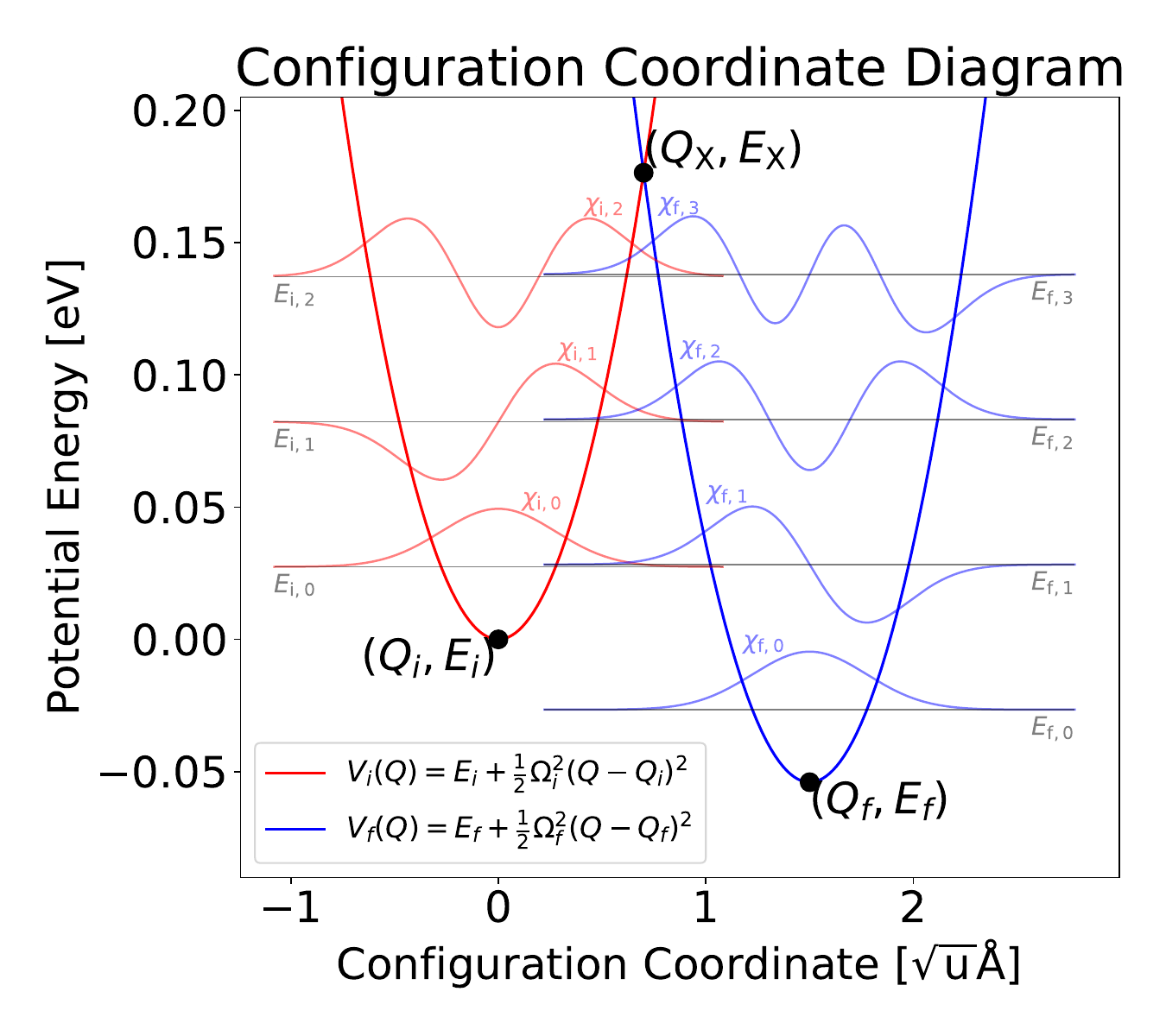} 
\end{subfigure} 
\caption{
Configuration-coordinate diagram highlighting the vibrational states $\chi_{i,m}$ with eigenenergies $E_{i,m}$ on the initial PES and the vibrational states $\chi_{f,n}$ with eigenenergies $E_{f,n}$ on the final PES. The two PES are shown for the equal-curvature case, $R = 1$, so that the corresponding vibrational wavefunctions have identical shapes and differ only by their displacement along the configuration coordinate. The minima of the two surfaces are labeled by $(Q_i,E_i)$ and $(Q_f,E_f)$, while the crossing point is labeled by $(Q_\mathrm{X},E_\mathrm{X})$.
}
\label{fig:phonon}
\end{figure}

Having established an effective one-dimensional Hamiltonian for the description of nonradiative charge transfer processes, we now turn to the calculation of the quantum-mechanical transition rates between the two charge states. To this end, the total Hamiltonian introduced in \Cref{eq:H_1D} is decomposed into an unperturbed part $H_0$, describing the uncoupled electronic and vibrational degrees of freedom, and a perturbation $H_\mathrm{int}$ describing the electron--phonon coupling,
\begin{equation}
H_0 =
|\phi_i\rangle\langle \phi_i|
\left(
\frac{P^2}{2} + V_i(Q)
\right)
+
|\phi_f\rangle\langle \phi_f|
\left(
\frac{P^2}{2} + V_f(Q)
\right), \quad 
H_\mathrm{int}
=
\left(
|\phi_f\rangle\langle\phi_i|
+
|\phi_i\rangle\langle\phi_f|
\right)
W_{if}Q.
\end{equation}
The eigenstates of the unperturbed Hamiltonian $H_0$ are given by direct products of electronic and vibrational states $
|\phi_\alpha\rangle \otimes |\chi_{\alpha,n}\rangle
$, which form the basis for the following perturbative treatment. Here, $|\phi_\alpha\rangle$ denotes the electronic wavefunction, while the vibrational wavefunctions $|\chi_{\alpha,n}\rangle$ are obtained from the effective 1D-Schrödinger equation
\begin{equation}
\left(
\frac{P^2}{2}
+
V_\alpha(Q)
\right)
\chi_{\alpha,n}(Q)
=
E_{\alpha,n} \, \chi_{\alpha,n}(Q),
\end{equation}
which describes the nuclear motion under the effective potential $V_\alpha(Q)$ associated with electronic state $\alpha$. Since the effective potentials are harmonic, the corresponding vibrational eigenstates are the well-known harmonic-oscillator wavefunctions. Specifically, since the minimum of the initial potential-energy surface is located at
\((Q_i,E_i)=(0,0)\), the vibrational spectrum of the initial charge state is given by
\begin{align}
E_{i,m}
&=
\hbar\Omega_i
\left(
m+\tfrac12
\right), \\
\chi_{i,m}
&=
\frac{1}{\sqrt{2^m m!}}
\left(
\frac{\Omega_i}{\pi\hbar}
\right)^{1/4}
H_m\!\left(
\sqrt{\frac{\Omega_i}{\hbar}}\,Q
\right)
\exp\!\left[
-\frac{\Omega_i}{2\hbar}Q^2
\right].
\end{align}
In contrast, the final potential-energy surface is characterized by the minimum
\((Q_f,E_f)=(\Delta Q,\Delta E)\), with the corresponding vibrational spectrum given by
\begin{align}
E_{f,n} &=
\Delta E
+
\hbar\Omega_f
\left(
n+\tfrac12
\right), \\
\chi_{f,n}
&=
\frac{1}{\sqrt{2^n n!}}
\left(
\frac{\Omega_f}{\pi\hbar}
\right)^{1/4}
H_n\!\left(
\sqrt{\frac{\Omega_f}{\hbar}}(Q-\Delta Q)
\right)
\exp\!\left[
-\frac{\Omega_f}{2\hbar}(Q-\Delta Q)^2
\right].
\end{align}
Here $H_m$ and $H_n$ denote Hermite polynomials of order $m$ and $n$, respectively. As illustrated in \Cref{fig:phonon}, $\chi_{i,m}$ and $\chi_{f,n}$ describe quantized nuclear vibrations about the equilibrium positions $Q_i=0$ and $Q_f=\Delta Q$ of the initial and final potential-energy surfaces, respectively. In this picture, every electronic charge transition \(i \rightarrow f\) is accompanied by a simultaneous vibrational transition \(m \rightarrow n\). Since the equilibrium atomic configurations of the two charge states generally differ, a purely electronic transition is not possible without a corresponding change in the vibrational state. Consequently, charge transfer proceeds through transitions between the combined vibronic states. \\

 We assume that, within each charge state, the vibrational subsystem stays close to thermal equilibrium. This assumption is generally well justified for nonradiative defect transitions because vibrational relaxation occurs on picosecond time scales (corresponding to typical phonon lifetimes) \cite{Ward2010, Ferry2021, Xiao2022}, whereas the charge trapping phenomena relevant for semiconductor reliability typically evolve on much longer time scales, ranging from approximately $10^{-6}\,\mathrm{s}$ to even $10^{6}\,\mathrm{s}$ \cite{Dobaczewski2004}. These slow capture and emission events are responsible for the gradual shifts of device characteristics under electrical stress. Although defects with much faster transition rates also exist, they are generally not the focus of reliability studies. Under this assumption, the nuclei can re-equilibrate between successive capture and emission events and the occupation probability of vibrational state $m$ of the corresponding electronic state $\alpha$ is therefore given by a Boltzmann weight
\begin{equation}
w_m^{(\alpha)} =
\frac{\mathrm{e}^{-\beta E_{\alpha,m}}}{Z_\alpha},
\qquad
Z_\alpha = \sum_m \mathrm{e}^{-\beta E_{\alpha,m}},
\qquad
\alpha \in \{i,f\}.
\end{equation}
with $\beta = (k_\mathrm{B}T)^{-1}$. For exceptionally fast defects, however, capture or emission may occur on time scales comparable to vibrational relaxation (i.e., on the order of picoseconds). In this regime, the assumption of thermal equilibrium within each charge state may no longer be valid, and a more general nonequilibrium treatment of the vibrational dynamics is required. \\

Importantly, the elementary transitions occur between the vibronic states
$
|\phi_i\rangle \otimes |\chi_{i,m}\rangle
\rightarrow
|\phi_f\rangle \otimes |\chi_{f,n}\rangle,
$
involving simultaneous changes in both the electronic and vibrational state. However, experiments typically detect only the change in the electronic charge state and not the accompanying vibrational transition. To compute these rates, which describe the reduced dynamics of the electronic subsystem observed experimentally, we treat the electron--phonon interaction perturbatively and trace out the vibrational degrees of freedom (see Supplementary Information~SI-6). This yields a thermally averaged form of Fermi's golden rule, in which the unresolved vibronic transitions are averaged over the equilibrium vibrational ensemble of the initially occupied charge state. Throughout this work, we adopt the convention that the localized defect state is labeled by $i$, while $f$ denotes the state in which the electron is no longer localized at the defect (e.g., a delocalized band state). The resulting electronic emission rate is
\begin{equation}
k_{if} =
\frac{2\pi}{\hbar}
|W_{if}|^2 \,\eta_{if},
\qquad \mathrm{with} \qquad
\eta_{if} =
\sum_m w_m^{(i)}
\sum_n
\left|
\langle \chi_{i,m}|Q|\chi_{f,n}\rangle
\right|^2
\delta(E_{i,m}-E_{f,n}),
\label{eq:k_if_final}
\end{equation}
while the corresponding electronic capture rate is
\begin{equation}
k_{fi} =
\frac{2\pi}{\hbar}
|W_{if}|^2 \,\eta_{fi},
\qquad \mathrm{with} \qquad
\eta_{fi} =
\sum_n w_n^{(f)}
\sum_m
\left|
\langle \chi_{f,n}|Q|\chi_{i,m}\rangle
\right|^2
\delta(E_{f,n}-E_{i,m}).
\label{eq:k_fi_final}
\end{equation}
\Cref{eq:k_if_final,eq:k_fi_final} are the central quantum-mechanical expressions for nonradiative transition rates in the NMP model. They are given by thermally weighted version of Fermi's golden rule accounting for all possible transition between initial vibronic states $
|\phi_i\rangle \otimes |\chi_{i,m}\rangle
$ and final vibronic states $
|\phi_f\rangle \otimes |\chi_{f,n}\rangle
$. The quantum mechanical transition rates are governed by two distinct parts: the electron--phonon coupling matrix element \(W_{if}\), which encodes the strength of the electronic coupling, and the line-shape functions \(\eta_{if}\) and \(\eta_{fi}\), which capture the vibrational contribution to the transition. The latter are determined entirely by the shapes and positions of the diabatic potential-energy surfaces and therefore depend on the parameters \(\Delta E\), \(\Delta Q\), \(E_\mathrm{R}^i\), and \(E_\mathrm{R}^f\). \\

Importantly, the capture and emission rates, as well as the corresponding line-shape
functions, are not independent. Instead, one finds that the ratio between
the forward and reverse transition rates is fixed by detailed balance (see
Supplementary Information SI-1),
\begin{equation}
\frac{k_{if}}{k_{fi}}
=
\frac{\eta_{if}}{\eta_{fi}}
=
\exp(-\beta\Delta F),
\qquad
\Delta F
=
\Delta E
-
T\Delta S_\mathrm{vib},
\qquad
\Delta S_\mathrm{vib}
=
-k_\mathrm{B}
\ln\!\left[
\frac{
\sinh\!\left(\frac{\beta\hbar\Omega_f}{2}\right)
}{
\sinh\!\left(\frac{\beta\hbar\Omega_i}{2}\right)
}
\right].
\label{eq:lsf_detailed_balance}
\end{equation}
where $\Delta F$ is the free-energy difference between
the two vibronic manifolds and and $\Delta S_\mathrm{vib}$ is the change in vibrational entropy arising from the different phonon frequencies of the two charge states. In many practical cases, this contribution is small compared to the energy difference $\Delta E$. For example, at room temperature,
$\hbar\Omega_i = 0.01$~eV and $\hbar\Omega_f = 0.02$~eV result in the vibrational entropy contribution of $T \Delta S_\mathrm{vib} \sim 0.01$~eV. In this limit, the free energy difference reduces to $\Delta F \approx \Delta E$ and the standard detailed-balance relation $k_{if}/k_{fi} = \exp(-\beta\Delta E)$ is recovered. Since the difference between $\Delta E$ and
$\Delta F$ is typically very small, we neglected it in the rest of this work for simplicity. Since capture and emission rates are connected by detailed balance, it is sufficient to evaluate one of them explicitly. \\

The line-shape functions given by \Cref{eq:k_if_final,eq:k_fi_final} can be evaluated explicitly by computing the double sum over the vibrational states of the initial and final potential-energy surfaces. 
At this point it is important to note that
\Cref{eq:k_if_final,eq:k_fi_final} consists of sums over Dirac delta distributions
enforcing energy conservation between discrete vibronic states. Consequently,
the resulting line-shape function forms a discrete Dirac comb rather than a
smooth finite function. This discrete structure is an artifact of reducing the full multidimensional phonon problem to a single configuration coordinate. Physically, the neglected orthogonal phonon modes act as an effective dissipative environment that induces finite vibronic lifetimes and consequently broadens the discrete energy levels. Their effect is commonly incorporated phenomenologically by replacing the delta distributions by Gaussian kernels \cite{May2011, Roos2014},
\begin{equation} 
\delta\left(E_{i,m}-E_{f,n}\right) \mapsto
K_\sigma(E_{i,m}-E_{f,n}) = 
\frac{1}{\sqrt{2\pi}\sigma}
\exp\!\left(-\frac{\left(E_{i,m}-E_{f,n}\right)^2}{2\sigma^2}\right),
\label{eq:k_if_broadened_2}
\end{equation}
This approach is a common practice in  spectroscopy to describe
broadening arising from environmental coupling and leads to physically meaningful finite transition rates. The broadening parameter $\sigma$ effectively encodes the complex influence of the orthogonal vibrational modes into a single phenomenological parameter, capturing their role in inducing finite lifetimes and energy uncertainty of the vibronic states \cite{Landi2024}. While a Gaussian broadening kernel is adopted throughout this work for
its simplicity, other broadening functions,
such as Lorentzian or Voigt profiles, may equally be employed depending
on the underlying broadening mechanism. \\

\begin{figure}[!htb]
    \centering

    \begin{subfigure}{0.32\textwidth}
        \centering
        \includegraphics[width=\textwidth]{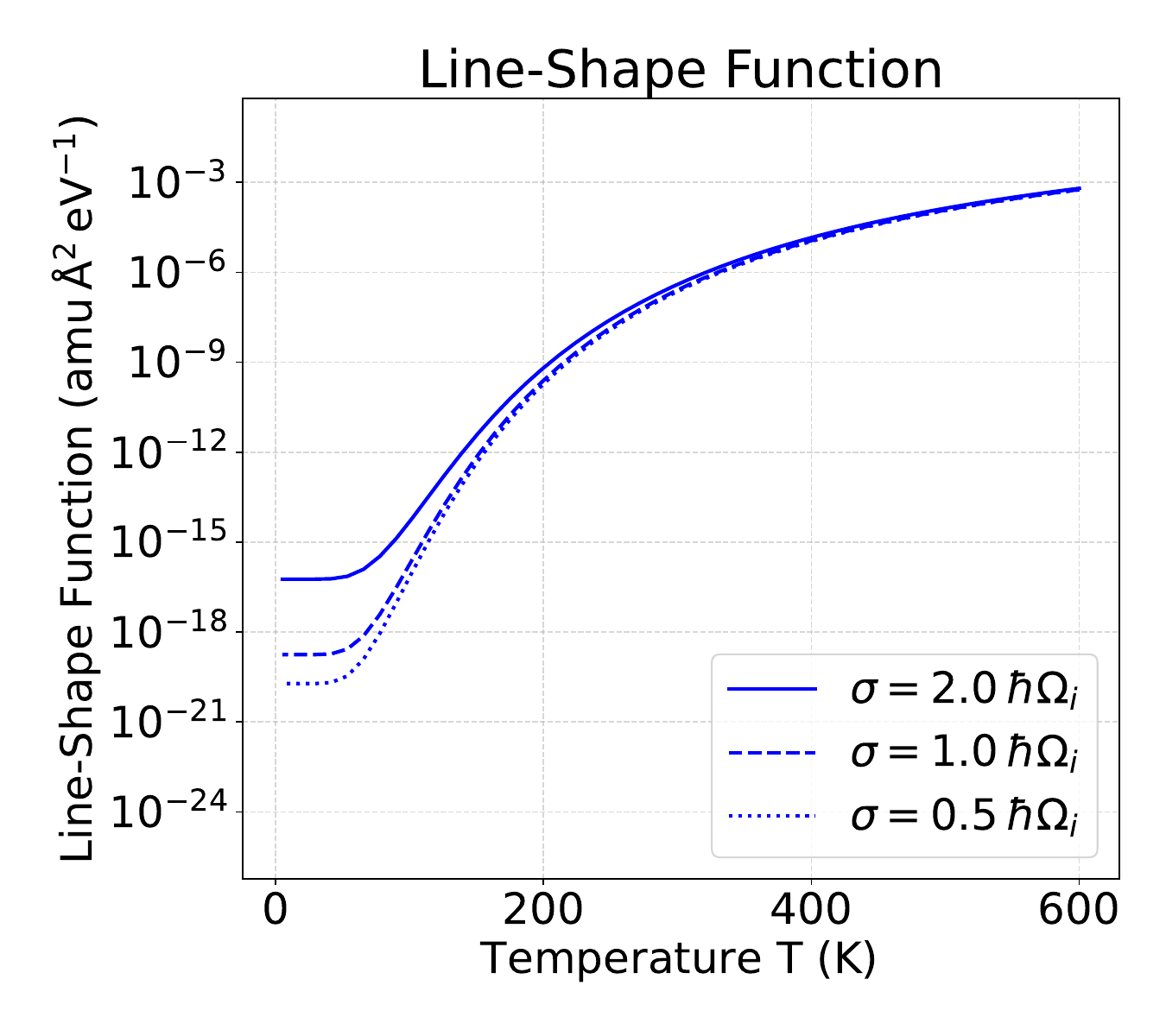}
        \caption{}
        \label{fig:lsf_temperature}
    \end{subfigure}
    \begin{subfigure}{0.32\textwidth}
        \centering
        \includegraphics[width=\textwidth]{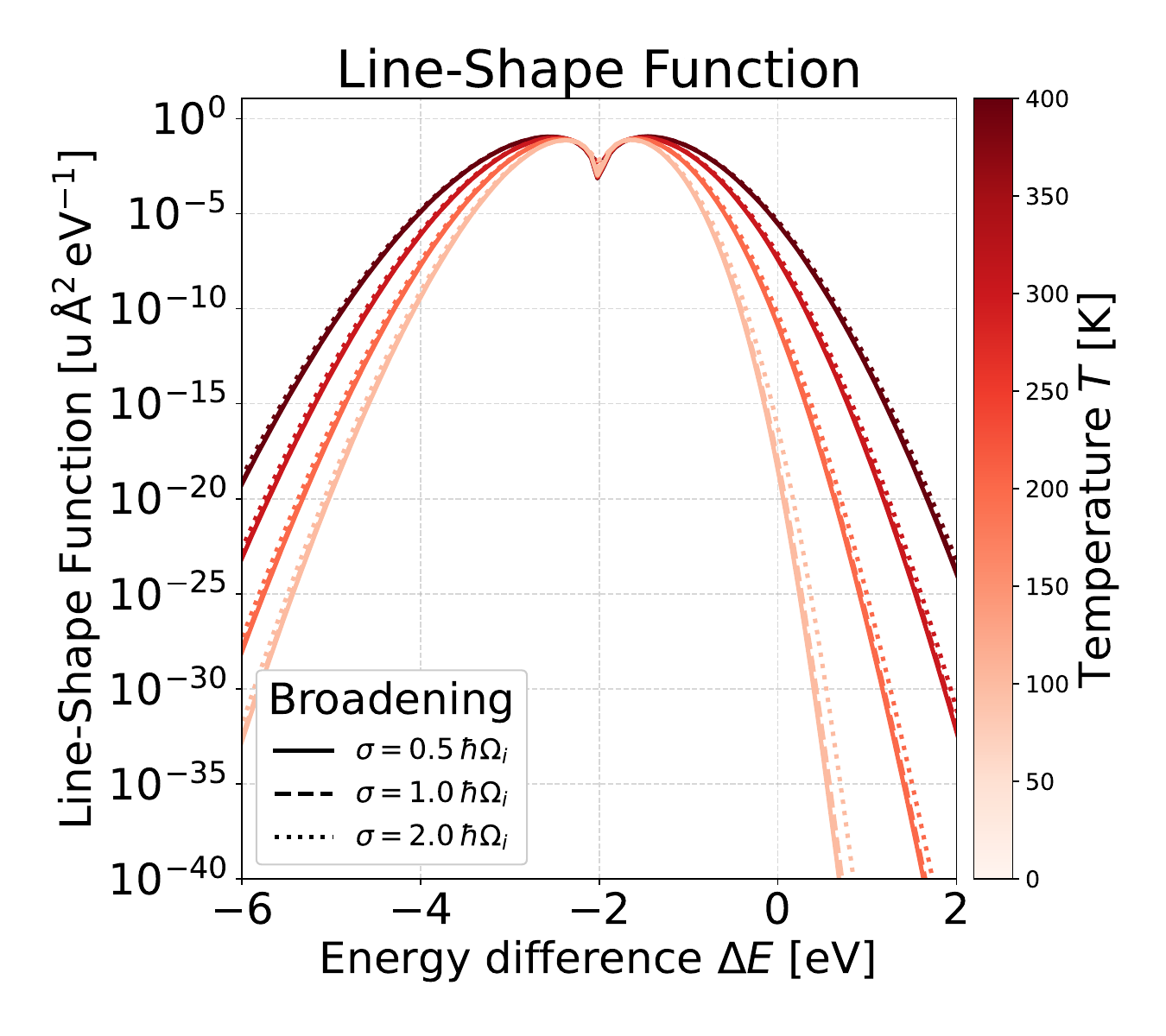}
        \caption{}
        \label{fig:lsf_energy}
    \end{subfigure}
    \caption{
    Line-shape function $\eta_{if}$ for different values of the Gaussian broadening parameter $\sigma$, calculated for $\Delta Q = 4\,\sqrt{\mathrm{u}}\,\text{\AA}$ and $E_\mathrm{R}^i = E_\mathrm{R}^f = 2\,\mathrm{eV}$.
    (a) Line-shape function as a function of temperature for $\Delta E = -1\,\mathrm{eV}$. The curves converge at high $T$, indicating a weak dependence on $\sigma$ due to thermal averaging over many initial vibrational states. At low temperatures, the line-shape function is strongly governed by $\sigma$, since the system predominantly occupies the vibrational ground state ($m=0$), and $\sigma$ controls the number of accessible final states.
    (b) Line-shape function as a function of energy difference $\Delta E$ at fixed temperatures $T = 300\,\mathrm{K}$, $T = 200\,\mathrm{K}$, and $T = 100\,\mathrm{K}$.
    }
    \label{fig:lsf_combined}
\end{figure}

The influence of $\sigma$ on the line-shape function $\eta_{if}$ is illustrated in \Cref{fig:lsf_temperature,fig:lsf_energy}. 
 At high temperatures, many initial vibrational states are thermally populated, and the sum over $m$ averages over a large number of transitions. In this regime, the detailed broadening of individual transitions becomes less important, and $\sigma$ has only a minor influence on the resulting line-shape function. In contrast, at low temperatures the system is predominantly in its vibrational ground state ($m=0$). Consequently, the transition is governed by a single initial state, and the broadening parameter $\sigma$ directly controls how many final vibrational states $n$ are energetically accessible. In the limit $T \to 0$, the broadening therefore plays a decisive role in determining both the magnitude and shape of $\eta_{if}$. For a reasonable description, typical values of the broadening parameter are chosen on the order of the vibrational level spacing, commonly in the range $\sigma \approx 0.5$--$2.0\,\hbar\Omega_i$  \cite{Alkauskas2014, Turiansky2021}. \\

\begin{figure*}[!htb]
    \centering

    \begin{subfigure}{0.32\textwidth}
        \centering
        \includegraphics[width=\textwidth]{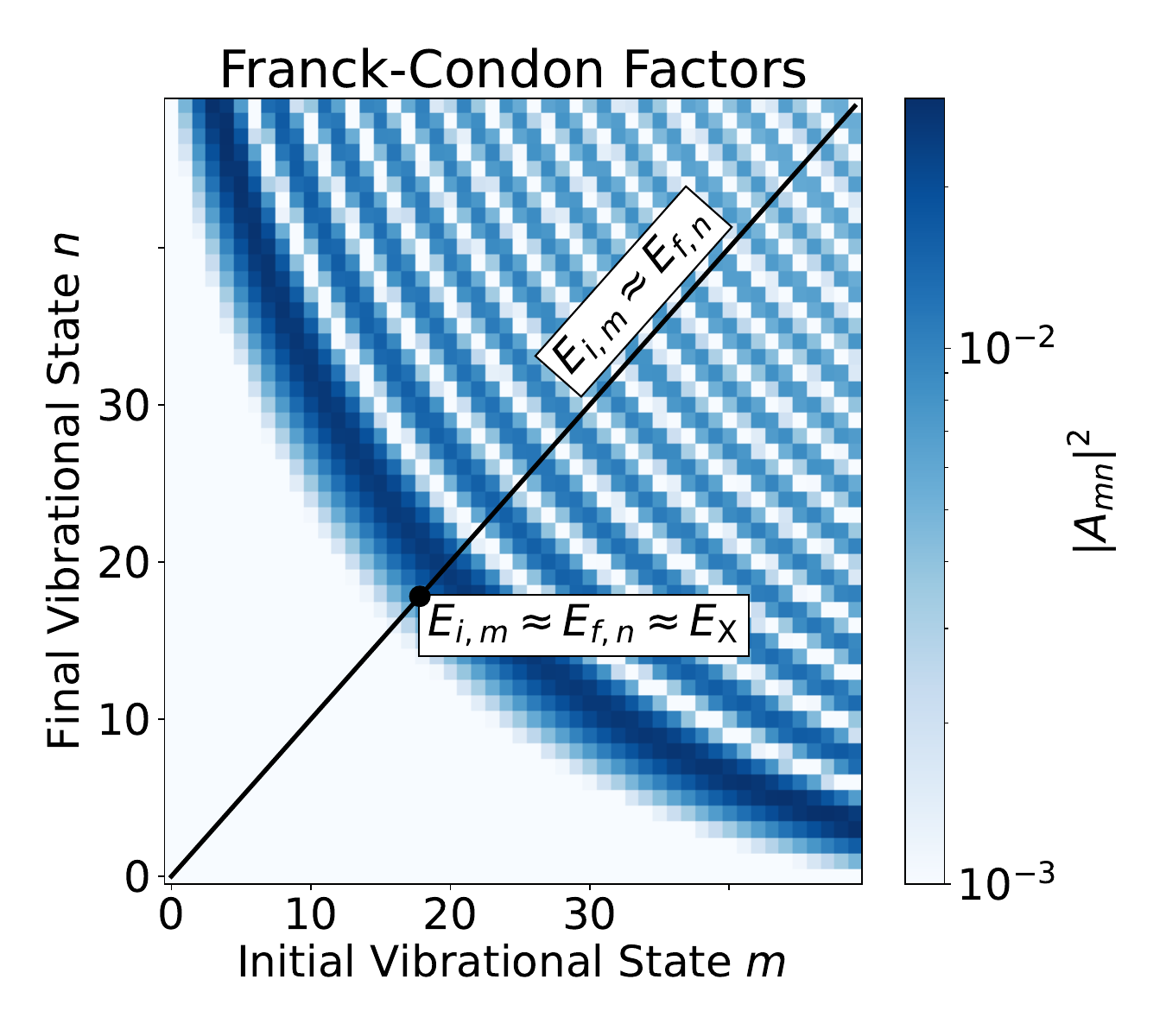}
        \caption{}
        \label{fig:fc_matrix_1}
    \end{subfigure}
    \hfill
    \begin{subfigure}{0.32\textwidth}
        \centering
        \includegraphics[width=\textwidth]{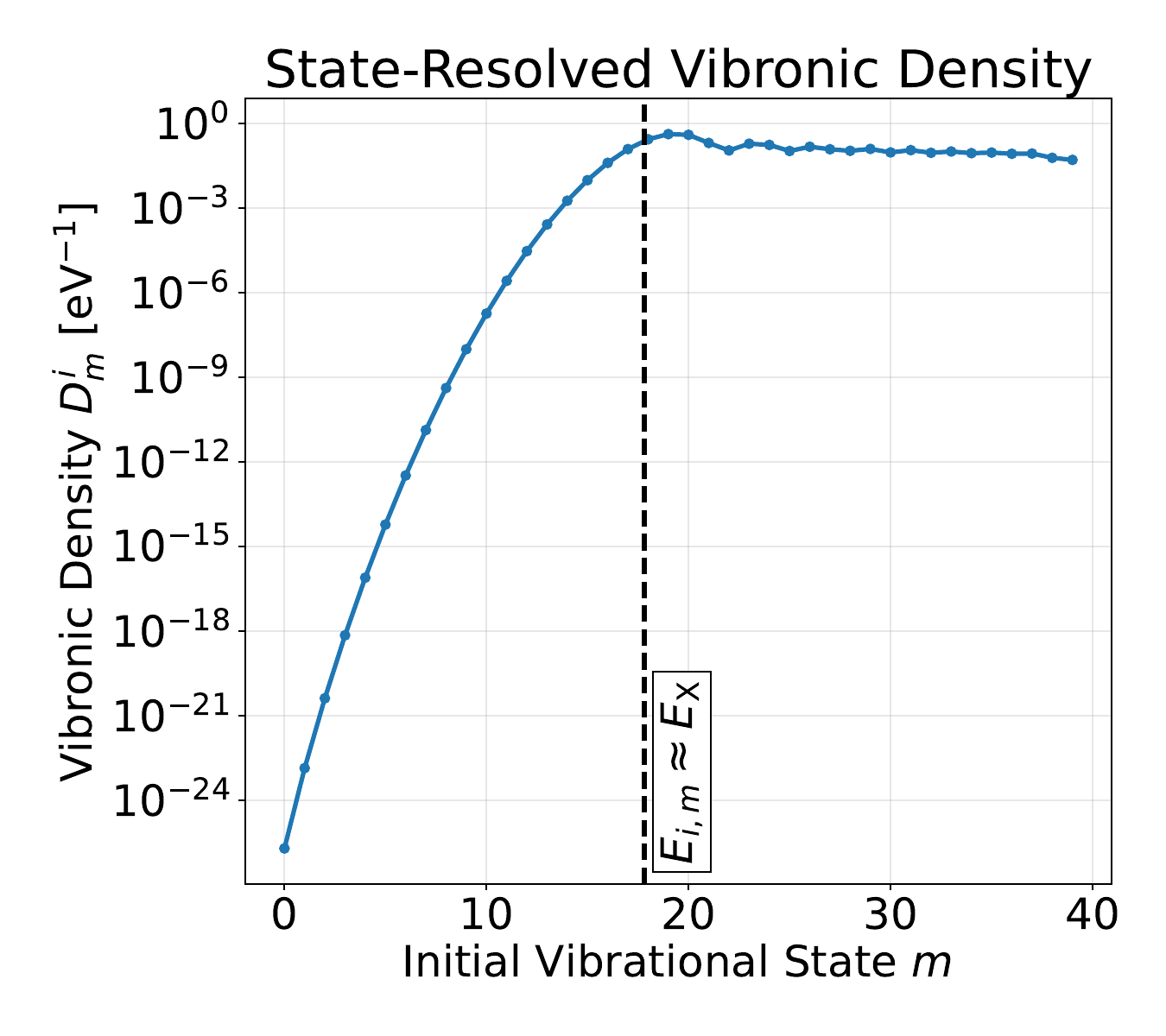}
        \caption{}
        \label{fig:fc_matrix_2}
    \end{subfigure}
    \hfill
    \begin{subfigure}{0.32\textwidth}
        \centering
        \includegraphics[width=\textwidth]{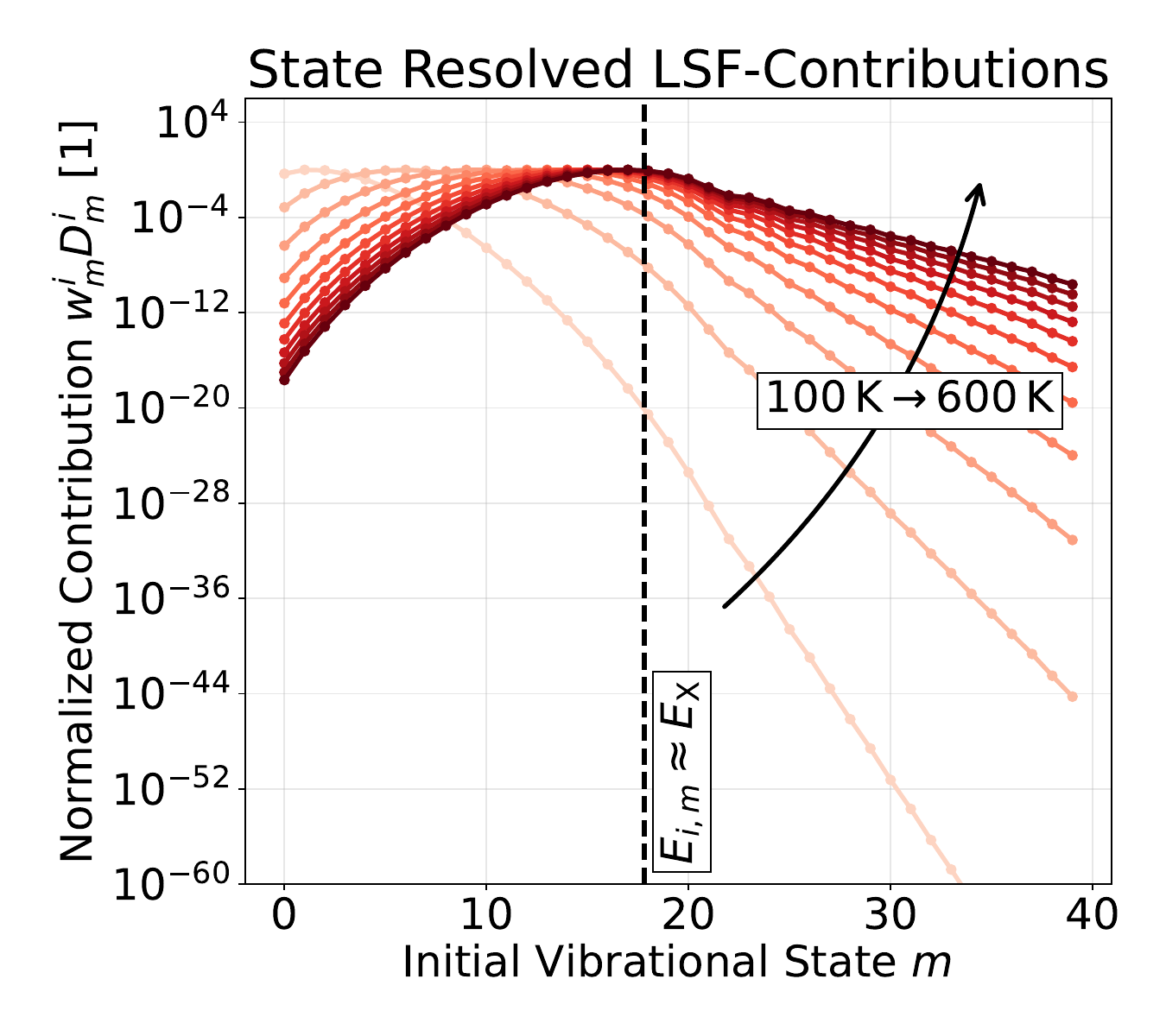}
        \caption{}
        \label{fig:fc_matrix_3}
    \end{subfigure}
    \caption{
    State-resolved contributions to the line-shape
    function for \(\Delta E=0\), \(\Delta Q=3.5\,\sqrt{\mathrm{u}}\,\mathrm{\AA}\),\(E_\mathrm{R}^i=E_\mathrm{R}^f=3.5\,\mathrm{eV}\) and $\sigma = 1.5 \hbar \Omega_i$.
    (a) Franck--Condon factors \(|A_{mn}|^2\) on a logarithmic color scale. The solid line indicates the condition
    \(E_{i,m} = E_{f,n}\)
    (b) State-resolved vibrational overlap density
    \(D_m^i\) as a function of the
    initial vibrational quantum number \(m\). For \(E_{i,m}\ll E_\mathrm{X}\),
    the density is exponentially suppressed because the corresponding vibrational
    wave functions overlap only through their evanescent tails. For
    \(E_{i,m}\gtrsim E_\mathrm{X}\), the density increases and saturates.
    (c) Normalized contribution \(w_m^{(i)} D_m^i\) for different
    temperatures. At high temperatures, the dominant contribution is located near
    the crossing energy \(E_\mathrm{X}\), whereas at lower temperatures the
    Boltzmann occupation shifts the dominant contribution toward smaller
    vibrational quantum numbers. 
    }
    \label{fig:fc_matrix}
\end{figure*}

With the broadening kernel specified, the remaining task for evaluating the
line-shape function is to evaluate the vibrational matrix elements
\(\langle \chi_{f,n}|Q|\chi_{i,m}\rangle\). Several analytical and numerical
strategies are available for this purpose. For direct evaluation, it is convenient
to express the coordinate operator \(Q\) in terms of the harmonic-oscillator
ladder operators $a_i$ and $a_i^\dagger$ of the initial potential-energy surface,
\[
Q=\sqrt{\frac{\hbar}{2\Omega_i}}\left(a_i+a_i^\dagger\right).
\]
This transforms the line-shape function into
\begin{equation}
\eta_{if}
=
\sum_m w_m^{(i)}
\sum_n
\left|
\sqrt{\frac{\hbar}{2\Omega_i}}
\left(
\sqrt{m+1}\,A_{m+1,n}
+
\sqrt{m}\,A_{m-1,n}
\right)
\right|^2
\delta(E_{i,m}-E_{f,n}),
\label{eq:lsf_fc_QQ0}
\end{equation}
where
\begin{equation}
A_{mn}
=
\langle \chi_{i,m} | \chi_{f,n} \rangle
=
\int \mathrm{d}Q\,
\chi_{i,m}(Q)\,
\chi_{f,n}(Q).
\end{equation}
are the Franck--Condon overlap matrix elements between the vibrational eigenstates. The overlap matrix elements \(A_{mn}\) are shown as a heat map in
\Cref{fig:fc_matrix_1}. Significant overlap occurs only for pairs of vibrational states \((m,n)\)
with sufficiently large quantum numbers, because low-lying states are
localized near the minima of their respective potential-energy surfaces and
overlap only weakly through their evanescent tails (see \Cref{fig:phonon}).
Accurate evaluation of the line-shape function therefore requires Franck--
Condon matrix elements \(A_{mn}\) for relatively high vibrational quantum
numbers. \\

Direct evaluation of the overlap integrals $A_{mn}$ is conceptually straightforward,
but numerically challenging. For large vibrational quantum numbers, the
wave functions become highly oscillatory, requiring a very dense integration
grid to resolve their structure accurately. As a result, direct numerical
integration is computationally expensive and impractical for large-scale
applications. Analytic closed-form expressions for $A_{mn}$ provide an alternative \cite{Chen1990, Iachello1998}, but often involve factorials that become numerically unstable for large vibrational quantum numbers. A more robust approach is to evaluate the Franck--Condon overlaps recursively \cite{Drallos1986, Schmidt2010}, avoiding ill-conditioned factors. In particular we recommend to use the scheme of Schmidt Schmidt~\cite{Schmidt2010} to evaluate the Franck--Condon overlaps, as presented in Supplementary Information SI-2. This recursion remains stable even for high vibrational quantum numbers and for unequal curvatures, \(R \neq 1\).\\

To gain deeper insight into the temperature dependence of the line-shape function, we
consider the special case of equal curvatures, \(R=1\). In this case, the line-shape function can be rewritten as (see Supplementary Information SI-4)
\begin{equation}
\eta_{if}
\propto \Delta Q_\mathrm{X}^2
\sum_m w_m^{(i)} \, D_m^i , \qquad D_m^i =
\sum_n
|A_{mn}|^2
K_\sigma(E_{i,m}-E_{f,n}) .
\label{eq:lsf_R1}
\end{equation}
where we have introduced the initial-state-resolved vibrational overlap density
\(D_m^i\). \Cref{eq:lsf_R1} shows that the line-shape function is governed
by the competition between two factors: the thermal occupation \(w_m^{(i)}\) of
the initial vibrational state and the vibrational overlap density \(D_m^i\). The behavior of \(D_m^i\) is shown in \Cref{fig:fc_matrix_2} as a function
of the initial vibrational state \(m\). For states with
\(E_{i,m} \ll E_\mathrm{X}\), \(D_m^i\) is exponentially suppressed because
the corresponding vibrational wave functions overlap only weakly through
their evanescent tails. In contrast, \(E_{i,m} \gtrsim E_\mathrm{X}\), \(D_m^i\) saturates and roughly stays constant. Thus,
\(D_m^i\) alone favors initial vibrational states with energies close to or
above the crossing point \(E_\mathrm{X}\). The thermal occupation
factor \(w_m^{(i)}\), however, decreases exponentially with increasing
vibrational energy and therefore favors low-lying vibrational states. \\

The competition between \(w_m^{(i)}\) and \(D_m^i\) is illustrated in
\Cref{fig:fc_matrix_3}, which shows the normalized product \(w_m^{(i)}D_m^i\)
for different temperatures. At high temperatures, the dominant contribution
arises from vibrational states with eigenenergies close to \(E_\mathrm{X}\).
The rate is therefore primarily controlled by the thermal occupation of states
near the crossing energy,
 leading to an Arrhenius-type
temperature dependence \(w_m^{(i)} \sim \exp[-\beta \Delta E_\mathrm{X}]\). This corresponds to the classical picture, in which a
transition occurs only when thermal fluctuations provide sufficient energy for
the nuclei to reach the crossing point and overcome the barrier
\(\Delta E_\mathrm{X}\). As the temperature decreases, this classical picture breaks down and the
dominant contribution shifts toward lower quantum numbers, including states
with eigenenergies below \(E_\mathrm{X}\). In the limit \(T\to0\), only the
ground state remains occupied, and the temperature dependence associated with
the Boltzmann weights vanishes. The transition rate therefore saturates to a
value determined by the overlap of the initial ground-state wave function with
the accessible final vibrational states. Thus, as shown in
\Cref{fig:lsf_temperature}, the transition rate crosses over from a pronounced Arrhenius-type behavior at high temperatures to a temperature-independent, nuclear-tunneling-dominated regime at low temperatures. \\

While the direct evaluation of the line-shape function is sufficiently efficient for the accurate description of individual defects, it becomes computationally too demanding when extended to large ensembles of defects, as required in TCAD simulations. For example, simulating a device containing $10^4$ defects using $100\times100$ vibrational states per defect already requires evaluating and storing about $10^8$ Franck--Condon matrix elements, corresponding to roughly $800\,\mathrm{MB}$ of memory in double precision. Even if these matrix elements are cached, the repeated access to such large memory blocks, together with the associated summations, significantly slows down the rate evaluation, typically to several tens of milliseconds. In TCAD applications, where transition rates must be evaluated repeatedly over a wide range of bias conditions, this computational cost renders the direct evaluation of the transition rates impractical. \\

Over the past decades, numerous methods have been proposed for evaluating quantum-mechanical line-shape functions while avoiding the explicit summation over Franck–Condon factors. Among them, the generating-function formalism, or equivalently the time-correlation-function formalism, is one of the most widely adopted approaches \cite{Lax1952,Kubo1955,Landi2024}. In this formalism, the line-shape function is expressed as the inverse Fourier transform of an analytically time correlation function, thereby avoiding the explicit summation over all vibrational transitions. Despite this reformulation, the numerical evaluation of the inverse Fourier integral remains computationally demanding and, for the exponentially small line-shape functions frequently encountered in nonradiative carrier capture, ultimately breaks down due to catastrophic cancellation between the highly oscillatory contributions of the time-correlation function. An alternative is to approximate the inverse Fourier integral using the method of steepest descent (saddle-point approximation). However, the time correlation function generally possesses multiple competing saddle points in the complex plane. Depending on the defect parameters and temperature, different saddle points may dominate the inverse Fourier integral, making a robust and automatic saddle-point approximation difficult to achieve over the entire parameter space. While advanced contour-deformation methods can identify the relevant saddle contributions, their computational cost makes them impractical for large-scale TCAD simulations. \\

These limitations motivate the development of alternative evaluation strategies that avoid the numerical difficulties associated with inverse Fourier transforms while remaining sufficiently efficient for large-scale TCAD simulations. In this paper, two complementary approaches are considered, differing in their level of approximation, computational cost, and suitability for device-level applications:
\begin{itemize}
\item \textbf{Classical high-temperature approximation:}
In the high-temperature limit, the quantum-mechanical rate expressions reduce to classical Arrhenius-type rates governed by the classical activation barriers, which are determined by the crossing point of the diabatic potential-energy surfaces. This approximation is computationally efficient and provides an intuitive physical picture, but it neglects nuclear tunneling. Consequently, it breaks down when the transition is dominated by tunneling rather than thermal activation, in particular at low temperatures, for small displacements $\Delta Q$, or for large relaxation energies $E_\mathrm{R}$.

\item \textbf{Effective crossing-preserving approximation:}
As an intermediate approach, the general unequal-curvature system is mapped onto an effective equal-curvature model that preserves the dominant crossing point of the potential-energy surfaces and thus retains the same classical transition rates. This leads to a closed-form analytic approximation to the quantum-mechanical line-shape functions while retaining essential quantum effects, making it well suited for TCAD applications.
\end{itemize}

\subsection{Classical High-Temperature Approximation}
The classical approximation can be derived from the quantum-mechanical lineshape function by taking the high-temperature limit of the vibrational degrees of freedom. Starting from \Cref{eq:k_if_final}, we consider the regime $k_\mathrm{B}T \gg \hbar\Omega_{i}, \hbar\Omega_{f} $, where many vibrational states are thermally populated and the discrete spectrum becomes quasi-continuous. In this limit, the sums over vibrational states can be replaced by an integral over nuclear configurations weighted by the classical Boltzmann distribution. At the same time, the quantum-mechanical matrix elements reduce to their classical counterpart, such that $|\langle \chi_{i,m}|Q|\chi_{f,n}\rangle|^2 \rightarrow Q^2$. This yields the classical lineshape function \cite{Goes2013}
\begin{equation}
\eta_{if}^{\mathrm{cl}}
=
\int  \mathrm{d}Q \, \rho_i^{\mathrm{cl}}(Q)\, Q^2\,
\delta \!\bigl(V_i(Q)-V_f(Q)\bigr) ,
\qquad
\rho_i^{\mathrm{cl}}(Q) 
=
\frac{1}{Z_i^{cl}}\, \mathrm{e}^{-\beta V_i(Q)},
\qquad
Z_i^\mathrm{cl} = \int \mathrm{d} Q\, \mathrm{e}^{-\beta V_i(Q)}.
\label{eq:eta_classical_broadened}
\end{equation}
This expression represents a classical average over nuclear configurations, where quantum effects such as zero-point motion, discrete vibrational levels, and tunneling are neglected. The delta function enforces energy conservation and restricts the integration to the crossing points $(Q_{\mathrm X}^{\alpha}, E_{\mathrm X}^{\alpha})$ of the potential-energy surfaces. As a result, the integral collapses into a sum over all crossing points  $\alpha$
\begin{equation}
k_{if}^{\mathrm{cl}}
=
\frac{2\pi}{\hbar}
|W_{if}|^2
\sum_{\alpha}
\frac{
\left| \Delta Q \right|
\left(
\Delta Q_{\mathrm X}^{\alpha}
\right)^2
}{
\left|
(E_\mathrm{R}^{i}-E_\mathrm{R}^{f})
\Delta Q_{\mathrm X}^{\alpha}
+
E_\mathrm{R}^{f}\Delta Q
\right|
}
\left(
\frac{\beta E_\mathrm{R}^{i}}{4\pi}
\right)^{1/2}
\exp\!\left[
-\beta
\Delta E_{\mathrm X}^{\alpha}
\right].
\label{eq:k_if_classical}
\end{equation}
Each crossing contributes an Arrhenius factor determined by its activation energy $\Delta E_\mathrm{X}^{\alpha}$. Since the contributions are exponentially
weighted, the rate is usually dominated by the crossing point with the lowest
crossing energy given by \Cref{eq:crossing_point_unequal_curvatures,eq:crossing_point_equal_curvatures}. \\

\begin{figure}[!hbt]
    \centering

    \begin{subfigure}{0.32\textwidth}
        \centering
        \includegraphics[width=\textwidth]{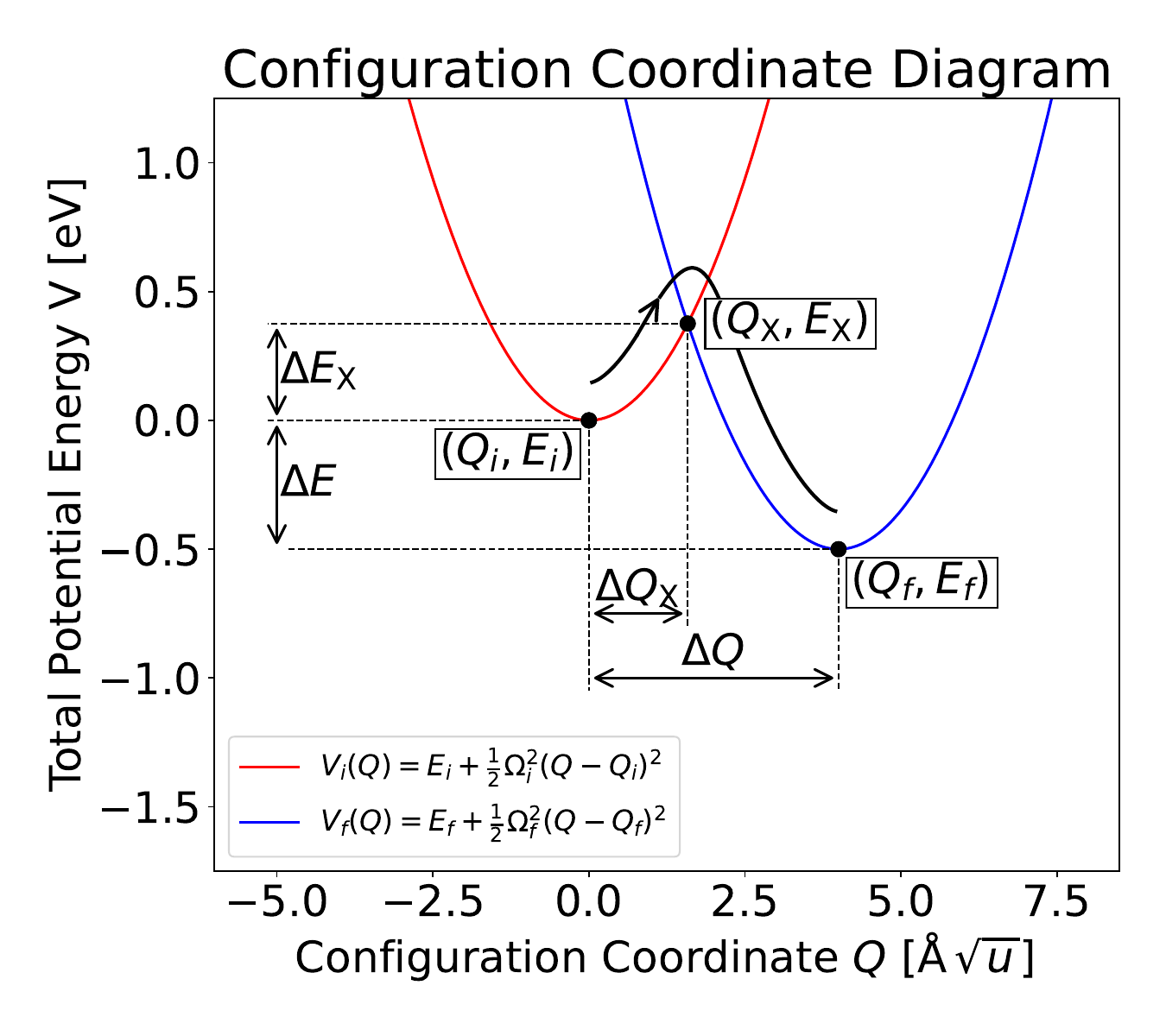}
        \caption{}
        \label{fig:lsf_classic_1}
    \end{subfigure}
    \begin{subfigure}{0.32\textwidth}
        \centering
        \includegraphics[width=\textwidth]{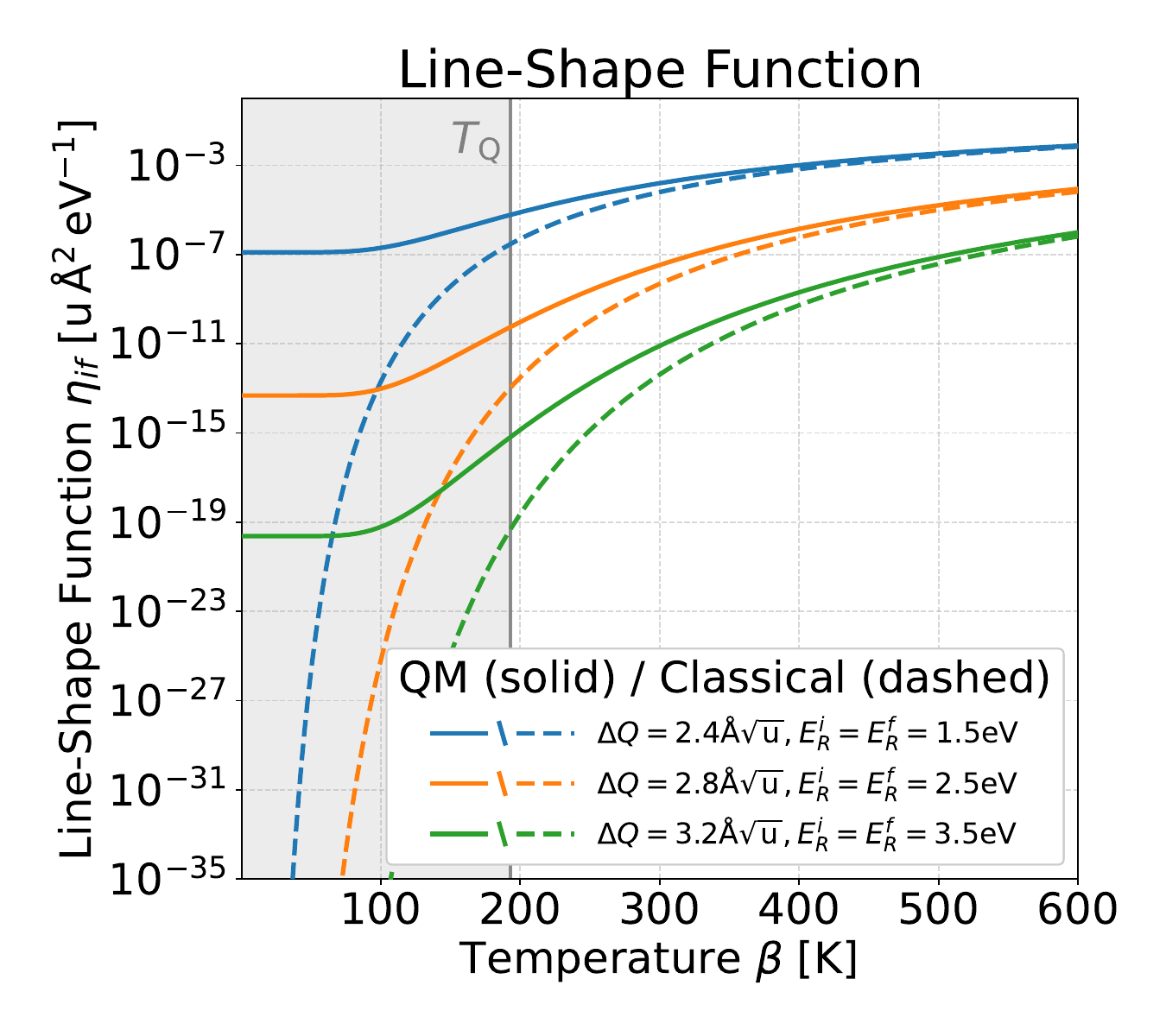}
        \caption{}
        \label{fig:lsf_classic_2}
    \end{subfigure}
    \begin{subfigure}{0.32\textwidth}
        \centering
        \includegraphics[width=\textwidth]{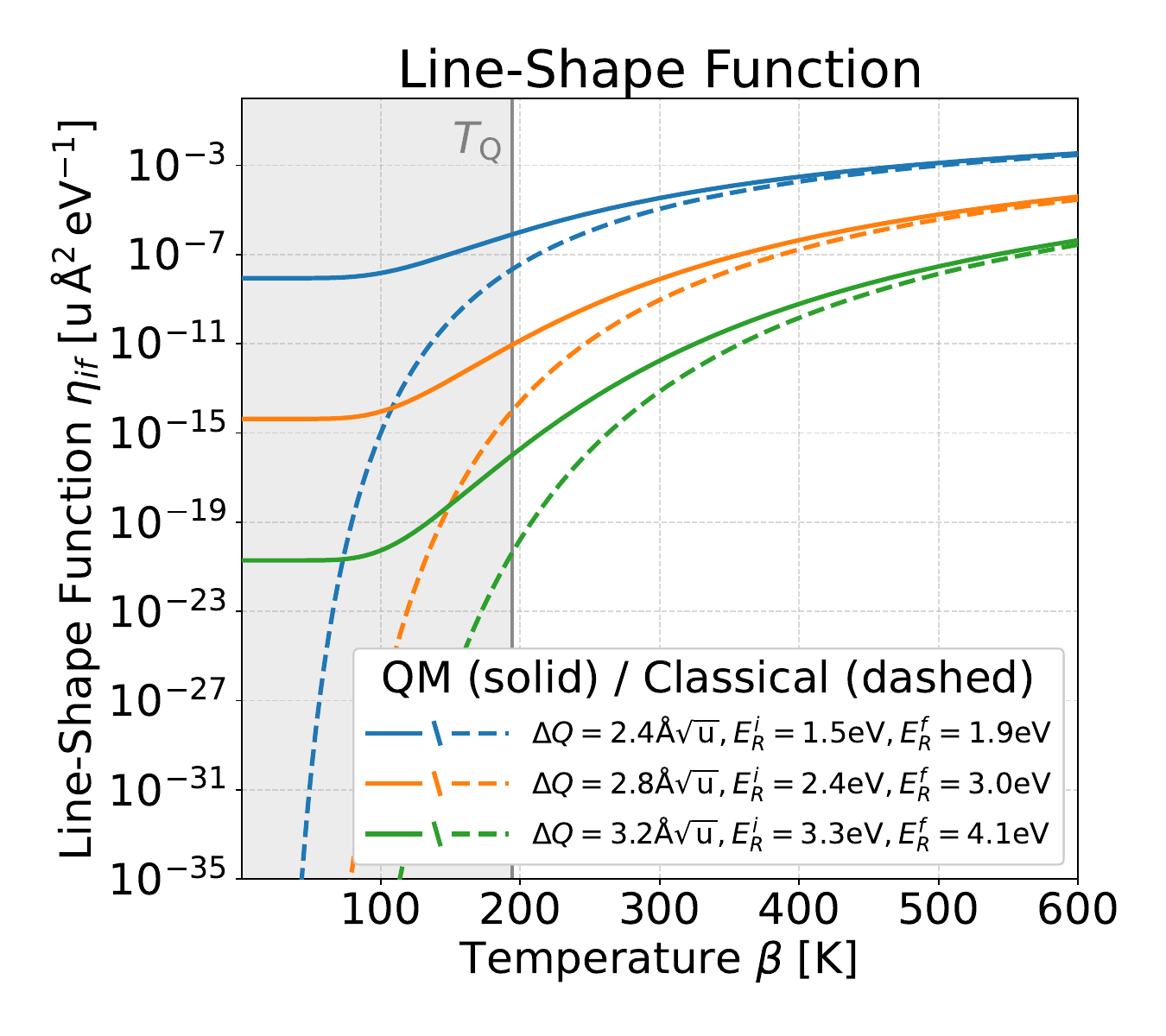}
        \caption{}
        \label{fig:lsf_classic_3}
    \end{subfigure}
    \caption{
    Line-shape function $\eta_{if}$ and configuration coordinate diagram illustrating the classical approximation and its breakdown.
    (a) Configuration coordinate diagram showing the initial and final potential-energy surfaces $V_i(Q)$ and $V_f(Q)$. The arrow highlights the classical transition pathway corresponding to a thermally activated transition over the barrier.
    (b--c) Line-shape function for different temperatures and NMP parameter sets at fixed $\Delta E = -0.25\,\mathrm{eV}$, comparing the full quantum-mechanical result and the classical approximation. Panel (b) shows the equal-curvature case, $E_\mathrm{R}^i = E_\mathrm{R}^f$, while panel (c) shows the unequal-curvature case, $E_\mathrm{R}^i \neq E_\mathrm{R}^f$. At high temperatures both approaches agree and show Arrhenius behavior, while at low temperatures the classical result freezes out whereas the quantum-mechanical one remains finite due to nuclear tunneling.
    }
    \label{fig:lsf_classic}
\end{figure}

\Cref{fig:lsf_classic_1} displays the classical transition path in the configuration-coordinate diagram. In this figure, $(Q_\mathrm{i},E_\mathrm{i})$ and $(Q_\mathrm{f},E_\mathrm{f})$ denote the initial and final equilibrium configurations, respectively, while $(Q_\mathrm{X},E_\mathrm{X})$ denotes the dominant crossing point. The classical picture allows for a simple physical interpretation: the crossing point separates the potential-energy surfaces and defines a classical activation barrier $\Delta E_\mathrm{X}$ for the forward transition. Since a transition can only occur when the system reaches this crossing point, the dynamics reduce to a thermally activated transition over this energy barrier, and the resulting rate takes the form of an Arrhenius law, $k_{if}^{\mathrm{cl}} \propto \exp(-\beta \Delta E_\mathrm{X})$. Importantly, the reverse process is also thermally activated: starting from the final potential-energy surface, the system must overcome the corresponding barrier $\Delta E_\mathrm{X}-\Delta E$, yielding $k_{fi}^{\mathrm{cl}} \propto \exp[-\beta(\Delta E_\mathrm{X}-\Delta E)]$. Thus, both transition directions are governed by activation barriers in the configuration-coordinate diagram. Taking the ratio of the forward and reverse rates immediately gives $k_{if}^{\mathrm{cl}}/k_{fi}^{\mathrm{cl}}=(E_\mathrm{R}^i/E_\mathrm{R}^f)^\frac{1}{2}\exp(-\beta\Delta E)$, which is classical limit of the detailed-balance relation given by \Cref{eq:lsf_detailed_balance}. \\

The classical approximation and its limitations are illustrated in \Cref{fig:lsf_classic_2,fig:lsf_classic_3}, which compares the quantum-mechanical and classical lineshape functions $\eta_{if}$ as a function of temperature for representative NMP parameter sets. At high temperatures, meaning $k_\mathrm{B}T \gg \hbar\Omega_{i}, \hbar\Omega_{f}$, many initial vibrational states are populated and the quantum-mechanical result reduces to the classical expression, exhibiting Arrhenius behavior. In contrast, at low temperatures the classical rate vanishes due to exponential suppression of thermal activation (“freeze-out”), while the full quantum-mechanical lineshape function remains finite. This breakdown arises because the classical approximation entirely neglects quantum-mechanical tunneling through the barrier; it therefore fails when the thermal transition over the barrier becomes improbable and tunneling provides the dominant transition pathway. \\

The crossover between thermally activated transitions over the barrier and quantum-mechanical tunneling through it can be estimated by matching the exponential factors governing both processes (see Supplementary Information~SI-3). This yields a characteristic temperature $T_{\mathrm{quant}}$ below which tunneling becomes the dominant transport mechanism and the line-shape function begins to saturate:
\begin{equation}
T_{\mathrm{quant}}
\sim
\frac{\hbar}{\sqrt{2}\,k_\mathrm{B}\,\Delta Q}
\frac{\sqrt{E_\mathrm{R}^{i}E_\mathrm{R}^{f}}}
{\sqrt{E_\mathrm{R}^{i}}+\sqrt{E_\mathrm{R}^{f}}}.
\label{eq:Tquant}
\end{equation}

\begin{figure}[!hbt]
    \centering
    \begin{subfigure}{0.32\textwidth}
        \centering
        \includegraphics[width=\textwidth]{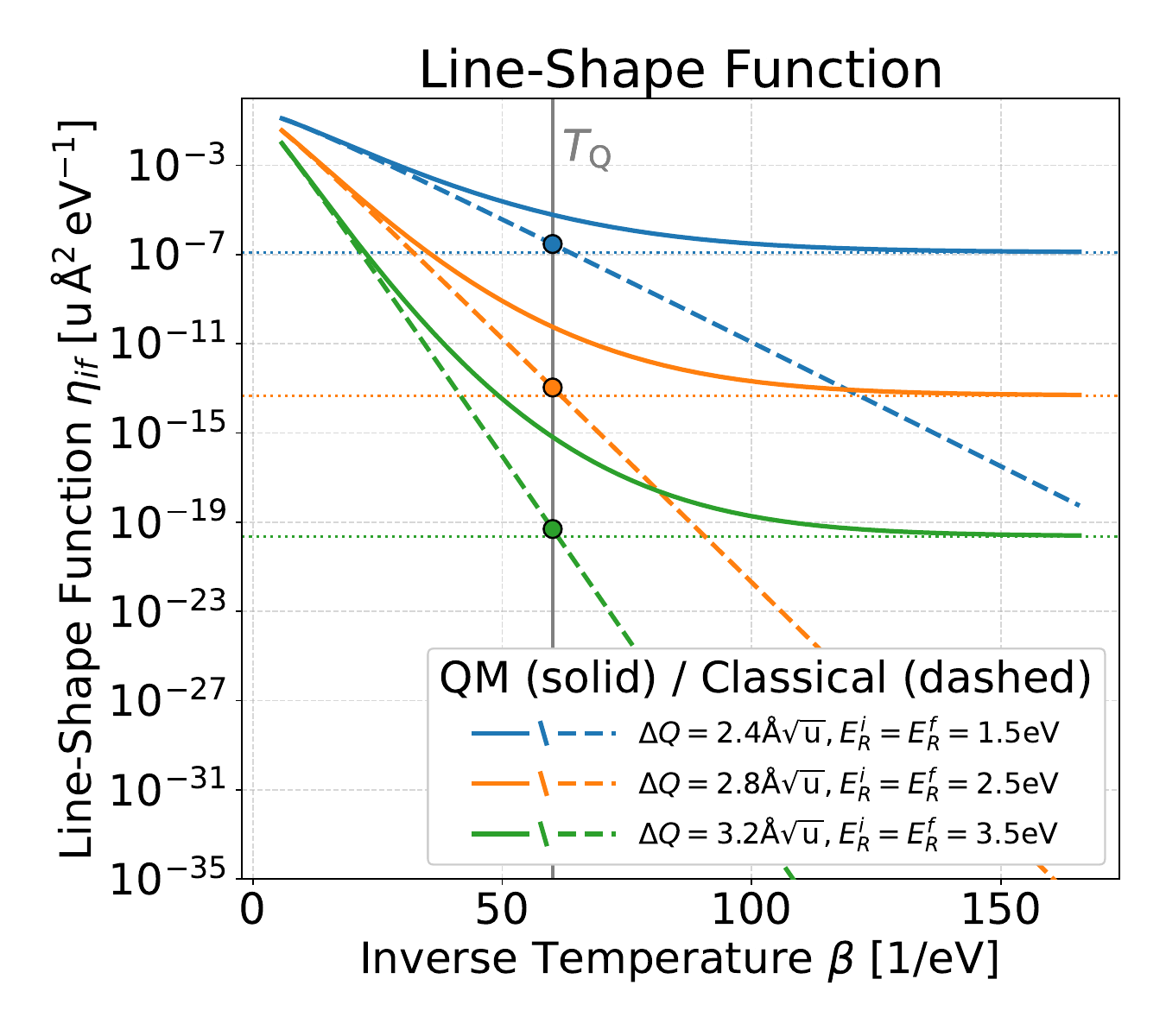}
        \caption{}
        \label{fig:lsf_classic_breakdown_1}
    \end{subfigure}
    \begin{subfigure}{0.32\textwidth}
        \centering
        \includegraphics[width=\textwidth]{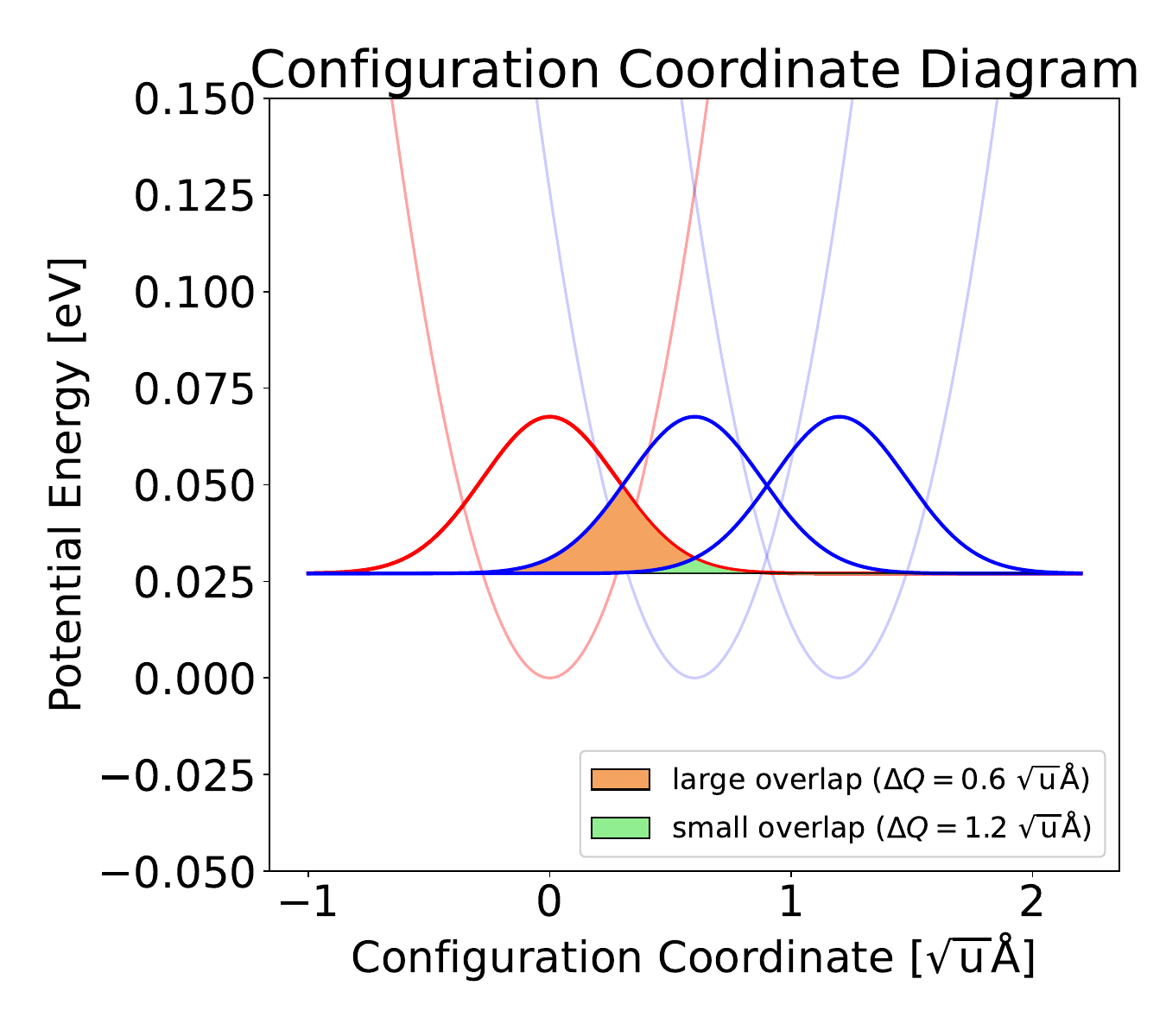}
        \caption{}
        \label{fig:lsf_classic_breakdown_2}
    \end{subfigure}
    \begin{subfigure}{0.32\textwidth}
        \centering
        \includegraphics[width=\textwidth]{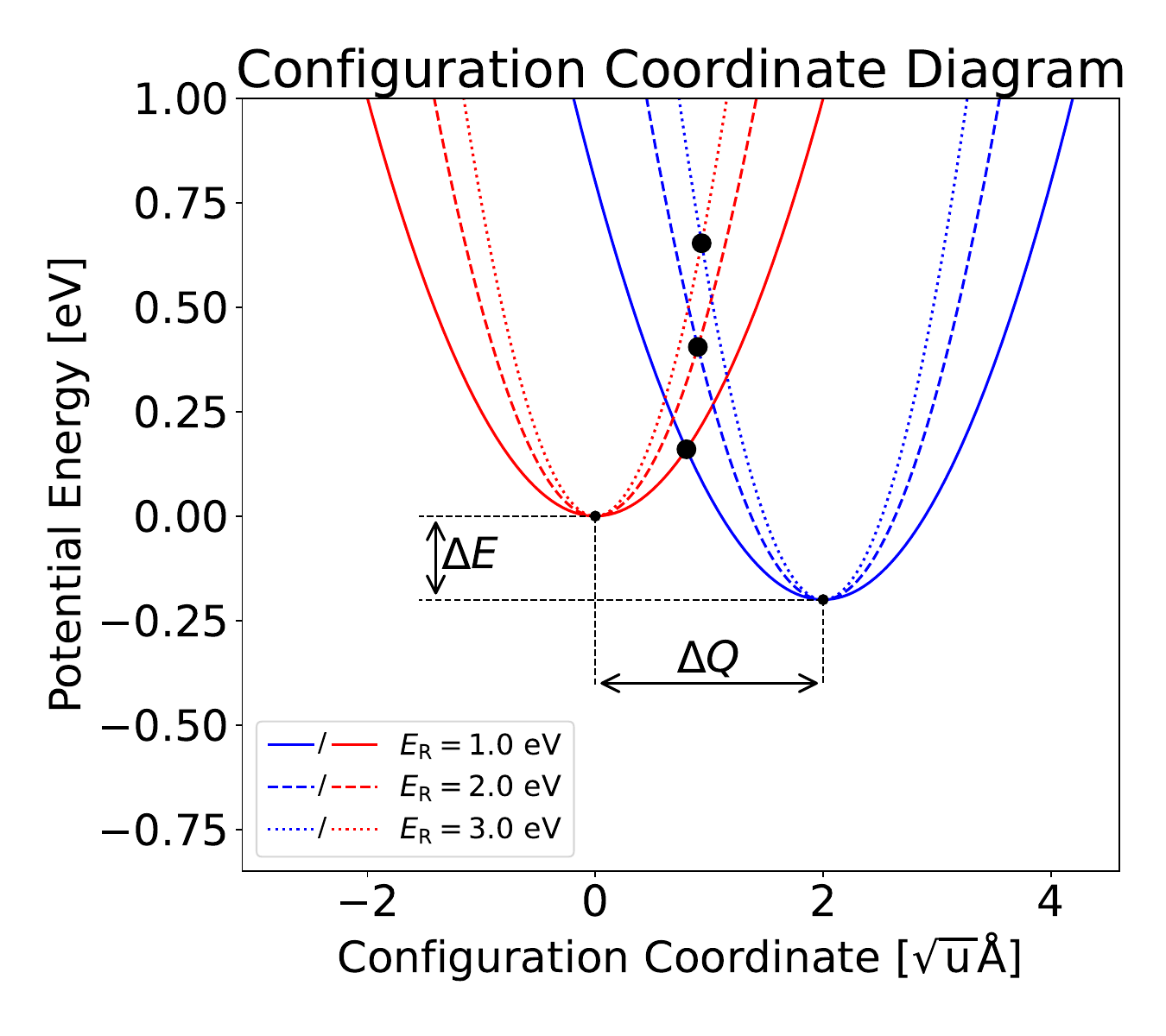}
        \caption{}
        \label{fig:lsf_classic_breakdown_3}
    \end{subfigure}

    \caption{
    Illustration of the breakdown of the classical approximation. (a) Arrhenius plot of the line-shape function comparing the exact quantum-mechanical result (solid), the classical approximation (dashed), and the low-temperature quantum limit (dotted). The characteristic crossover temperature $T_{\mathrm{quant}}$ is indicated by the temperature at which the classical line-shape function reaches the low-temperature limit of the quantum-mechanical result, marking the onset of significant nuclear tunneling. (b) Overlap of the ground-state vibrational wavefunctions for two configurations with $\Delta Q = 0.6~\sqrt{\mathrm{u}}\,\mathrm{\AA}$ (orange) and $\Delta Q = 1.2~\sqrt{\mathrm{u}}\,\mathrm{\AA}$ (green). The shaded region indicates the wavefunction overlap, demonstrating that the tunneling contribution increases with decreasing displacement $\Delta Q$. (c) Configuration-coordinate diagram showing the crossing points for relaxation energies $E_\mathrm{R}=1$~eV (solid), $2$~eV (dashed), and $3$~eV (dotted) at fixed $\Delta E=-0.2$~eV and $\Delta Q=2.0~\sqrt{\mathrm{u}}\,\mathrm{\AA}$. Increasing $E_\mathrm{R}$ raises the classical barrier height, thereby suppressing thermally activated transitions relative to tunneling.
    }
    \label{fig:lsf_classic_breakdown}
\end{figure}

This estimate is an order-of-magnitude criterion and is
valid only in the near-resonant regime, \(\Delta E \approx 0\), and when the
classical crossing barrier is large compared to the vibrational quantum, i.e. 
\(\Delta E_\mathrm{X} \gg \hbar\Omega_i\), so that zero-point corrections can be
neglected. As illustrated in the Arrhenius representation of the line-shape function in \Cref{fig:lsf_classic_breakdown_1}, $T_{\mathrm{quant}}$ approximately corresponds to the temperature at which the classical line-shape function falls below the low-temperature limit of the exact quantum-mechanical result. Below this crossover, the classical approximation increasingly underestimates the transition rate because it neglects nuclear tunneling. In general, $T_{\mathrm{quant}}$ increases as the displacement $\Delta Q$ decreases. This occurs because smaller separations between the potential minima increase the overlap of the vibrational wavefunctions, thereby enhancing tunneling contributions (see \Cref{fig:lsf_classic_breakdown_2}). Additionally, $T_{\mathrm{quant}}$ also increases for larger relaxation energies $E_\mathrm{R}^{i}$ and $E_\mathrm{R}^{f}$, because they raise the classical activation barrier and thereby suppress thermally activated transition relative to tunneling (see \Cref{fig:lsf_classic_breakdown_3}). \\

\Cref{eq:Tquant} demonstrates that, depending on the defect parameters $\Delta Q$, $E_\mathrm{R}^{i}$ and $E_\mathrm{R}^{f}$, the crossover temperature $T_{\mathrm{quant}}$ can reach values well within the temperature range of practical interest rendering the classical approximation inadequate. This is particularly important in device simulations involving broad statistical distributions of defect parameters, where some defects will inevitably fall into parameter regimes in which quantum tunneling remains relevant. Moreover, even in the nominal regime $T > T_{\mathrm{quant}}$, the classical approximation can still deviate from the full quantum-mechanical result by several orders of magnitude. This highlights that the breakdown of the classical description is not abrupt but rather gradual, and that significant quantitative errors may persist well above the estimated crossover temperature. Consequently, the classical approximation should be used only with care and, whenever possible, replaced by the more accurate approximation to the transition rate developed in the next section.

\subsection{Crossing-Preserving Approximation}

In the following, we derive an improved approximation that retains the essential
quantum-mechanical features while remaining suitable for practical TCAD
applications. The general quantum-mechanical rate expressions involve a double
sum over vibrational states (see \Cref{eq:k_if_final,eq:k_fi_final}). While this
double sum is generally not analytically tractable for unequal curvatures of the
potential-energy surfaces, the equal-curvature case admits a closed-form
analytic treatment. Fortunately, first-principles calculations indicate that, the curvatures
of the initial and final potential-energy surfaces are similar for many defects
\cite{Alkauskas2014,Turiansky2021,WaldhoerDISS}. Motivated by this observation, we treat the deviation
from the equal-curvature limit perturbatively, by mapping the original unequal-curvature system with \(R  \approx 1\)  onto an effective equal-curvature model with \(R=1\). The mapping is
constructed such that key quantities governing the transition rate are
preserved. Once this
mapping is established, the transition rates of the effective model can be
evaluated analytically, as shown in the subsequent part of this section. \\

\begin{figure}[!hbt]
    \centering

    \begin{subfigure}{0.33\textwidth}
        \centering
        \includegraphics[width=\textwidth]{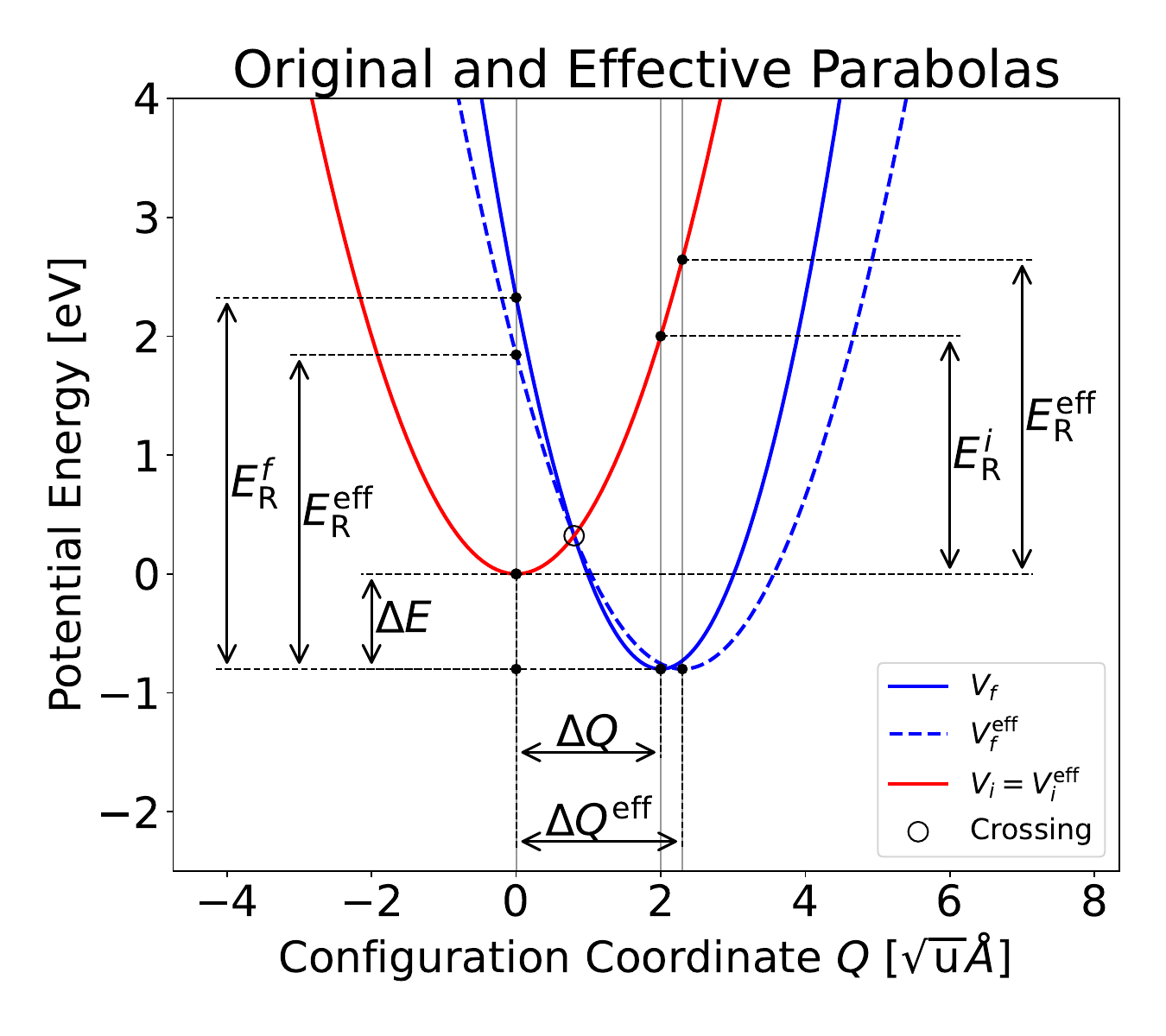}
        \caption{}
        \label{fig:effective_parabolas_2}
    \end{subfigure}
\caption{
Crossing-preserving approximation (CPA). Panel (a) shows the geometric construction of the CPA
for a system with \(\Delta E=-0.4\,\mathrm{eV}\),
\(\Delta Q=2.0~\sqrt{\mathrm{u}}\,\mathrm{\AA}\),
\(E_\mathrm{R}^{i}=1.0\,\mathrm{eV}\), and
\(E_\mathrm{R}^{f}=1.5\,\mathrm{eV}\). The mapping preserves the dominant crossing point
\((\Delta Q_\mathrm{X}, \Delta E_\mathrm{X})\) by replacing the final PES $V_f(Q)$ by an effective equal-curvature surrogate $V_f^\mathrm{eff}(Q)$.
}
    \label{fig:effective_parabolas}
\end{figure}

Although many defects exhibit only moderate curvature mismatch, a naive reduction to an
equal-curvature model is generally not sufficient. For example, simply replacing
the two relaxation energies \(E_\mathrm{R}^i\) and \(E_\mathrm{R}^f\) by their
average value can lead to large errors in the line-shape function \(\eta_{if}\). This sensitivity
arises because \(\eta_{if}\) depends exponentially on the dominant crossing
point \((\Delta Q_\mathrm{X},\Delta E_\mathrm{X})\). In the classical high-temperature limit,the crossing point determines the activation barrier that must be overcome
for a transition to occur, whereas in the quantum-mechanical tunneling regime it
governs the attenuation of the vibrational wave functions and therefore the tunneling probability. It is therefore natural to construct the effective model such that this crossing
point is preserved. This choice ensures that the effective equal-curvature model
retains the dominant nuclear configuration and energy scale that control the
transition rate. As
illustrated in \Cref{fig:effective_parabolas}, requiring both
\(\Delta Q_\mathrm{X}\) and \(\Delta E_\mathrm{X}\) to remain unchanged fixes
the mapping uniquely: the initial PES remains unchanged, whereas the final PES
is renormalized. This asymmetry
follows directly from preserving the crossing point measured relative to the
initial equilibrium configuration. Alternatively, one could construct an
analogous mapping in which the crossing point is preserved with respect to the
final equilibrium configuration. Because the effective
model is constructed from this crossing-preserving geometric mapping, we refer
to it in the following as the crossing-preserving approximation (CPA). \\

In practice, the CPA is constructed in two steps. First, the classical crossing point \((\Delta Q_\mathrm{X}, \Delta E_\mathrm{X})\) is obtained from \Cref{eq:crossing_point_unequal_curvatures} using the parameters \((\Delta E,\Delta Q,E_\mathrm{R}^\mathrm{i},E_\mathrm{R}^\mathrm{f})\) of the original system. Then the effective equal-curvature model is introduced with parameters
\((\Delta E,\Delta Q^\mathrm{eff},E_\mathrm{R}^\mathrm{eff})\), chosen such
that this crossing point is preserved. The required effective displacement $\Delta Q_{\mathrm{eff}}$ follows from inverting the equal-curvature crossing condition in \Cref{eq:crossing_point_equal_curvatures}. For a given crossing configuration, this inversion yields two formal solutions. The physically relevant branch is selected by requiring that the geometric location of the crossing point relative to the two minima is preserved under the mapping, i.e. a crossing located between the minima in the original system remains between the minima in the effective model, while a crossing outside this interval remains outside. This yields
\begin{equation}
\Delta Q_{\mathrm{eff}}=
\begin{cases}
\Delta Q_\mathrm{X}
\left(
1+\sqrt{1-\dfrac{\Delta E}{\Delta E_\mathrm{X}}}
\right),
& 0\le \Delta Q_\mathrm{X}\le \Delta Q, \\[10pt]
\Delta Q_\mathrm{X}
\left(
1-\sqrt{1-\dfrac{\Delta E}{\Delta E_\mathrm{X}}}
\right),
& \Delta Q_\mathrm{X}<0
\;\text{or}\;
\Delta Q_\mathrm{X}>\Delta Q .
\end{cases}
\label{eq:dQeff_mapping}
\end{equation}
Once $\Delta Q^{\mathrm{eff}}$ is fixed, preservation of the crossing energy determines the required effective relaxation energy as
\begin{equation}
E_\mathrm{R}^{\mathrm{eff}}
=
E_\mathrm{R}^{i}
\left(
\frac{\Delta Q^\mathrm{eff}}{\Delta Q}
\right)^2.
\label{eq:EReff_mapping_1}
\end{equation}
This construction preserves not only the crossing configuration itself but also the topology of the potential-energy surfaces along the configuration coordinate, ensuring a physically meaningful effective model.\\

A special situation arises in the limit $(\Delta Q_\mathrm{X}, \Delta E_\mathrm{X})\to(0,0)$, where the inversion formula given by \Cref{eq:EReff_mapping_1} becomes singular and breaks down. This reflects the fact that the two potential-energy surfaces intersect exactly at the minimum of the initial surface, such that the crossing condition fixes only the vertical alignment of the two surfaces but does not uniquely determine their effective displacement. The crossing geometry alone is therefore insufficient to define the mapping in this singular limit, and an additional condition must be imposed. We require that the mapping is continuous as $\Delta Q_\mathrm{X}\to0$, i.e. that no discontinuous jump in the effective parameters occurs when approaching the singular point from finite $\Delta Q_\mathrm{X}$. This is achieved by replacing \Cref{eq:dQeff_mapping} in the singular limit by,
\begin{equation}
\Delta Q_{\mathrm{eff}}
=
\Delta Q
\sqrt{\frac{E_\mathrm{R}^{f}}{E_\mathrm{R}^{i}}}.
\label{eq:EReff_mapping_2}
\end{equation}
This prescription regularizes the singular case and provides a smooth continuation of the general mapping.  In practical implementations, we found that switching
to the singular-limit expression for
\(|\Delta Q_\mathrm{X}| \lesssim 10^{-6}\sqrt{\mathrm{u}}\,\mathrm{\AA}\)
ensures a smooth transition between both branches of the mapping and avoids
numerical artifacts, thereby providing stable and robust simulations.  \\ 

\begin{figure*}[!hbt]
    \centering

    %

    \begin{subfigure}{0.33\textwidth}
        \centering
        \includegraphics[width=\textwidth]{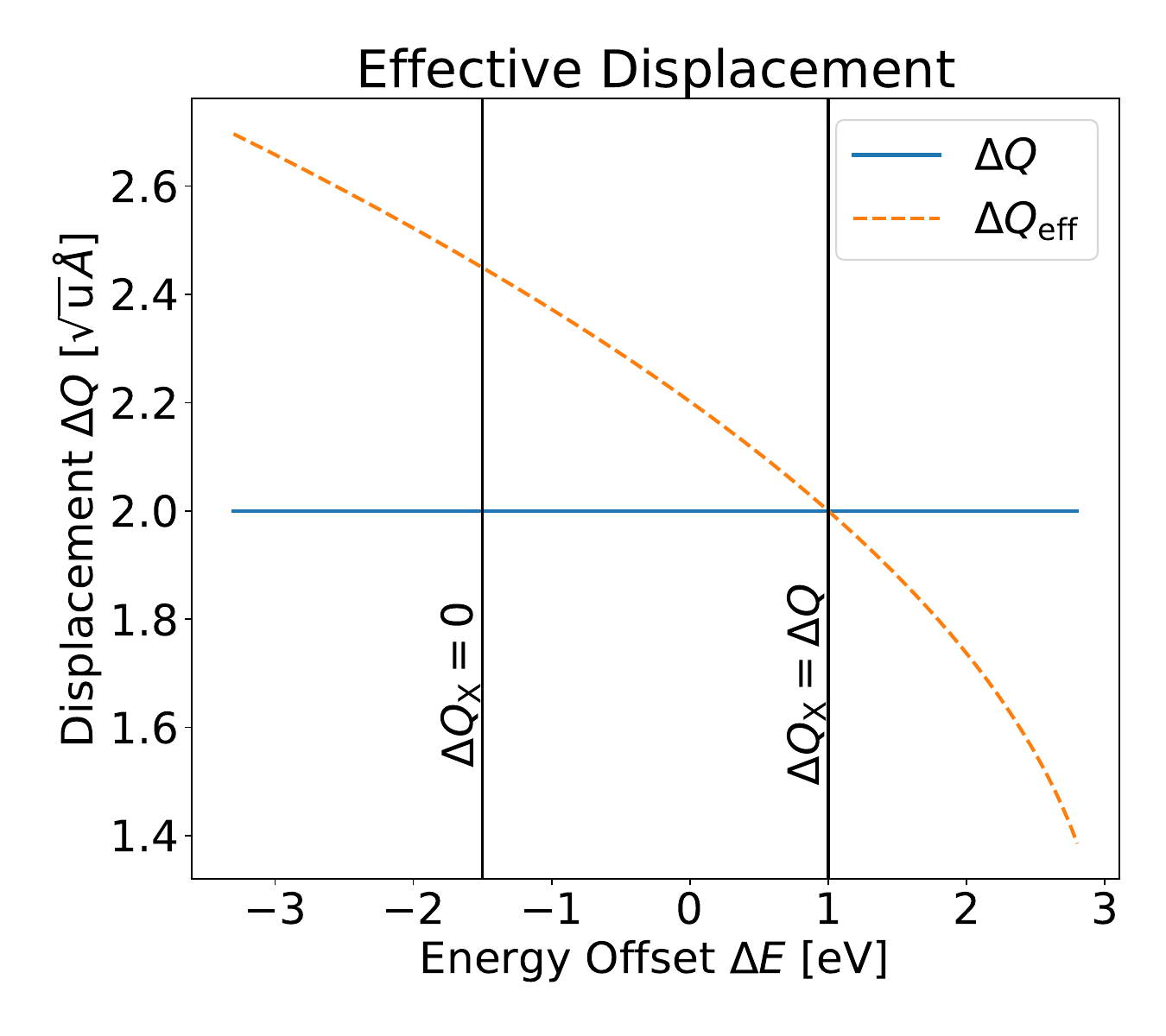}
        \caption{}
        \label{fig:pmapping_3}
    \end{subfigure}
    \begin{subfigure}{0.33\textwidth}
        \centering
        \includegraphics[width=\textwidth]{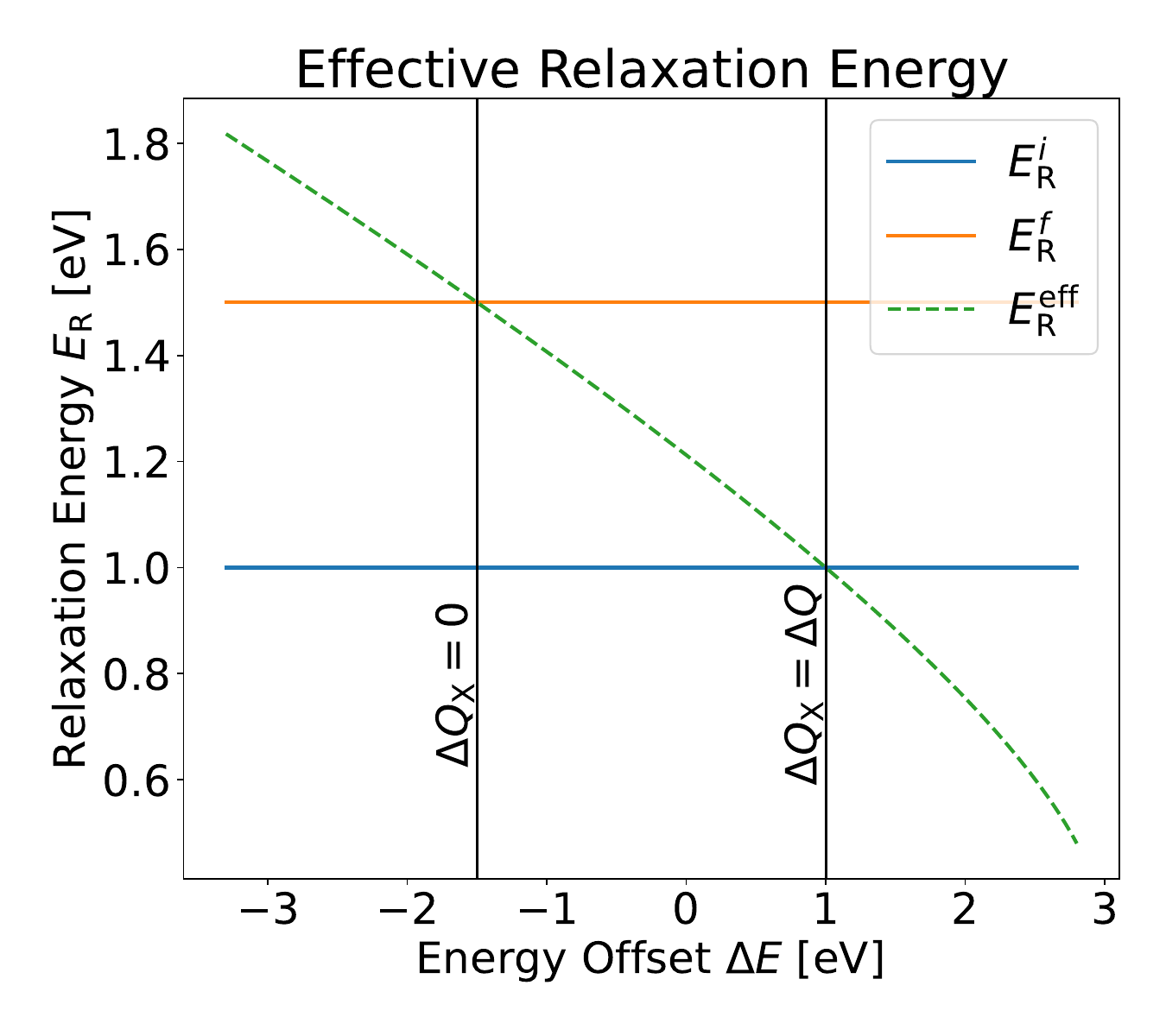}
        \caption{}
        \label{fig:pmapping_4}
    \end{subfigure}

    \caption{
    Mapping of an unequal-curvature system onto an effective equal-curvature model for $\Delta Q=2.0~\sqrt{\mathrm{u}}\,\mathrm{\AA}$, $E_\mathrm{R}^{i}=1.2$ eV, and $E_\mathrm{R}^{f}=1.5$ eV as a function of the energy offset $\Delta E$.  Panels (a) and (b) show the resulting effective displacement $\Delta Q^\mathrm{eff}$ and effective relaxation energy $E_\mathrm{R}^{\mathrm{eff}}$.
    The vertical lines indicate the special limits $\Delta Q_\mathrm{X}=0$ and $\Delta Q_\mathrm{X}=\Delta Q$, which separate the different mapping branches.
    The smooth variation of $\Delta Q^\mathrm{eff}$ and $E_\mathrm{R}^{\mathrm{eff}}$ illustrates the continuity of the mapping, including the regularization of the $\Delta Q_\mathrm{X}\rightarrow0$ case.
    }
    \label{fig:pmapping}
\end{figure*}

The mapping given by \Cref{eq:dQeff_mapping,eq:EReff_mapping_1,eq:EReff_mapping_2} should be understood as a perturbative correction to the
equal-curvature limit. In the special case \(R=1\), where the original system
already has identical curvatures, no renormalization is required and the mapping
reduces to the identity. The effective model then coincides exactly with the
original model, such that
\(\Delta Q^\mathrm{eff}=\Delta Q\) and
\(E_\mathrm{R}^{\mathrm{eff}}=E_\mathrm{R}^i=E_\mathrm{R}^f\). \Cref{fig:pmapping} shows how the mapped parameters $\Delta Q^\mathrm{eff}$ and $E_\mathrm{R}^{\mathrm{eff}}$ vary as a function of $\Delta E$ in order to preserve the dominant crossing point ($\Delta Q_\mathrm{X}$, $\Delta E_\mathrm{X}$). Importantly, preserving the crossing point requires significant renormalization of the effective equal-curvature parameters. Nevertheless, these parameter adjustments occur continuously, ensuring that the effective model evolves smoothly as the original unequal-curvature system is varied. This continuity is essential for avoiding artificial discontinuities in the resulting transition rates. \\

Having constructed the mapping from the original unequal-curvature system onto an effective equal-curvature model, we now proceed to the second step: the analytic evaluation of the transition rate of the resulting effective model. To obtain a practical closed-form expression, the discrete sum over final vibrational states is replaced by a continuum integral, corresponding to the continuum limit of the vibronic spectrum. This replaces the discrete vibronic comb by a smooth spectral envelope and allows the remaining series to be summed analytically (see Supplementary Information SI-4). The approximation is expected to be accurate when the vibronic spectrum is sufficiently dense such that individual vibronic peaks cannot be resolved. Within this continuum approximation one obtains the closed-form expression \cite{Stoneham1981}
\begin{equation}
\eta_{if}^\mathrm{CPA} = 
\frac{\Delta Q_\mathrm X^2}{\hbar\Omega^\mathrm{eff}}
\exp\!\left[-S^\mathrm{eff}(1+2\bar n)\right]
\left(
\frac{\bar n}{1+\bar n}
\right)^{p/2}
I_{|p|}
\!\left(
2S^\mathrm{eff}
\sqrt{\bar n(1+\bar n)}
\right),
\label{eq:LSF_effective_bessel}
\end{equation}
where \(I_{|p|}\) denotes the modified Bessel function of the first kind of order \(|p|\). \\

The parameters entering this expression are
\begin{equation}
\Omega^\mathrm{eff}
=
\frac{\sqrt{2E_\mathrm R^\mathrm{eff}}}
{\Delta Q^\mathrm{eff}},
\qquad
S^\mathrm{eff}
=
\frac{E_\mathrm R^\mathrm{eff}}
{\hbar\Omega^\mathrm{eff}},
\qquad
p=
\frac{\Delta E}{\hbar\Omega^\mathrm{eff}},
\qquad
\bar n=
\frac{1}{e^{\beta\hbar\Omega^\mathrm{eff}}-1},
\end{equation}
where $\Omega^\mathrm{eff}$ is the vibrational frequency of the effective equal-curvature model, which equals $\Omega_i$ by construction; $S^\mathrm{eff}$ is the so-called Huang--Rhys factor characterizing the strength of the structural relaxation; $p$ denotes the energy offset measured in units of the phonon energy; and $\bar n$ is the thermal phonon occupation number. \Cref{eq:LSF_effective_bessel} provides a fully analytic approximation to the original unequal-curvature line-shape function, with all unequal-curvature effects absorbed into the renormalized parameters
$(\Delta E,\Delta Q^\mathrm{eff},E_\mathrm R^\mathrm{eff})$.
\\

\begin{figure*}[!htb]
\centering
\begin{subfigure}{0.32\textwidth}
    \centering
    \includegraphics[width=\textwidth]{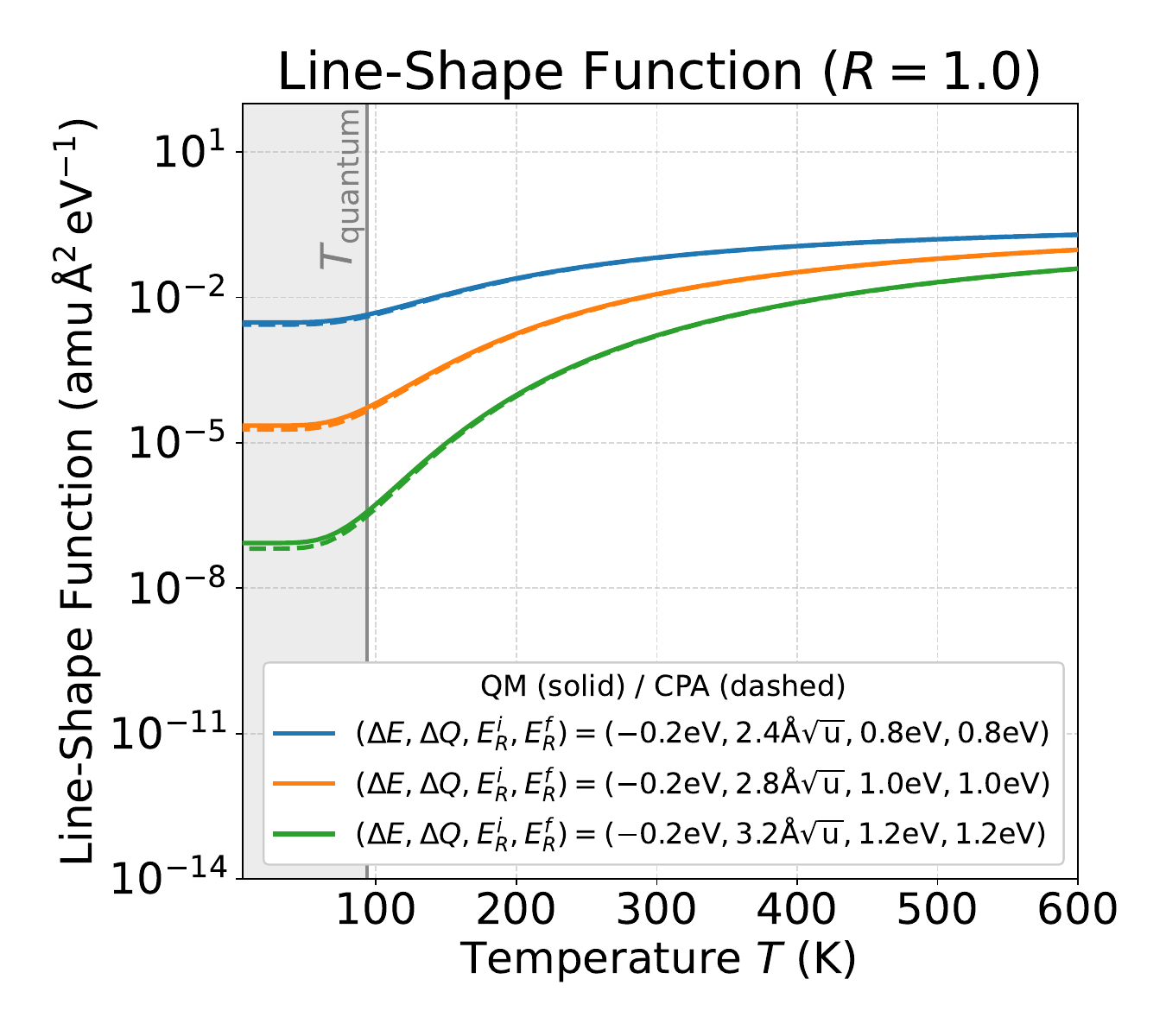}
    \caption{ }
    \label{fig:mapping_1}
\end{subfigure}
\hfill
\begin{subfigure}{0.32\textwidth}
    \centering
    \includegraphics[width=\textwidth]{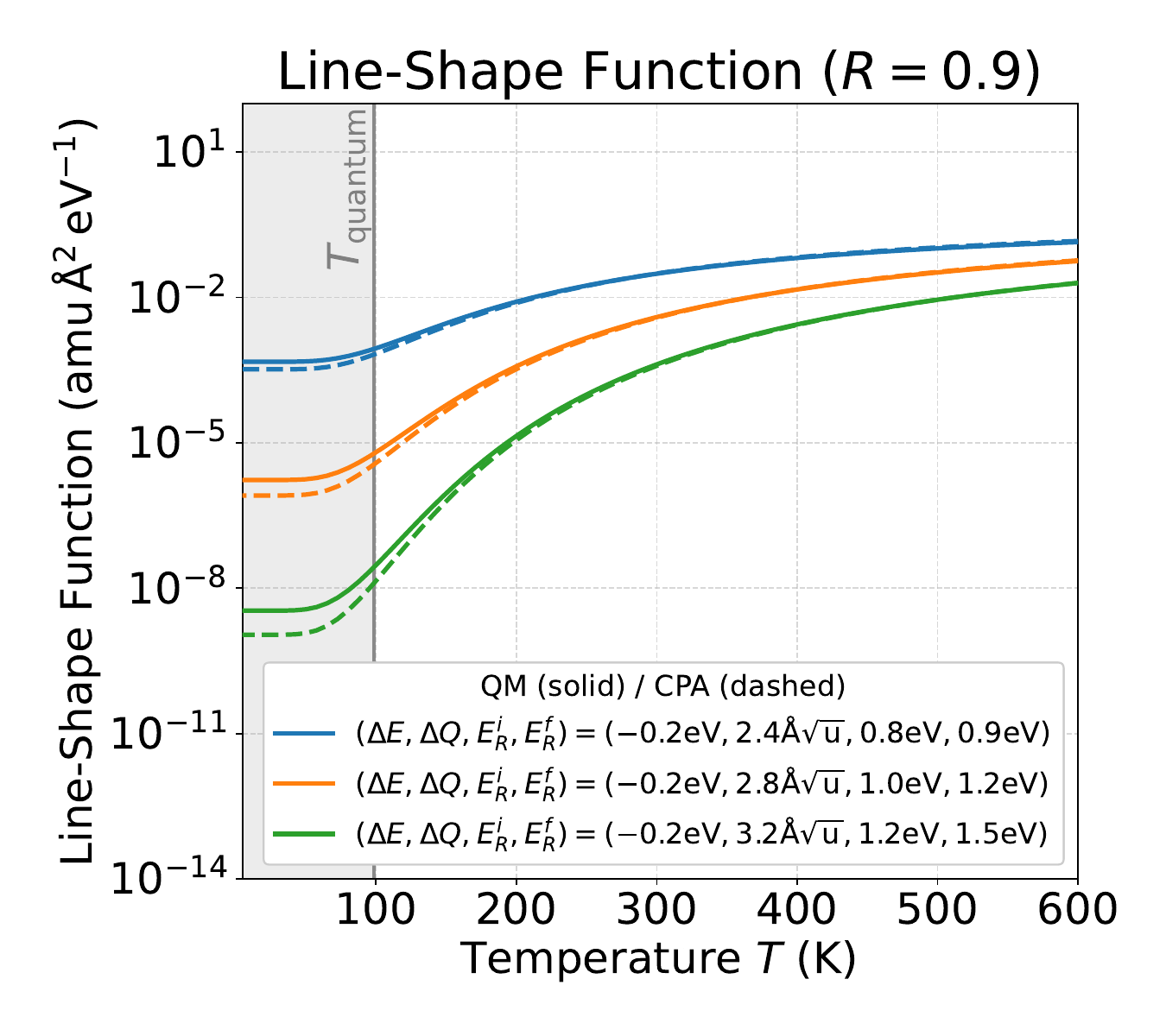}
    \caption{ }
    \label{fig:mapping_2}
\end{subfigure}
\hfill
\begin{subfigure}{0.32\textwidth}
    \centering
    \includegraphics[width=\textwidth]{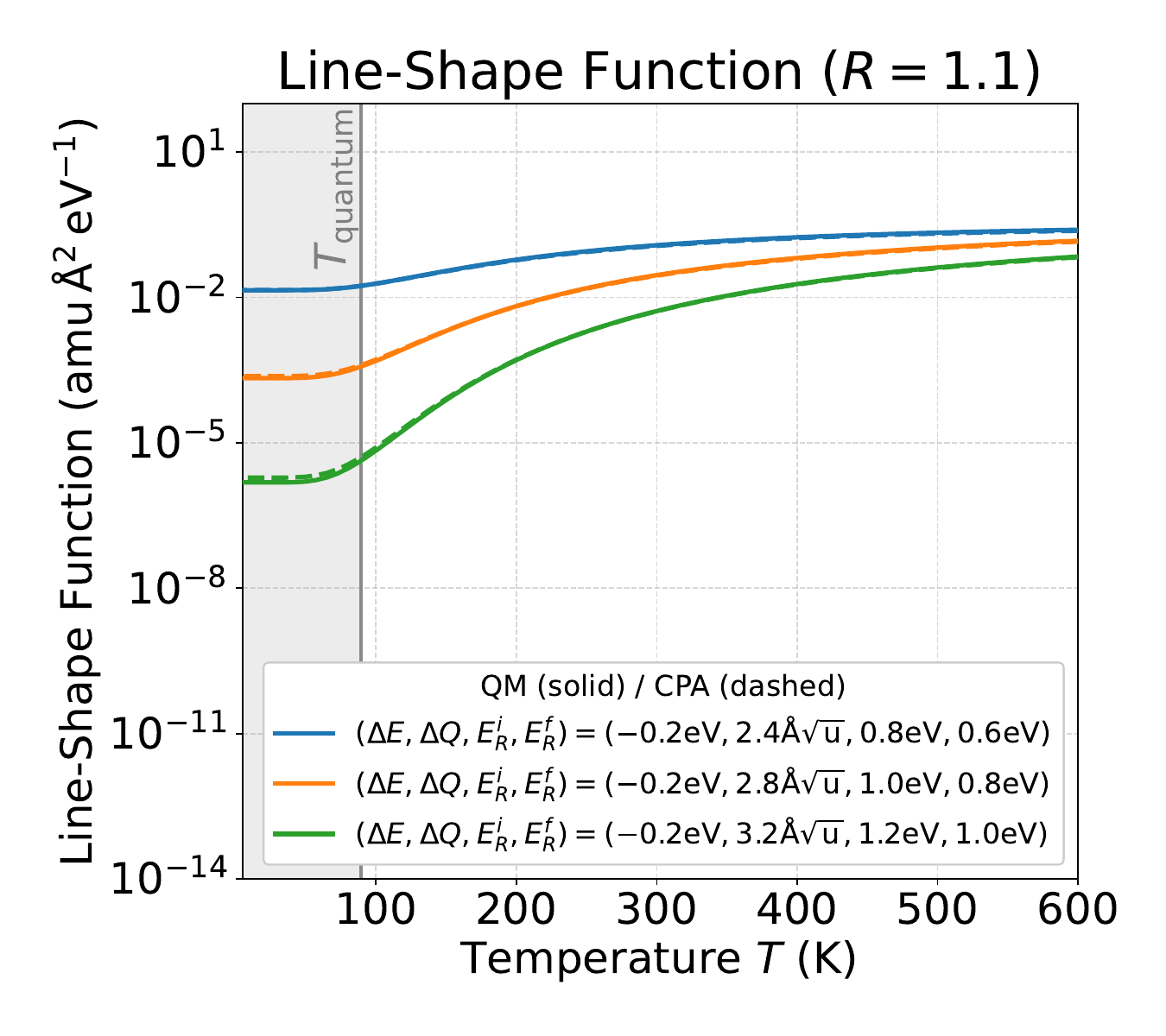}
    \caption{ }
    \label{fig:mapping_3}
\end{subfigure}

\vspace{0.4cm}

\begin{subfigure}{0.32\textwidth}
    \centering
    \includegraphics[width=\textwidth]{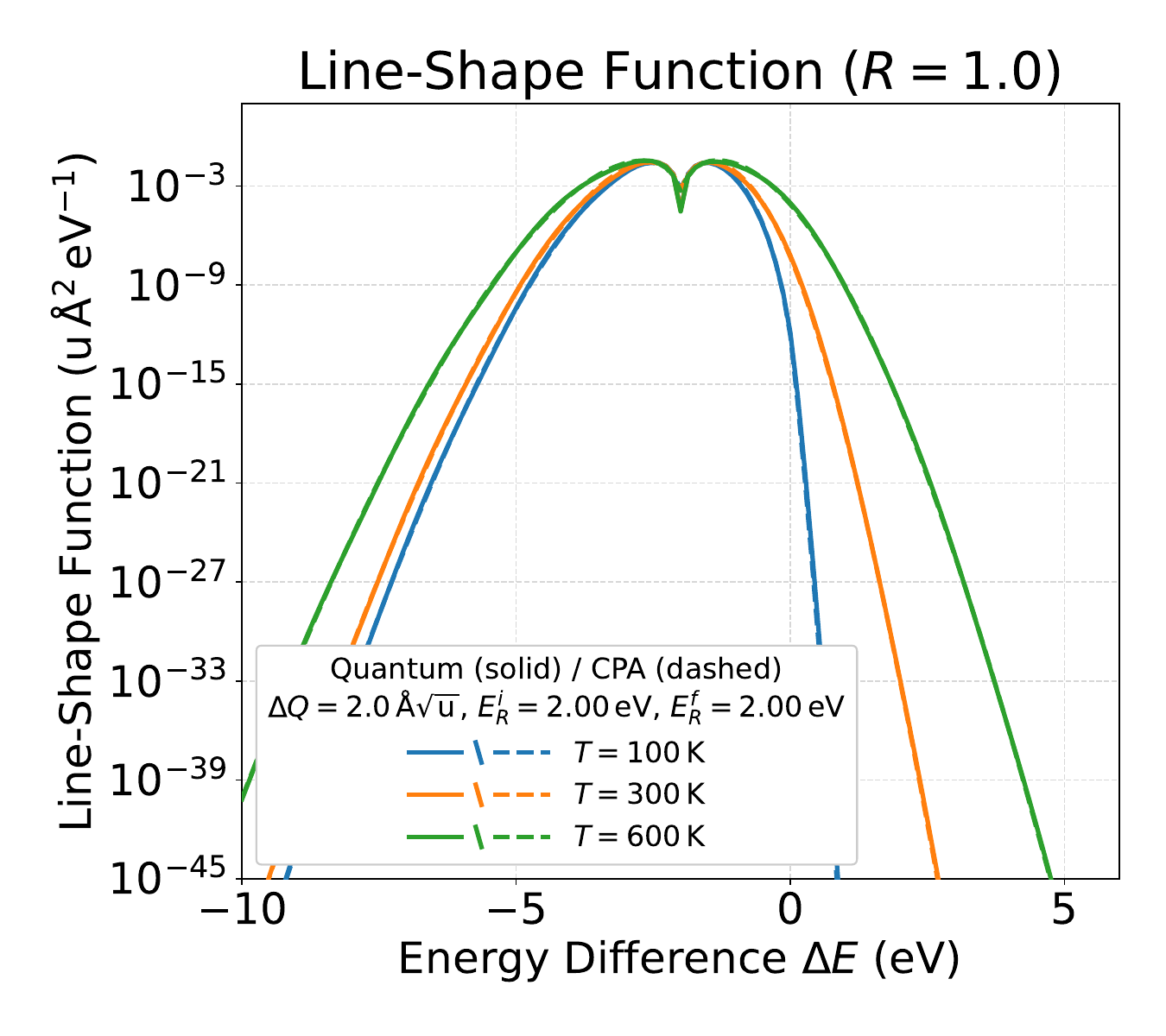}
    \caption{ }
    \label{fig:mapping_4}
\end{subfigure}
\hfill
\begin{subfigure}{0.32\textwidth}
    \centering
    \includegraphics[width=\textwidth]{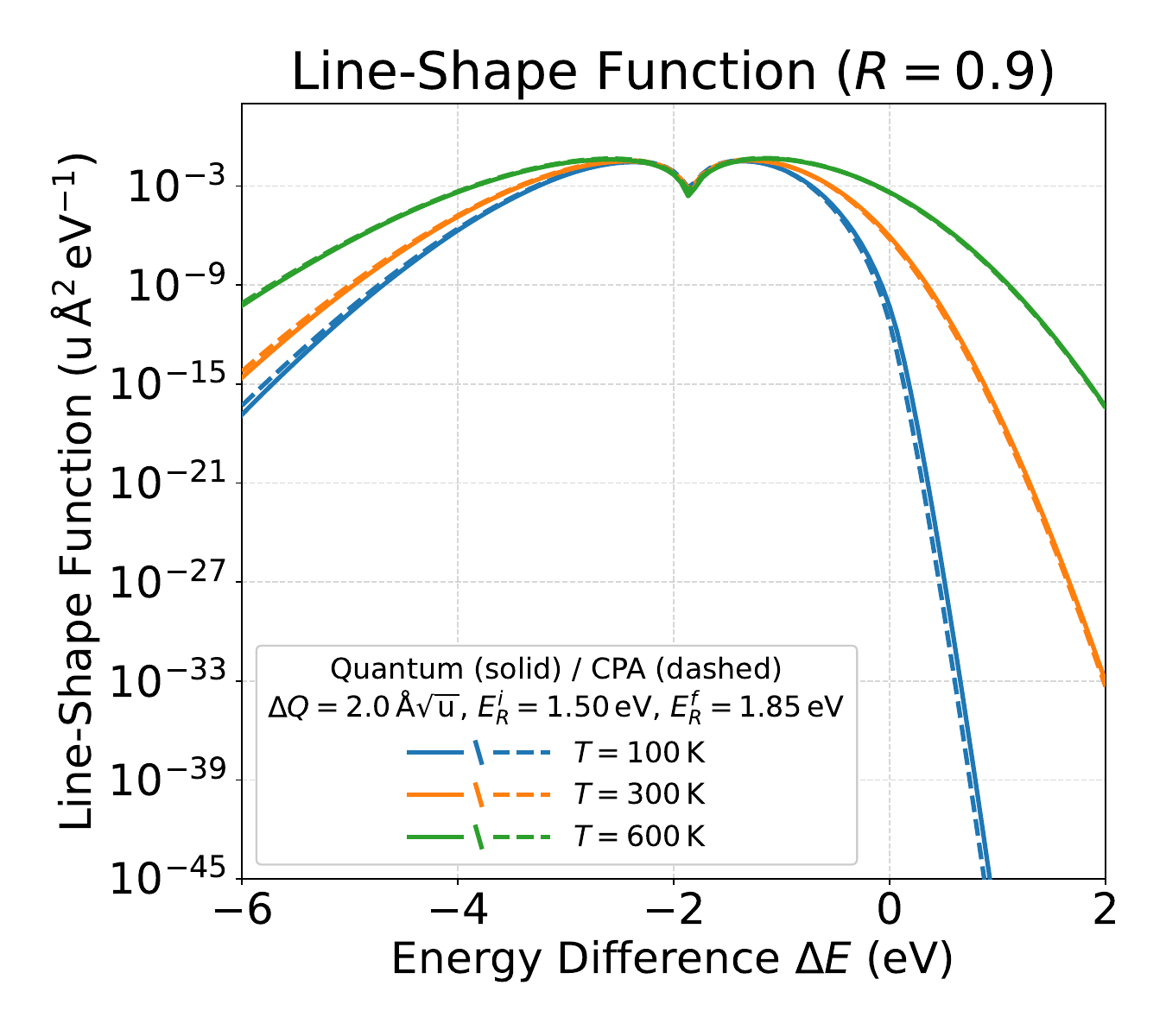}
    \caption{ }
    \label{fig:mapping_5}
\end{subfigure}
\hfill
\begin{subfigure}{0.32\textwidth}
    \centering
    \includegraphics[width=\textwidth]{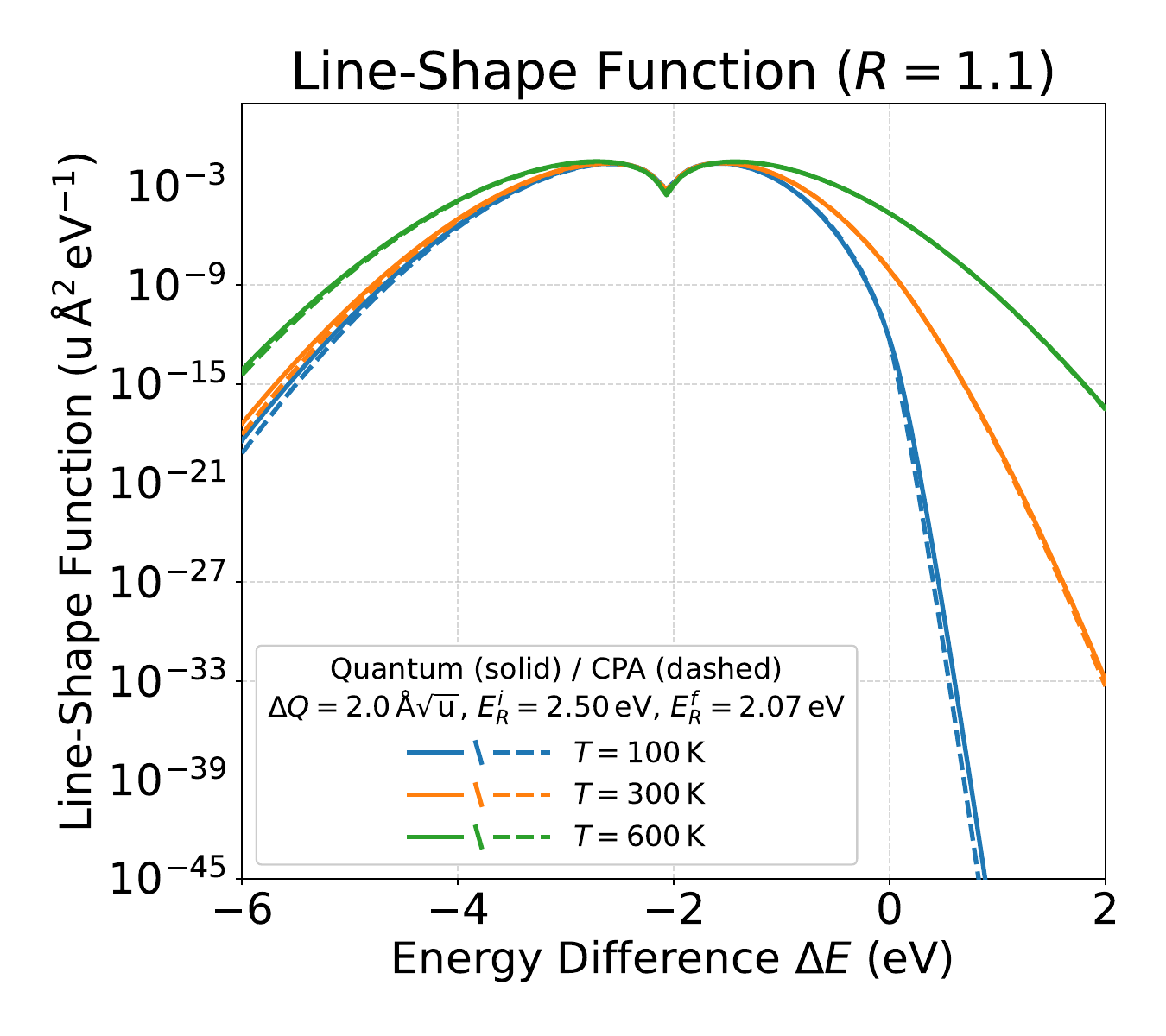}
    \caption{ }
    \label{fig:mapping_6}
\end{subfigure}

\caption{
Line-shape function (LSF) calculated for different sets of model parameters $\{\Delta E,\Delta Q ,E_\mathrm{R}^{i},E_\mathrm{R}^{f}\}$. Panels (a)--(c) show the temperature dependence of the LSF for several parameter sets. Panels (d)--(f) show the dependence of the LSF on the energy offset $\Delta E$
at fixed temperatures $T=100$ K, 300 K, and 600 K for several parameter sets.
}
\label{fig:mapping}
\end{figure*}

The accuracy of the CPA is assessed in \Cref{fig:mapping}, where it is compared with the numerically exact quantum-mechanical result for both equal- and unequal-curvature systems. Panels (a)--(c) show the temperature dependence of the line-shape function for several representative parameter sets, while panels (d)--(f) show the corresponding dependence on the energy offset \(\Delta E\) at fixed temperatures. In all cases, the CPA closely reproduces the exact line-shape function over a broad range of parameters and, unlike the classical approximation, remains well behaved at low temperatures. The largest deviations are observed in the low-temperature limit, where the numerical line-shape function is more sensitive to the empirical broadening parameter \(\sigma\). Depending on the chosen value of \(\sigma\), the exact result exhibits more or less deviation from the CPA. Moreover, the CPA remains accurate for moderate curvature mismatch,
\[
0.8 \lesssim R \lesssim 1.2 ,
\]
covering the majority of technologically relevant defect candidates identified in first-principles studies. Within this regime, preserving the dominant crossing configuration captures the essential unequal-curvature physics governing the transition rates. \\

\begin{figure}[!hbt]
    \centering
    \begin{subfigure}{0.32\textwidth}
        \centering
        \includegraphics[width=\textwidth]{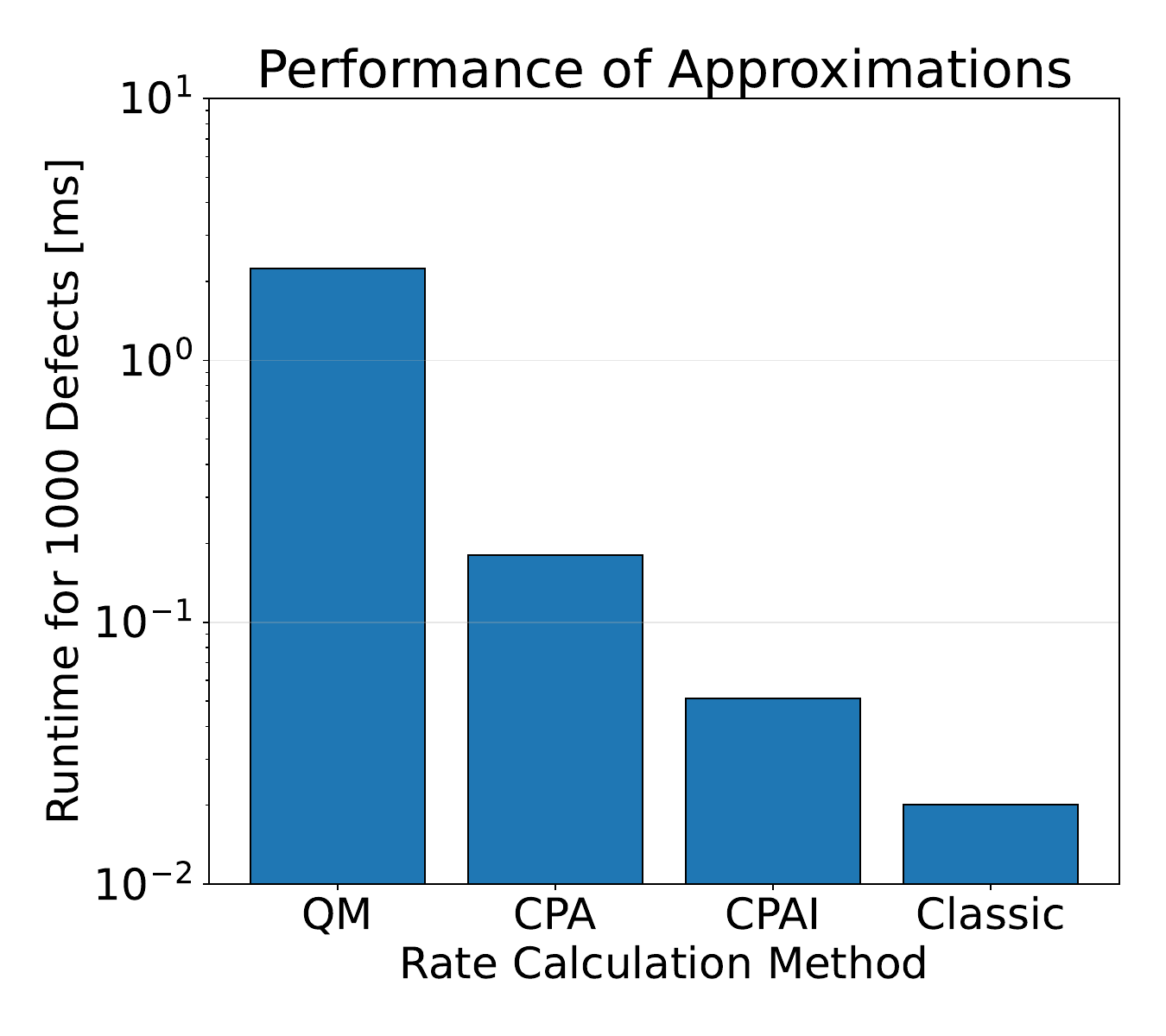}

    \end{subfigure}
    \caption{
    Performance benchmark of the different approaches used to evaluate the transition rates. Shown are the full quantum-mechanical calculation (QM), the crossing-preserving approximation with direct on-the-fly evaluation of the Bessel function (CPA), the corresponding implementation employing interpolation from a precomputed Bessel-function lookup table (CPAI), and the classical approximation (Classic).
    }
    \label{fig:performance_benchmark}
\end{figure}

Beyond its accuracy, a central advantage of the CPA is its computational simplicity. Once the effective parameters are obtained, the line-shape function reduces to evaluation of a closed-form expression involving only the evaluation of a modified Bessel function, avoiding explicit summation over vibrational states. This makes the approximation orders of magnitude faster than direct quantum-mechanical evaluation of the transition rates. \Cref{fig:performance_benchmark} quantifies  the runtimes of the different approaches for the evaluation of transition rates for 1000 defects. The direct quantum-mechanical calculation is the most computationally expensive approach, requiring several \(\mathrm{ms}\) per evaluation. Evaluating the CPA with the modified Bessel function computed on the fly reduces the runtime by about one order of magnitude. If the Bessel function is instead obtained by interpolation from a precomputed lookup table, the runtime decreases by another order of magnitude. This demonstrates that the CPA provides an analytic, numerically stable, and computationally inexpensive quantum-mechanical description of the transition, making it well suited for large-scale TCAD workflows where capture coefficients must be evaluated repeatedly over broad multidimensional parameter spaces.

\section{Charge Exchange Between Defects and Electronic Bands}
\label{sec:continuum-transitions}
In the previous chapter, we discussed charge transitions between two discrete electronic states. A typical example of such a process is trap-to-trap tunneling, where an electron tunnels between two spatially separated defects, each described by a localized electronic state. We now extend this description to the case in which the defect interacts not with a single discrete state, but with a continuum of electronic states, such as the conduction or valence band of a semiconductor. Such transitions are particularly important for semiconductor-device applications, as they provide the basis for simulating charge exchange between defects and electronic bands within device-level models. The objective of this chapter is therefore to derive the corresponding carrier capture and emission rates for the interaction of a localized defect state with a continuum of electronic band states.

\subsection{Continuum Capture and Emission Rates}

\begin{figure}[!hbt]
    \centering

    \begin{subfigure}{0.33\textwidth}
        \centering
        \includegraphics[width=\textwidth]{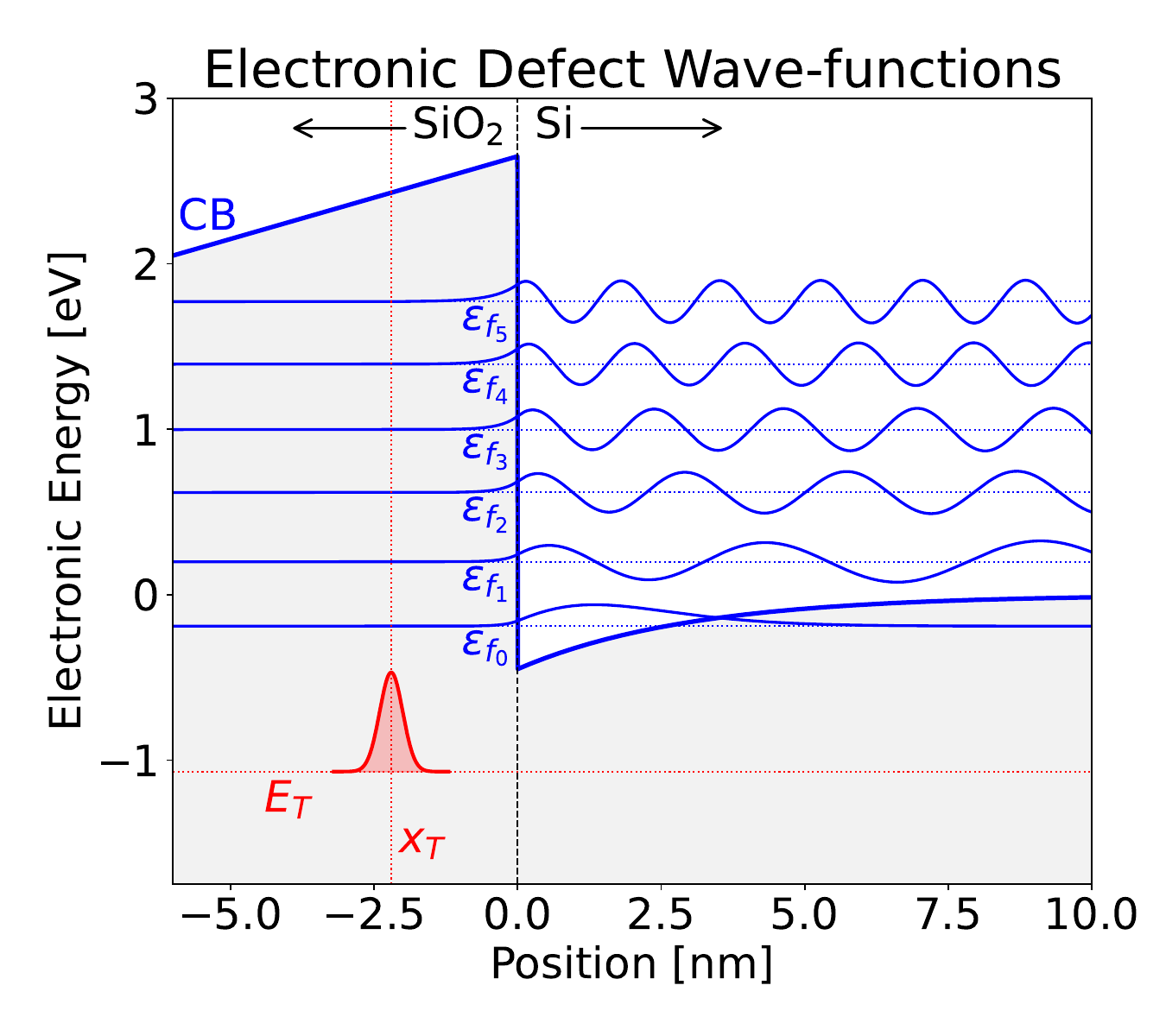}
        \caption{ }
        \label{fig:continuum_transition_1}
    \end{subfigure}
    \begin{subfigure}{0.33\textwidth}
        \centering
        \includegraphics[width=\textwidth]{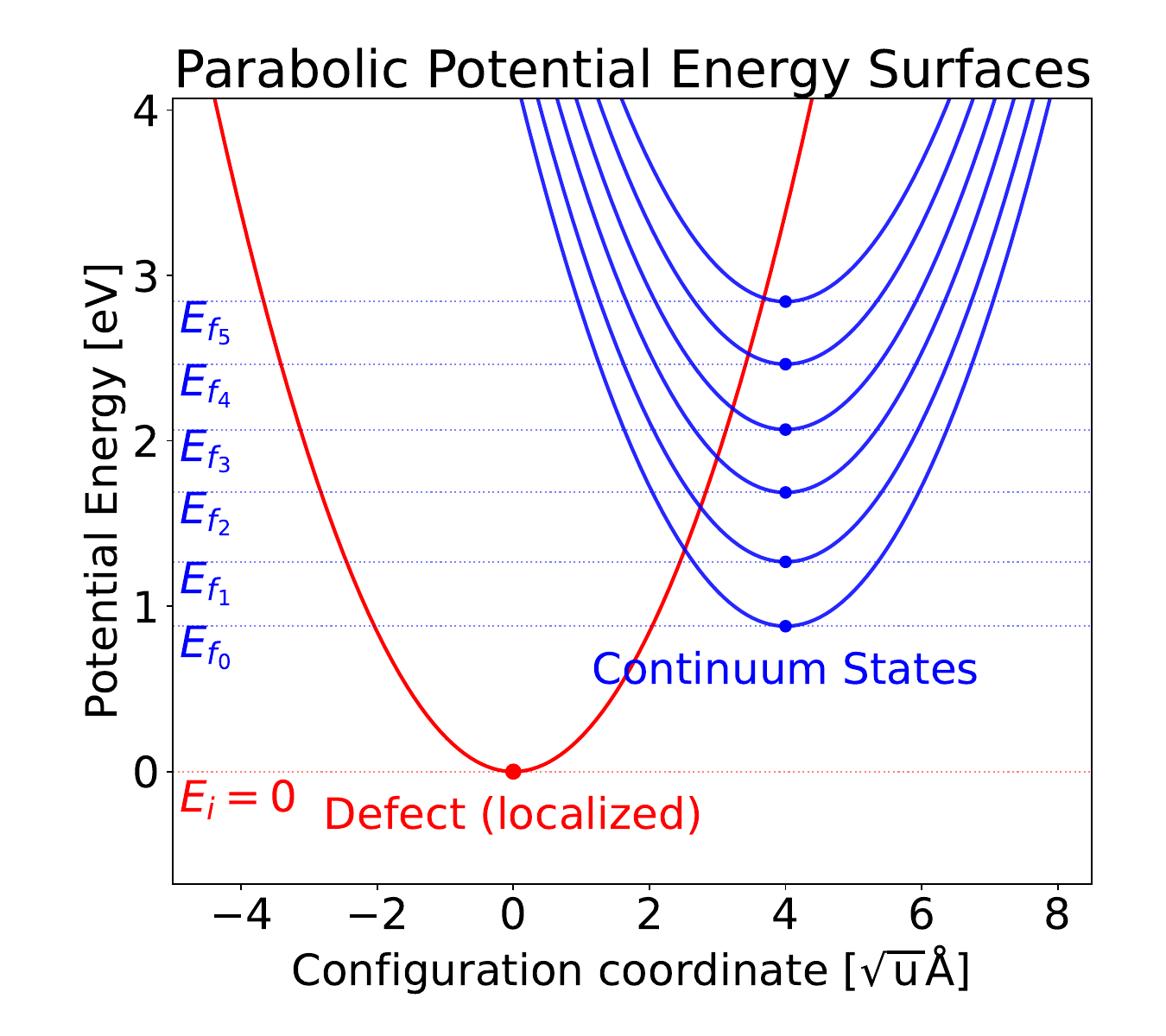}
        \caption{ }
        \label{fig:continuum_transition_2}
    \end{subfigure}
    \caption{
    Charge transitions between a localized defect state and a continuum of band states at \ce{Si}/\ce{SiO2} surface. 
    (a) Electronic structure at a semiconductor interface: a localized defect state at position $x_\mathrm{T}$ with charge transition level $E_\mathrm{T}$ coexists with a quasi-continuum of extended band states $\{\nrg_{f_k}\}$. The conduction-band states shown were calculated numerically for the plotted conduction-band profile.
    (b) Corresponding configuration-coordinate diagram: the defect is described by a single PES $V_i(Q)$, while all band states share the same structural relaxation and are represented by a family of parabolic final-state surfaces $V_{f_k}(Q)$ that differ only by a vertical energy shift $\Delta E_k = \nrg_{f_k} - E_\mathrm{T}$. This representation illustrates that the nuclear relaxation is defect-specific, whereas the continuum of final states enters solely through the electronic energy offset.}
    \label{fig:continuum_transition}
\end{figure}

\Cref{fig:continuum_transition_1} shows a prototypical case where a defect located in the oxide region of a Si/SiO$_2$ structure interacts with the extended states in the silicon conduction band across the interface barrier. The conduction band contains a dense set of delocalized electronic states that can be treated as a quasi-continuum. We denote the corresponding electronic states and their electronic energies by
\(\{(\phi_{f_k}, \nrg_{f_k})\}\), where the index \(k\) labels the \(k\)-th delocalized band state. \Cref{fig:continuum_transition_2} shows the corresponding configuration coordinate diagram for the defect. While the initial charge state is described by a single potential-energy surface \(V_i(Q)\), each final band state \(\phi_{f_k}\) gives rise to a corresponding final potential-energy surface \(V_{f_k}(Q)\). \\

Since the final electronic states are delocalized band states, the electronic density associated with any individual band state is distributed throughout the solid and therefore contributes only negligibly to the charge density in the immediate vicinity of the defect. As a result, occupying different conduction-band states perturbs the defect only weakly. As illustrated in \Cref{fig:continuum_transition_2}, this has two important consequences: \textbf{First}, the local forces acting on the defect in the final charge state, are essentially independent of the particular final band state into which the electron is emitted. As a result, the final potential-energy surfaces \(V_{f_k}(Q)\) share the same equilibrium configuration, curvature, and relaxation energy. \textbf{Second}, because the interaction between the localized defect and the delocalized conduction-band electron is negligible, the total energy in the final charge state can be expressed as a sum of a defect contribution and the energy of the occupied band state. Consequently, the minimum energy of the final potential-energy surface can be written as
$E_{f_k}=\nrg_{f_k} + \mathrm{constant}$, where the constant comprises all contributions to the total energy that are independent of the particular occupied band state. Thus, the minima of the final potential-energy surfaces shift linearly with the electronic energy \(\nrg_{f_k}\) of the occupied band state. As a result of these two properties, the individual configuration-coordinate diagrams shown in \Cref{fig:continuum_transition_2} differ only through their vertical energy offset
\(\Delta E_k = E_{f_k}-E_i\), whereas the structural-relaxation parameters
\(\Delta Q\), \(E_\mathrm{R}^{\mathrm i}\), and \(E_\mathrm{R}^{\mathrm f}\) are independent of the final band state. \\

Generalizing the transition rates of \Cref{eq:k_if_final,eq:k_fi_final} to the continuum of band states, the transition rates for electron emission into the \(k\)-th band state and for electron capture from the \(k\)-th band state are given by
\begin{equation}
k_{\mathrm e,k}
=
\frac{2\pi}{\hbar}
\left|W_{if_k}\right|^2
\eta_{\mathrm e}(\Delta E_k),
\qquad \mathrm{with} \qquad
\eta_{\mathrm e}(\Delta E_k)
\equiv
\eta_{if}\!\left(
\Delta E_k;
\Delta Q,E_\mathrm{R}^i,E_\mathrm{R}^f
\right),
\end{equation}

\begin{equation}
k_{\mathrm c,k}
=
\frac{2\pi}{\hbar}
\left|W_{if_k}\right|^2
\eta_{\mathrm c}(\Delta E_k),
\qquad \mathrm{with} \qquad
\eta_{\mathrm c}(\Delta E_k)
\equiv
\eta_{fi}\!\left(
\Delta E_k;
\Delta Q,E_\mathrm{R}^i,E_\mathrm{R}^f
\right).
\end{equation}
In order to relate the offset $\Delta E_k$ entering the lineshape functions to the electronic energy $\nrg_{f_k}$, it is convenient to introduce the thermodynamic charge transition level \(E_\mathrm{T}\), 
\begin{equation}
\Delta E_k = E_{f_k}-E_i
\equiv
\nrg_{f_k} - E_\mathrm{T}.
\label{eq:charge_Transition_level_definition}
\end{equation}
Physically, the thermodynamic charge-transition level (CTL) is the electronic energy at which the initial and final charge states have equal total (free) energy. It therefore provides the constant energy reference that relates the electronic energies of the band states in the band diagram to the total potential energies of the corresponding charge states in the configuration-coordinate diagram. As illustrated in \Cref{fig:continuum_transition}, it thus serves as the natural reference for the energy offsets. \\

To calculate the total transition rates involving the continuum of band states,
the occupation of the available electronic states has to be specified. For near-equilibrium simulations, we assume that the band is occupied according to the Fermi--Dirac distribution \(f_\mathrm{FD}(\nrg_{f_k}, E_\mathrm{F})\), where \(E_\mathrm{F}\) represents the
corresponding (quasi-)Fermi level in the band. For situations far from
equilibrium, this assumption can be relaxed by replacing \(f_\mathrm{FD}(\nrg_{f_k},E_\mathrm{F})\) with an energy-resolved carrier distribution obtained, for example,
from a Boltzmann transport equation (BTE) solver. A typical example is
hot-carrier degradation, where the carrier energy distribution no longer follows
a Fermi--Dirac distribution but instead acquires a pronounced high-energy tail \cite{Jech2021}. \\

Finally, the total emission and capture rates are obtained by summing the state-resolved
transition rates over all band states. Since electron emission into the
\(k\)-th band state is only possible if that state is unoccupied, the
corresponding contribution is weighted by the Pauli blocking factor
\(1-f_{\mathrm{FD}}(\nrg_{f_k}, E_\mathrm{F})\). Conversely, electron capture from the
\(k\)-th band state requires that the state is occupied and is therefore
weighted by the Fermi--Dirac occupation factor
\(f_{\mathrm{FD}}(\nrg_{f_k}, E_\mathrm{F})\). The resulting total rates are
\begin{align}
k_{\mathrm{e}}
&= \sum_k k_{\mathrm{e}, k}  =
\frac{2\pi}{\hbar}
\sum_k
|W_{if_k}|^2
\eta_{\mathrm{e}}\!\left( \nrg_{f_k} - E_\mathrm{T} \right)
\left[1-f_{\mathrm{FD}}\left(\nrg_{f_k}, E_\mathrm{F}\right)\right],
\label{eq:band_emission_sum}
\\
k_{\mathrm{c}}
&= \sum_k k_{\mathrm{c}, k}  =
\frac{2\pi}{\hbar}
\sum_k
|W_{if_k}|^2
\eta_{\mathrm{c}}\!\left( \nrg_{f_k} - E_\mathrm{T}\right)
f_{\mathrm{FD}}(\nrg_{f_k}, E_\mathrm{F}).
\label{eq:band_capture_sum}
\end{align}
Since the localized defect wave function is typically confined to a small region around the defect position \(\bm{x}_T\), whereas the delocalized band-state wave functions vary only on a much larger length scale (see \Cref{fig:continuum_transition_1}), the band-state wave functions can be regarded as approximately constant over the spatial extent of the defect wave function. Consequently, the electronic coupling matrix may be approximated as
\begin{equation}
W_{if_k}
=
\int \mathrm{d}\bm{x}\,
\phi_{f_k}(\bm{x}) 
\left[
\left.
\frac{\partial H_\mathrm{e}}{\partial Q}
\right|_{\bm{Q}_0}
\phi_i(\bm{x})
\right]^*
\simeq
\phi_{f_k}(\bm{x}_T) \mathcal{W}_i 
\label{eq:Wif_factorization}
\end{equation}
where the effective coupling constant is given by,
\begin{equation}
\mathcal{W}_i =
\int \mathrm{d}\bm{x}\,
\left[
\left.
\frac{\partial H_\mathrm{e}}{\partial Q}
\right|_{\bm{Q}_0}
\phi_i(\bm{x})
\right]^*,
\label{eq:Wif_factorization_2}
\end{equation}
and carries an additional
factor of $\sqrt{\mathrm{volume}}$ compared to the
coupling matrix element $W_{if_k}$. This approximation is valid for strongly localized defect states, whose spatial extent is much smaller than the characteristic length scale over which the delocalized band-state wave functions vary. Such localization is characteristic of deep defects with charge-transition levels well within the band gap. By contrast, shallow donor and acceptor states are only weakly bound and possess spatially extended wave functions, rendering the approximation inapplicable \cite{Freysoldt2014}. In the following, we therefore restrict our discussion to strongly localized deep defects. Substituting \Cref{eq:Wif_factorization} into the band rates and inserting the identity
\(1=\int \mathrm{d}\nrg\,\delta(\nrg-\nrg_{f_k})\) allows us to rewrite the state sum as an integral: 
\begin{equation}
k_{\mathrm{e}}
=
\frac{2\pi}{\hbar} |\mathcal{W}_i|^2
\int 
\eta_\mathrm{e}\!\left(\nrg - E_\mathrm{T}\right) \,
\left[1-f_{\mathrm{FD}}(\nrg, E_\mathrm{F})\right] 
g_f(\nrg,\bm{x}_T) \, \mathrm{d}\nrg,
\label{eq:band_emission_ldos}
\end{equation}
\begin{equation}
k_{\mathrm{c}}
=
\frac{2\pi}{\hbar} |\mathcal{W}_i|^2
\int 
\eta_\mathrm{c}\!\left(\nrg - E_\mathrm{T} \right) \,
f_{\mathrm{FD}}(\nrg, E_\mathrm{F}) \,
g_f(\nrg,\bm{x}_T) \, \mathrm{d}\nrg,
\label{eq:band_capture_ldos}
\end{equation}
where the sum over band states has been absorbed in the local density of states (LDOS) of the band evaluated at the defect position,
\begin{equation}
g_f(\nrg,\bm{x}_T)
=
\sum_k
|\phi_{f_k}(\bm{x}_T)|^2
\delta(\nrg-\nrg_{f_k}).
\label{eq:ldos_definition}
\end{equation}
\Cref{eq:band_emission_ldos,eq:band_capture_ldos} express the total transition rates as energy integrals over the LDOS at the defect position, thereby accounting for the local availability of band states accessible to the defect.
In the case shown in \Cref{fig:continuum_transition_1}, the silicon conduction-band wave functions decay evanescently into the oxide. As a result, the \ce{Si}-LDOS decays approximately exponentially with distance from the \ce{Si}/\ce{SiO2} interface. The \ce{Si}-LDOS is therefore commonly approximated in the oxide region by its bulk value $g_f^{\mathrm{Si}}(\nrg)$ multiplied by a tunneling factor \(T_\mathrm{WKB}(\nrg,\bm{x}_\mathrm{T})\), which accounts for the exponential attenuation of the conduction-band wave functions in the oxide. This approximation avoids the explicit calculation of the \ce{Si}-LDOS inside the oxide, which would otherwise require a self-consistent Schrödinger--Poisson simulation. \\

The relative position of the charge-transition level $E_\mathrm{T}$ with respect to the Fermi level $E_\mathrm{F}$ determines the balance between capture and emission. This follows from the detailed-balance relation, which carries over directly to the corresponding continuum transition rates:
\begin{equation}
\frac{
\eta_{\mathrm{e}}\!\left(\nrg-E_\mathrm{T}\right)
}{
\eta_{\mathrm{c}}\!\left(\nrg-E_\mathrm{T}\right)
}
=
\exp[-\beta(\nrg-E_\mathrm{T})]
\quad \Rightarrow \quad
\frac{k_{\mathrm{e}}}{k_{\mathrm{c}}}
=
\exp[\beta(E_\mathrm{T}-E_\mathrm{F})] .
\label{eq:detailed_balance_continuum}
\end{equation}
Correspondingly, when the Fermi level rises above \(E_\mathrm{T}\), capture dominates over emission and an initially empty defect starts to capture an electron on the characteristic timescale \(1/k_{\mathrm{c}}\). Conversely, when the Fermi level falls below \(E_\mathrm{T}\), emission dominates over capture and an occupied defect starts to emit its electron on the characteristic timescale \(1/k_{\mathrm{e}}\).
Thus, the relative position \(E_\mathrm{F}-E_\mathrm{T}\) fixes the equilibrium
occupation: when \(E_\mathrm{T}>E_\mathrm{F}\), the empty charge state is
thermodynamically favored, whereas when \(E_\mathrm{T}<E_\mathrm{F}\), the
occupied charge state is favored. \\

\begin{figure}[!hbt]
    \centering

    \begin{subfigure}{0.33\textwidth}
        \centering
        \includegraphics[width=\textwidth]{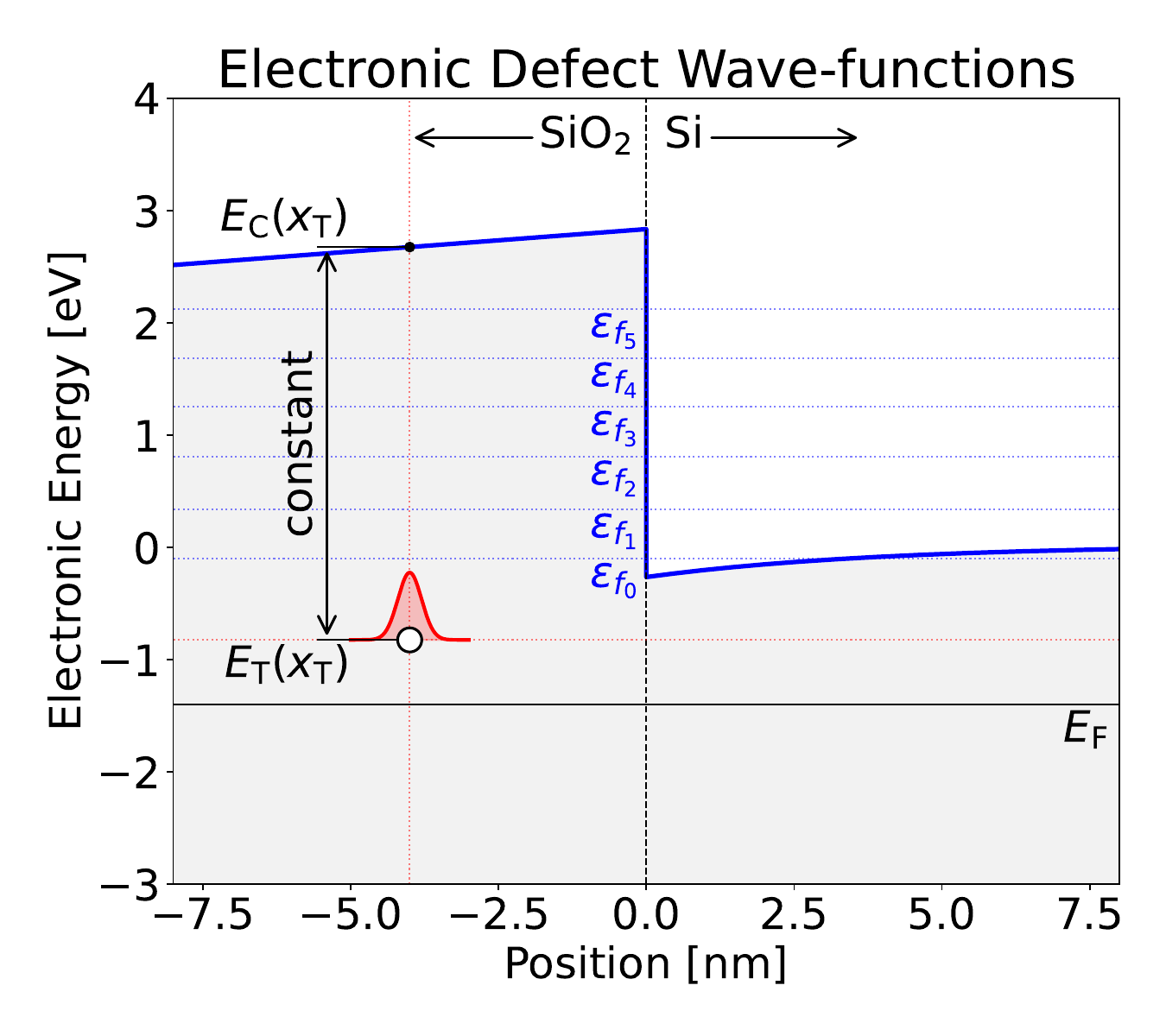}
        \caption{ }
    \label{fig:charge_transition_level_1}
    \end{subfigure}
    \begin{subfigure}{0.33\textwidth}
        \centering
        \includegraphics[width=\textwidth]{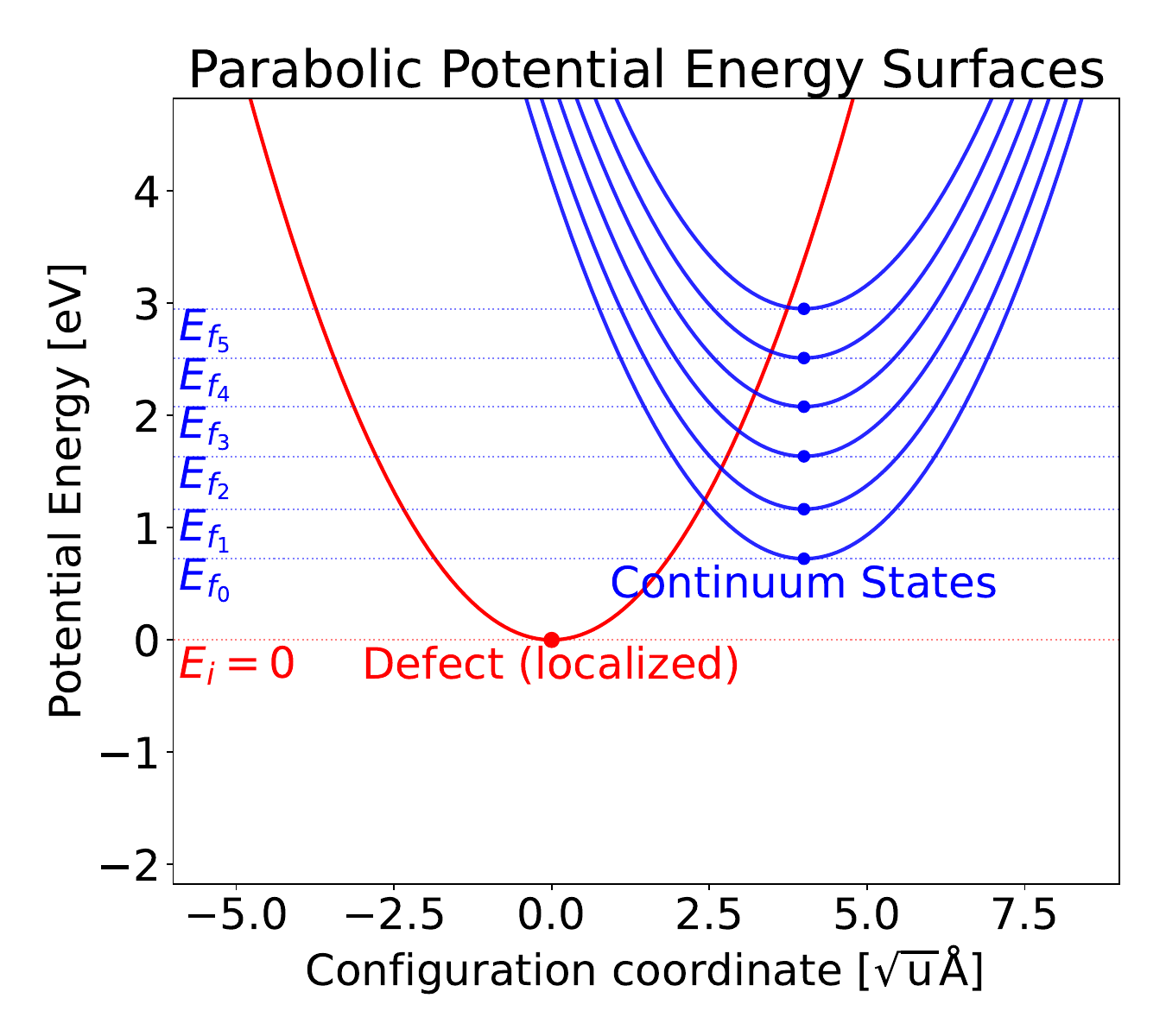}
        \caption{ }
        \label{fig:charge_transition_level_2}
    \end{subfigure}

    \hfill

    \begin{subfigure}{0.33\textwidth}
        \centering
        \includegraphics[width=\textwidth]{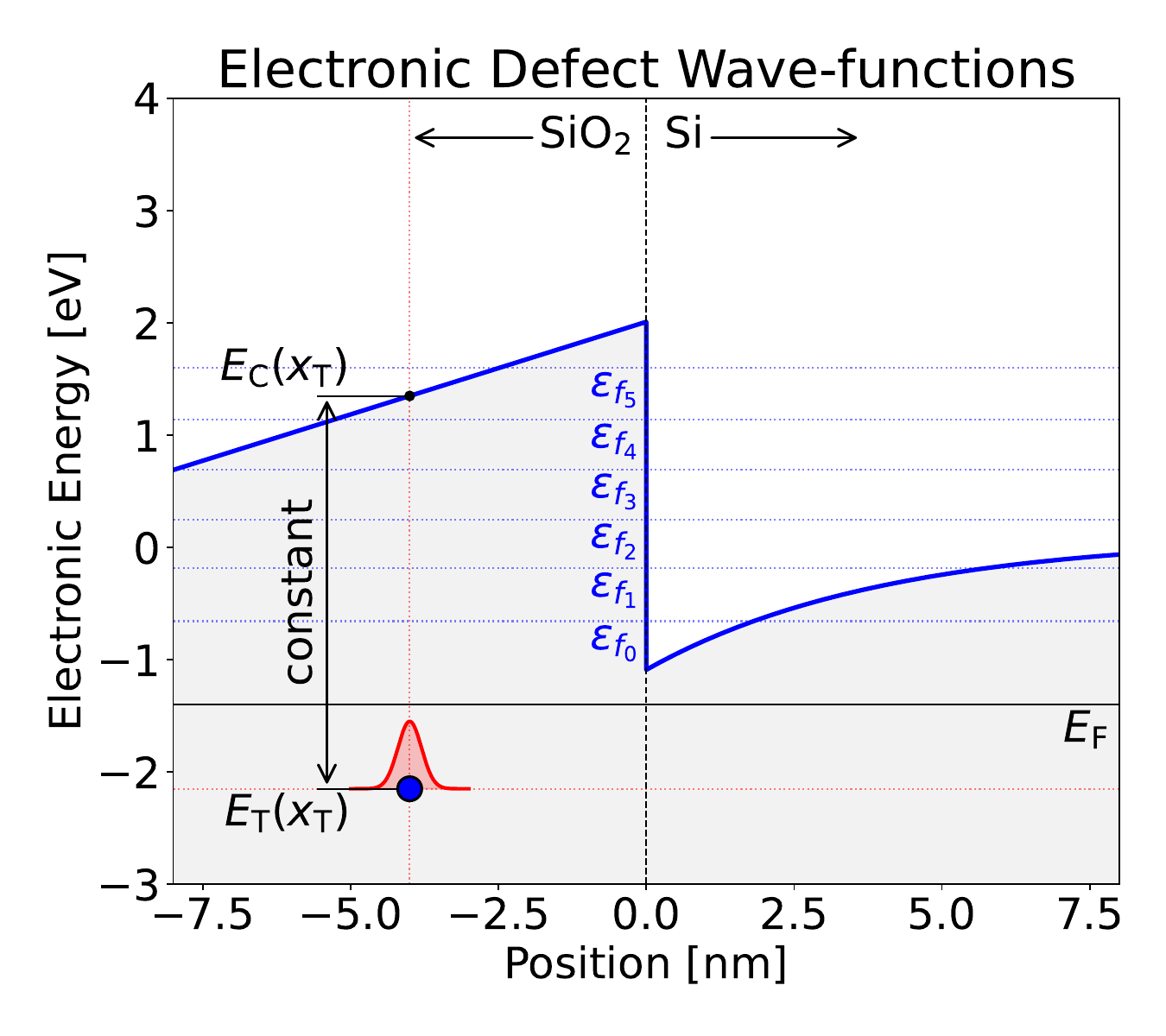}
        \caption{ }
    \label{fig:charge_transition_level_3}
    \end{subfigure}
    \begin{subfigure}{0.33\textwidth}
        \centering
        \includegraphics[width=\textwidth]{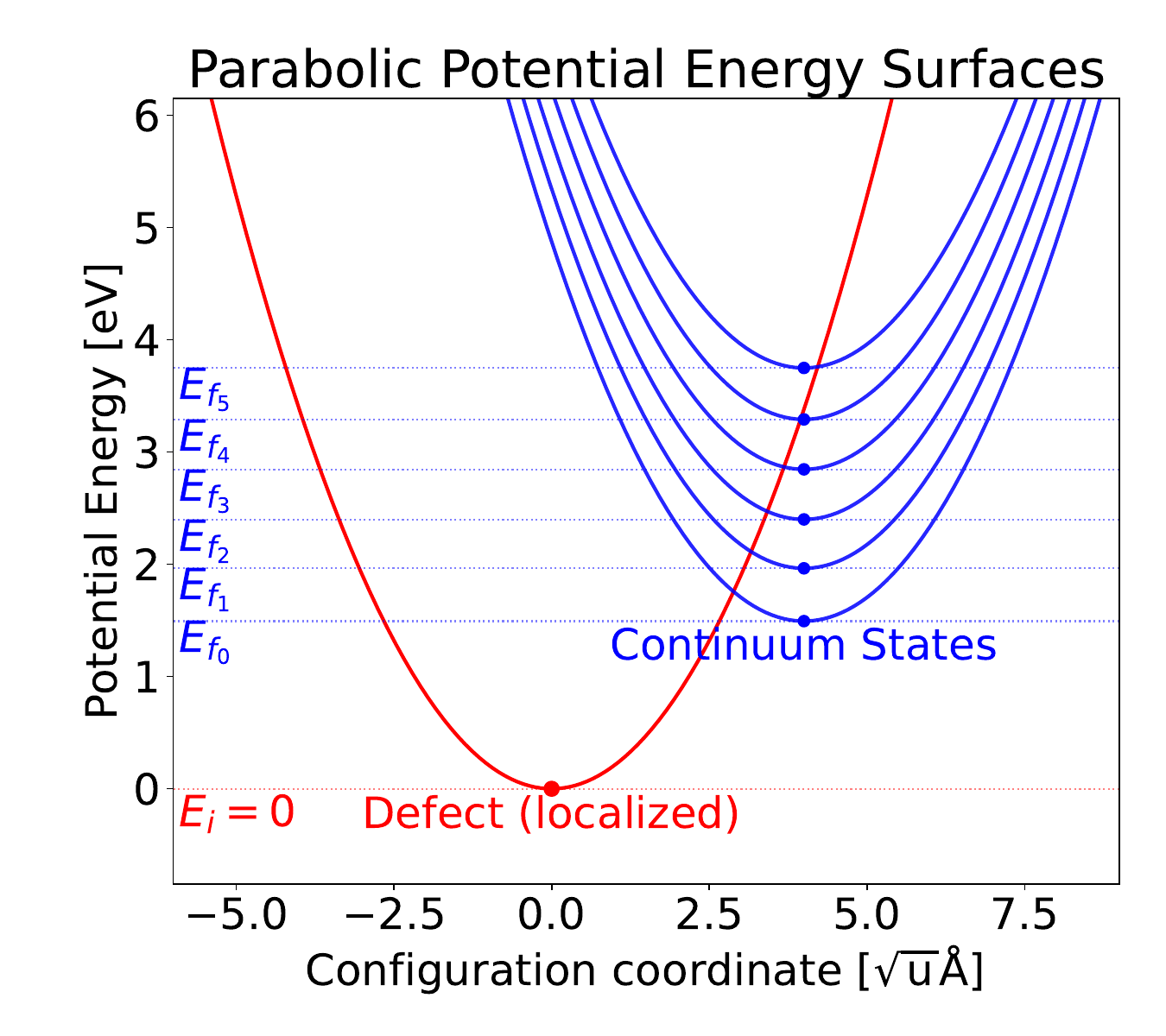}
        \caption{ }
        \label{fig:charge_transition_level_4}
    \end{subfigure}
    
 \caption{
    Effect of the gate bias on charge trapping at \ce{Si}/\ce{SiO2} interface in MOSFET. 
    (a-b) MOSFET under a small bias. The defect charge-transition level $E_\mathrm{T}$ lies above the Fermi level $E_\mathrm{F}$, so the defect is energetically favored to remain empty, indicated by the white circle. 
    (c-d) MOSFET under a large bias. The electrostatic potential shifts all electronic energies locally, including the charge-transition level $E_\mathrm{T}$, such that $E_\mathrm{T}$ moves below the Fermi level $E_\mathrm{F}$. Electron capture therefore becomes energetically favorable, leading to an occupied defect, indicated by the blue circle.
    In both cases, the charge-transition level remains fixed relative to the local conduction-band edge, while its position relative to the Fermi level changes with the applied bias.
    }
\label{fig:charge_transition_level}

\end{figure}

It is important to note that, in a device, the electrostatic potential shifts all
electronic energies locally, including the charge-transition level
\(E_\mathrm{T}\), as illustrated in \Cref{fig:charge_transition_level}. Since the local band edges $E_\mathrm{C}$ and $E_\mathrm{V}$ are shifted by the same electrostatic potential as
\(E_\mathrm{T}\), the offsets \(E_\mathrm{C}-E_\mathrm{T}\) and
\(E_\mathrm{T}-E_\mathrm{V}\) remain fixed. For this reason, charge-transition levels are commonly reported
relative to the conduction- or valence-band edge of the host material. 
The electrostatic shift has several consequences for the charge-transition rates. First, the
electrostatic potential modifies the barrier separating the extended conduction-band states from the defect (see ~\Cref{fig:charge_transition_level_1,fig:charge_transition_level_3}). It therefore controls the penetration of the extended conduction-band states into the barrier and thereby modulates the band LDOS at the defect position. Second, and typically more importantly, the electrostatic potential changes the relative alignment of the band states and the CTL. In the configuration-coordinate picture (see ~\Cref{fig:charge_transition_level_2,fig:charge_transition_level_4}), this corresponds to a shift of the relative alignment of the potential-energy
surfaces, thereby modifying the NMP transition barriers and the line-shape functions. As a result of these combined effects, the capture and emission rates acquire a pronounced bias dependence. This bias dependence is also reflected in the equilibrium charge state of the
defect, as shown in \Cref{fig:charge_transition_level_2,fig:charge_transition_level_4}.
At low bias, \(E_\mathrm{T}\) lies above \(E_\mathrm{F}\), and the defect
therefore thermodynamically favors the empty charge state, indicated by the white
circle. Under high bias, \(E_\mathrm{T}\) is shifted below \(E_\mathrm{F}\), so
that the occupied charge state, indicated by the blue circle, becomes
thermodynamically favored.

\subsection{Band-Edge Approximation for Continuum Rates}
\Cref{eq:band_emission_ldos,eq:band_capture_ldos} provide the most general device-level NMP transition rates for carrier exchange with a continuum of band states. These expressions require an explicit integration over the full energy range of the LDOS. To obtain a compact and computationally efficient representation of the transition rates suitable for TCAD applications, we approximate these integrals in terms of a small number of evaluations of the line-shape function in the vicinity of the band edge. In the lowest-order approximation, this reduces to a single evaluation of the line-shape function, which significantly accelerates the continuum-rate calculation. \\

\begin{figure}[!hbt]
    \centering

    \begin{subfigure}{0.33\textwidth}
        \centering
        \includegraphics[width=\textwidth]{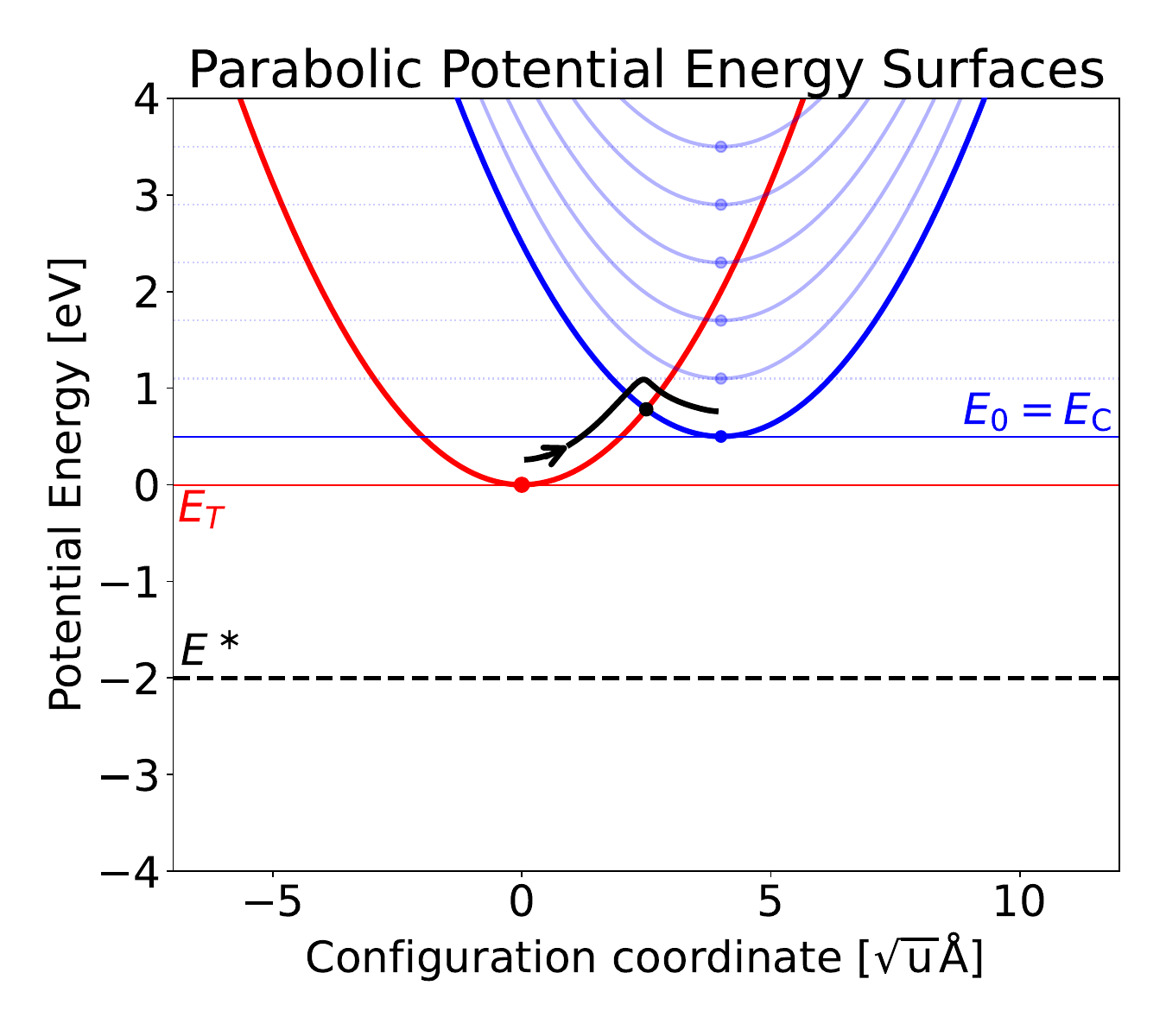}
        \caption{ }
        \label{fig:band_edge_approximation_1}
    \end{subfigure}
    \begin{subfigure}{0.33\textwidth}
        \centering
        \includegraphics[width=\textwidth]{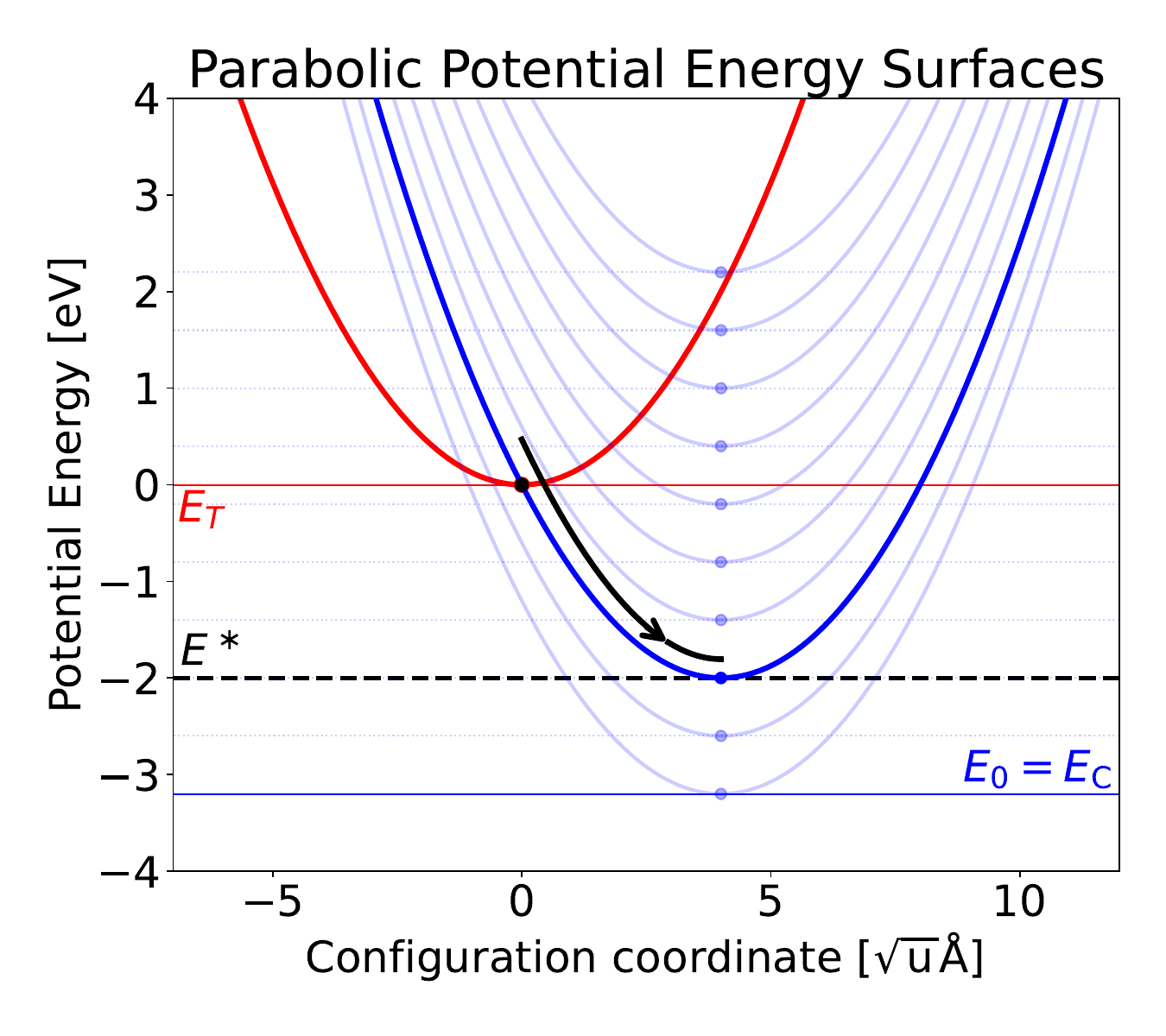}
        \caption{ }
        \label{fig:band_edge_approximation_2}
    \end{subfigure}
  \caption{
      Visualization of the break down of the band edge approximation for the non-degenerate case, \(E_0=E_\mathrm{C}\). The defect is described by the
    initial PES \(V_i(Q)\), shown in red, while the continuum states are represented
    by a family of parabolas \(V_{f_k}(Q)\), shown in blue. The black arrows
    indicate the dominant classical transition path in each case.
    (a) \(E_\mathrm{C} \gg E^\ast\): the continuum
    integral is dominated by states with \(\nrg \approx E_\mathrm{C}\), which have the
    smallest classical transition barrier among the available states, while the
    barrier increases rapidly for excited band states.
    (b) \(E_\mathrm{C}  \ll E^\ast \): the
    continuum integral is dominated by excited states  with \(\nrg \approx E^\ast\), where the classical transition barrier becomes small.}
    \label{fig:band_edge_approximation}
\end{figure}

In the following, we demonstrate the procedure for electron exchange with the conduction band. The same derivation can be applied to electron exchange with the valence band or metallic electrodes by replacing the corresponding LDOS. Let
\(g_\mathrm{CB}(\nrg,E_\mathrm{C})\) denote the conduction-band LDOS
evaluated at the defect position and let \(E_\mathrm{C}\)
denote the corresponding conduction band edge. For emission, the band state must be empty. In the degenerate case, meaning \(E_\mathrm{F}>E_\mathrm{C}\), states between \(E_\mathrm{C}\) and \(E_\mathrm{F}\) are mostly occupied and therefore Pauli blocked. The lowest available final states are then located near \(E_\mathrm{F}\). This motivates replacing the lower integration boundary of the emission rate by an effective lower boundary \(E_0=\max(E_\mathrm{C},E_\mathrm{F})\), yielding
\begin{align}
    k_\mathrm{e}^{\mathrm{CB}}
    &=
    \frac{2\pi}{\hbar}
    |\mathcal W_i|^2
    \int_{E_\mathrm{C}}^{\infty}
    d\nrg\,
    g_\mathrm{CB}(\nrg,E_\mathrm{C})
    \left[
        1-f_\mathrm{FD}(\nrg)
    \right]
    \eta_\mathrm{e}(\nrg-E_\mathrm{T}) \\
    &\approx
    \frac{2\pi}{\hbar}
    |\mathcal W_i|^2
    \int_{E_0}^{\infty}
    d\nrg\,
    g_\mathrm{CB}(\nrg,E_\mathrm{C})
    \left[
        1-f_\mathrm{FD}(\nrg)
    \right]
    \eta_\mathrm{e}(\nrg-E_\mathrm{T}) ,
\label{eq:effective_rate_E0}
\end{align}
For \(E_\mathrm{F}<E_\mathrm{C}\), this replacement is exact and for \(E_\mathrm{F}>E_\mathrm{C}\), it represents an
approximation that neglects the exponentially small density of
available empty states below \(E_\mathrm{F}\). 
To derive practical approximations for the continuum rates, we proceed by rewriting the Pauli-blocking factor as
\begin{equation}
    1-f_{\mathrm{FD}}(\nrg, E_\mathrm{F})
    =
    f_{\mathrm{FD}}(\nrg, E_\mathrm{F})
    \exp[\beta(\nrg-E_\mathrm{F})] ,
\end{equation}
resulting in
\begin{align}
    k_{\mathrm{e}}
    &\approx
    \frac{2\pi}{\hbar}
    |\mathcal W_i|^2
    \int_{E_0}^{\infty}
    \mathrm{d}\nrg\,
    g_\mathrm{CB}(\nrg,E_\mathrm{C})
    f_{\mathrm{FD}}(\nrg, E_\mathrm{F})
    \exp[\beta(\nrg-E_\mathrm{F})]
    \eta_{\mathrm{e}}(\nrg-E_\mathrm{T}).
\end{align}
The emission rate can then be rewritten compactly as
\begin{equation}
    k_{\mathrm{e}}
    =
    \frac{2\pi}{\hbar}
    |\mathcal W_i|^2
    n_0(\bm x_T)
    \left\langle
        F_{\mathrm{e}}(\nrg)
    \right\rangle,
    \qquad
    F_{\mathrm{e}}(\nrg)
    =
    \exp[\beta(\nrg-E_\mathrm{F})]
    \eta_{\mathrm{e}}(\nrg-E_\mathrm{T}) .
\end{equation}
The corresponding capture rate follows from detailed balance and is given by
\begin{equation}
    k_{\mathrm{c}}
    =
    \frac{2\pi}{\hbar}
    |\mathcal W_i|^2
    n_0(\bm x_T)
    \left\langle
        F_{\mathrm{c}}(\nrg)
    \right\rangle,
    \qquad
    F_{\mathrm{c}}(\nrg)
    =
    \eta_{\mathrm{c}}(\nrg-E_\mathrm{T}) .
\end{equation}
Here, the averages are taken with respect to the normalized carrier measure
\begin{equation}
\mathrm d\mu(\nrg)
=
\frac{
g_\mathrm{CB}(\nrg,E_\mathrm{C})
f_{\mathrm{FD}}(\nrg,E_\mathrm{F})
}{
n_0 \qquad 
}
\,\mathrm d\nrg ,    \qquad n_0
    =
    \int_{E_0}^{\infty}
    \mathrm{d}\nrg\,
    g_\mathrm{CB}(\nrg, E_\mathrm{C})
    f_{\mathrm{FD}}(\nrg, E_\mathrm{F}),
\end{equation}
which describes the energetic distribution of the carriers in the band. To arrive at a computationally efficient approximation we expand the functions \(F_{\mathrm e}(\nrg)\) and \(F_{\mathrm c}(\nrg)\) in a Taylor series about \(E_0\). Substituting this expansion into the average and interchanging summation and integration yields 
\begin{equation}
\left\langle
F_{\mathrm{e/c}}(\nrg)
\right\rangle
=
\sum_{j=0}^{\infty}
\frac{1}{j!}
\left.
\frac{d^jF_{\mathrm{e/c}}}{d\nrg^j}
\right|_{\nrg=E_0}
\left\langle
(\nrg-E_0)^j
\right\rangle .
\end{equation}
The expansion coefficients in this expression are thus fully determined by the centered moments of the carrier measure. We denote these moments by
\begin{equation}
    \mu_{j}(E_0)
    =
    \left\langle
        (\nrg-E_0)^j
    \right\rangle
    =
    \frac{
        \int_{E_0}^{\infty}
        \mathrm{d}\nrg\,
        g_\mathrm{CB}(\nrg,E_\mathrm{C})
        f_{\mathrm{FD}}(\nrg,E_\mathrm{F})
        (\nrg-E_0)^j
    }{
        n_0
    } .
\end{equation}
Finally, the expansion can be written compactly as
\begin{equation}
    k_\mathrm{e}^{\mathrm{CB}}
    \approx
    \frac{2\pi}{\hbar}
    |\mathcal W_i|^2
    n_0
    \sum_{j=0}^{N}
    \frac{\mu_j(E_0)}{j!}
    \left.
    \frac{
        d^jF_{\mathrm{e}}
    }{
        d\nrg^j
    }
    \right|_{\nrg=E_0},
    \label{eq:boundary_expansion_em}
\end{equation}

\begin{equation}
    k_\mathrm{c}^{\mathrm{CB}}
    \approx
    \frac{2\pi}{\hbar}
    |\mathcal W_i|^2
    n_0
    \sum_{j=0}^{N}
    \frac{\mu_j(E_0)}{j!}
    \left.
    \frac{
        d^jF_{\mathrm{c}}
    }{
        d\nrg^j
    }
    \right|_{\nrg=E_0}.
    \label{eq:boundary_expansion_cap}
\end{equation}
where \(N\) denotes the truncation order. At lowest order, only a single evaluation of the line-shape function at the effective edge $E_0$ is required, resulting in a particularly compact and computationally efficient approximation for the continuum rates
\begin{equation}
    k_\mathrm{e}^{\mathrm{CB}}
    \approx
    \frac{2\pi}{\hbar}
    |\mathcal W_i|^2
    \eta_{\mathrm{e}}(E_0-E_\mathrm{T})
    \exp[\beta(E_0-E_\mathrm{F})]
    n_0 .
    \label{eq:emission_leading}
\end{equation}
\begin{equation}
    k_\mathrm{c}^{\mathrm{CB}}
    \approx
    \frac{2\pi}{\hbar}
    |\mathcal W_i|^2
    \eta_{\mathrm{c}}(E_0-E_\mathrm{T})
    n_0 .
    \label{eq:capture_leading}
\end{equation}
This band-edge approximation is valid whenever the dominant contribution to the continuum integral originates from states close to the effective lower integration boundary \(E_0=\max(E_\mathrm{C},E_\mathrm{F})\),  i.e., when the integrand peaks near \(E_0\) and decays rapidly for higher-energy states in the band. However, the line-shape function generally possesses an intrinsic maximum at an energy \(E^\ast\). As shown in Supplementary Information SI-5 the precise position of  \(E^\ast\) depends on the chosen line-shape model. Nevertheless, \(E^\ast\) generally lies close to the energy at which the
classical emission barrier vanishes. This reflects the fact that, within the classical picture, the emission rate follows the Arrhenius law \(k_\mathrm{e}\propto\exp(-\beta \Delta E_\mathrm{X})\), so that states with small emission barriers contribute most strongly to the transition rate. Once the intrinsic maximum $E^\ast$ enters the integration domain, the line-shape function preferentially weights states close to its intrinsic maximum \(E^\ast\), shifting the dominant contribution of the continuum integral away from the band edge and thereby violating the central assumption of the band-edge approximation. Consequently, the regime \(E_0<E^\ast\) marks the most pronounced and readily identifiable breakdown of the band-edge expansion. \Cref{fig:band_edge_approximation_1} illustrates the boundary-dominated case \(E^\ast<E_0\). Here, states near \(E_0\) dominate because they have the smallest available classical emission barrier, whereas transitions into higher-energy states require the system to overcome progressively larger barriers and are therefore exponentially suppressed. Conversely, \Cref{fig:band_edge_approximation_2} shows the regime \(E^\ast>E_0\), where conduction-band states near the intrinsic maximum \(E^\ast\) dominate due to their small emission barriers. \\

Fortunately, the breakdown of the band-edge approximation for
\(E_0 < E^\ast\) can be repaired in a simple way. Once \(E^\ast\) lies within the integration domain, the dominant contributing band states are already available. The emission rate therefore cannot continue to increase significantly when the band states are lowered further relative to the defect state and the emission rate starts to saturate. We account for this behavior by clamping the band-edge approximation for the emission rate to its value at \(E_0=E^\ast\) whenever \(E_0<E^\ast\). This provides a continuous crossover from the \(E_0\)-dominated regime to the \(E^\ast\)-dominated regime. Operationally, the clamping is implemented by shifting the band to the configuration at which this crossover occurs:
\begin{equation}
E_\mathrm{F} \to E_\mathrm{F}^\ast
=
E^\ast-(E_0-E_\mathrm{F}),
\qquad
E_\mathrm{C} \to E_\mathrm{C}^\ast
=
E^\ast-(E_0-E_\mathrm{C}), \qquad
E_0 \to 
E^\ast, 
\end{equation}
In the following, thus all starred quantities denote the corresponding band-edge-expansion quantities evaluated at the crossover point, i.e. with the effective lower integration boundary fixed to $E_0=E^\ast$ and the band-edge and Fermi-level positions replaced by $E_\mathrm{C}=E_\mathrm{C}^\ast$ and $E_\mathrm{F}=E_\mathrm{F}^\ast$, respectively. The emission rate is thus obtained by,
\begin{equation}
k_\mathrm{e}^{\mathrm{CB}}
\approx
\frac{2\pi}{\hbar}
|\mathcal W_i|^2
n_0^\ast
\sum_{j=0}^{N}
\frac{\mu_j^\ast(E^\ast)}{j!}
\left.
\frac{
\mathrm{d}^j F_\mathrm{e}^*
}{
\mathrm{d}\nrg^j
}
\right|_{\nrg=E^\ast}.
\label{eq:clamped_emission}
\end{equation}
whereas the corresponding capture rate follows from detailed balance from the clamped expression,
\begin{equation}
k_\mathrm{c}^{\mathrm{CB}}
\approx
\exp[\beta(E_0-E^\ast)]
\frac{2\pi}{\hbar}
|\mathcal W_i|^2
n_0^\ast
\sum_{j=0}^{N}
\frac{\mu_j^\ast(E^\ast)}{j!}
\left.
\frac{
\mathrm{d}^jF_\mathrm{c}^*
}{
\mathrm{d}\nrg^j
}
\right|_{\nrg=E^\ast}.
\label{eq:clamped_capture}
\end{equation}
At leading order, $N=0$, the fully clamped expressions reduce to
\begin{equation}
    k_\mathrm{e}^{\mathrm{CB}}
    \approx
    \frac{2\pi}{\hbar}
    |\mathcal W_i|^2
    n_0^\ast
    \exp[\beta(E^\ast-E_\mathrm{F}^\ast)]
    \eta_{\mathrm{e}}(E^\ast-E_\mathrm{T}) .
    \label{eq:clamped_emission_leading}
\end{equation}
\begin{equation}
    k_\mathrm{c}^{\mathrm{CB}}
    \approx
    \frac{2\pi}{\hbar}
    |\mathcal W_i|^2
    n_0^\ast
    \exp[\beta(E_0-E^\ast)]
    \eta_{\mathrm{c}}(E^\ast-E_\mathrm{T}) .
    \label{eq:clamped_capture_leading}
\end{equation}
By construction, these expressions remain exactly constant throughout the clamped regime and reproduce the value of the boundary expansion continuously at the crossover point $E_0=E^\ast$.\\

\begin{figure}[!hbt]
    \centering
    \begin{subfigure}{0.33\textwidth}
        \centering
        \includegraphics[width=\textwidth]{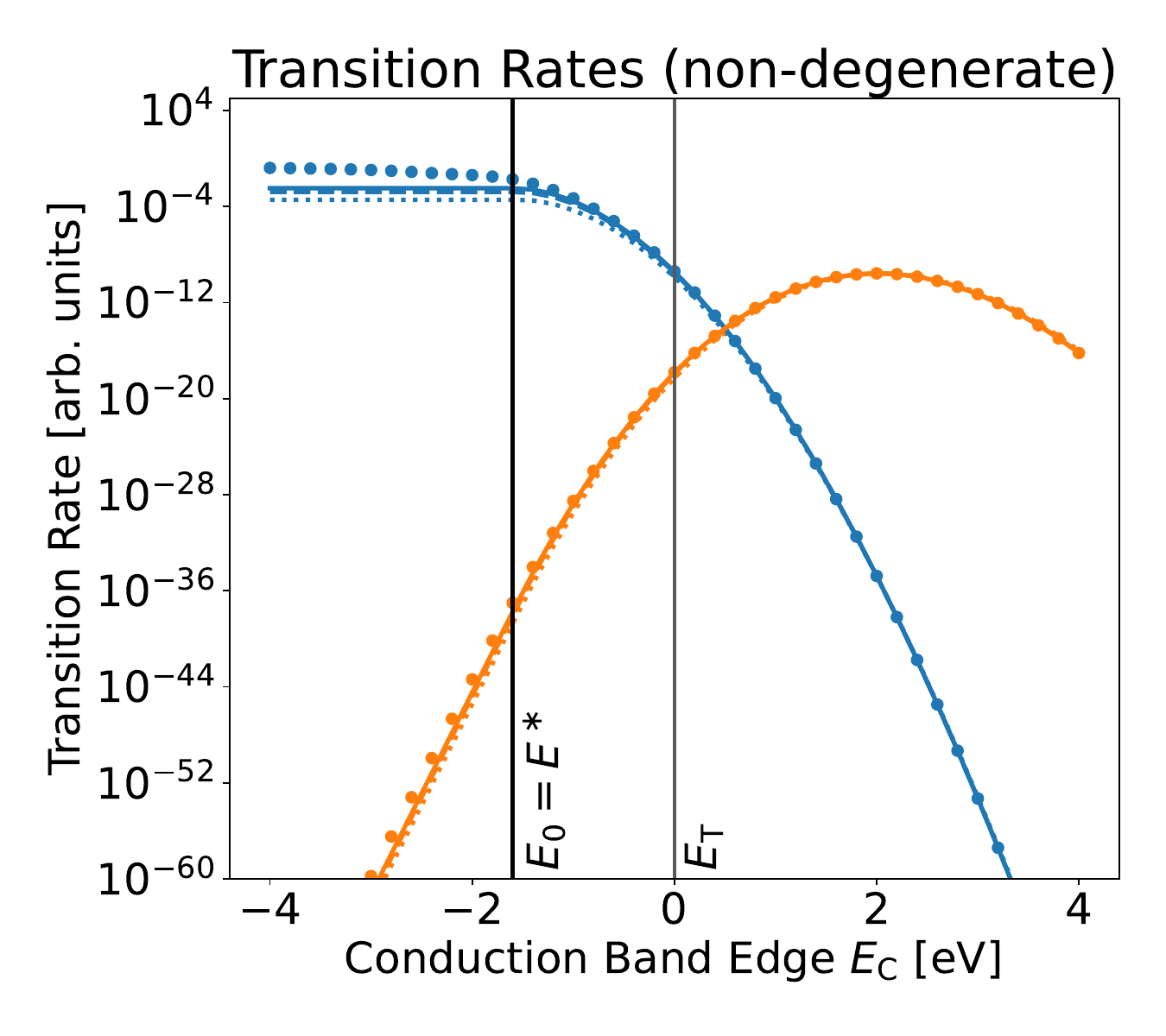}
        \caption{ }
        \label{fig:band_edge_approximation_rates_1}
    \end{subfigure}
    \begin{subfigure}{0.33\textwidth}
        \centering
        \includegraphics[width=\textwidth]{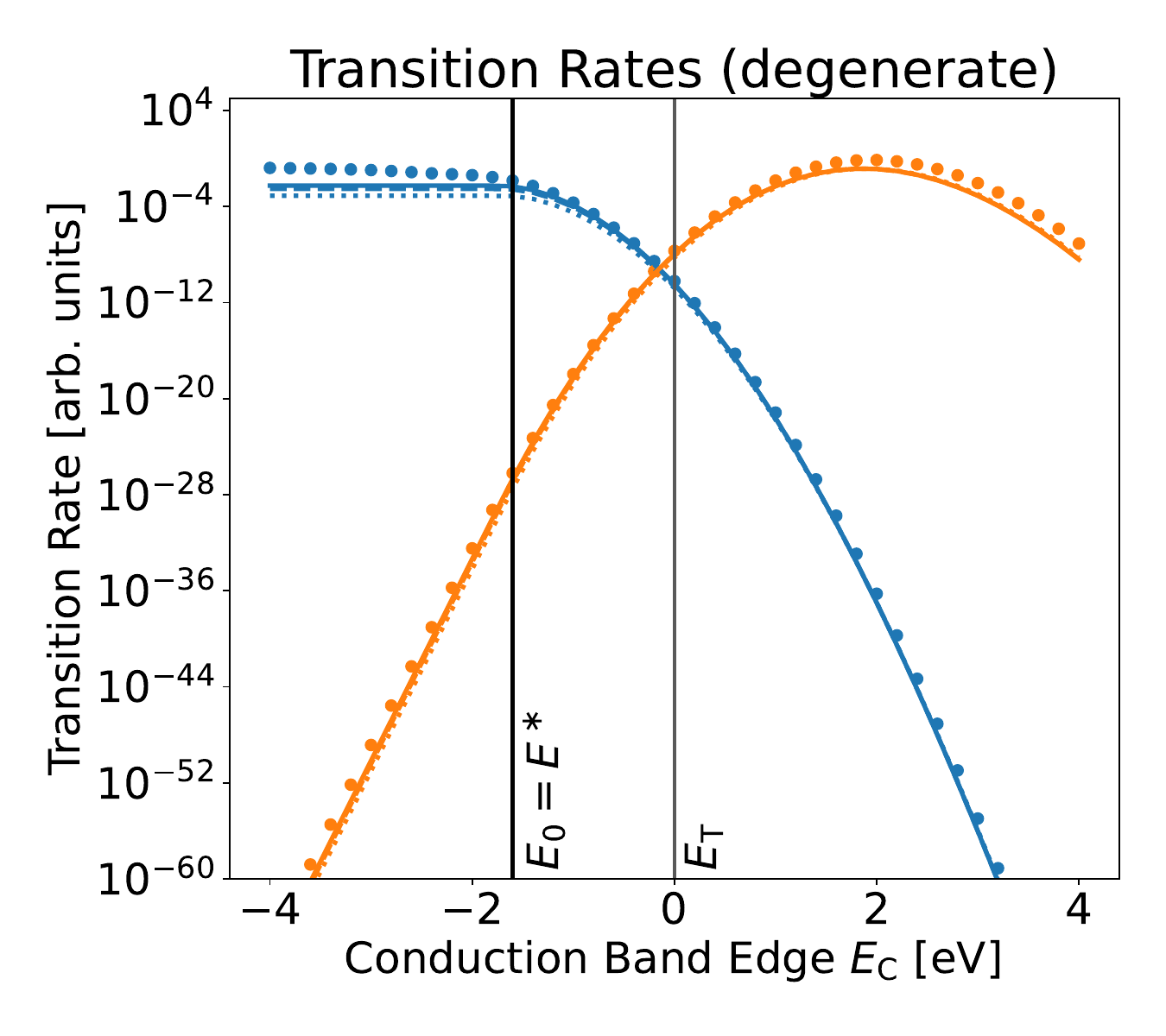}
        \caption{ }
    \label{fig:band_edge_approximation_rates_2}
    \end{subfigure}
    \caption{
    Comparison of the emission and capture rates obtained from the exact numerical evaluation and the band edge approximation. The NMP parameters are \(E_\mathrm{R}^i=E_\mathrm{R}^f=2.0\,\mathrm{eV}\), \(\Delta Q=4.0\sqrt{\mathrm{u}}\,\text{\AA}\), \(R=1.0\), \(T=300\,\mathrm{K}\). (a) shows the rates for the non-degenerate case $E_\mathrm{C} - E_\mathrm{F} = 0.5 \, \mathrm{eV}$ (b) shows the rates for the degenerate case $E_\mathrm{F} - E_\mathrm{C} = 0.2 \, \mathrm{eV}$. In both panels, the exact integration is shown as points, Different line styles correspond to different truncation orders \(N\) of the band edge approximation, with \(N=0\) shown as a dotted line, \(N=2\) as a dashed line, and \(N=4\) as a solid line.
    }
    \label{fig:band_edge_approximation_rates}
\end{figure}

\Cref{eq:boundary_expansion_em,eq:boundary_expansion_cap}
and \Cref{eq:clamped_emission,eq:clamped_capture} constitute the generalized band-edge approximation for continuum NMP transition rates. The quality of the approximation is assessed in \Cref{fig:band_edge_approximation_rates}, where the emission and capture rates are compared to exact numerical integration as a function of \(E_\mathrm{C}\). The figure shows the band-edge approximation for different truncation orders \(N\) of the asymptotic expansion. A visible deviation remains in the \(E^\ast\)-dominated regime for both the non-degenerate and degenerate case, where the clamped continuation necessarily introduces an error. In the degenerate case  a visible deviation also appears for large values of \(E_\mathrm{C}-E_\mathrm{T}\), where the combined effect of the line-shape function and the Fermi occupation develops a second interior maximum violating the band edge approximation. This deviation is typically less pronounced, since in degenerate operation the offset \(E_\mathrm{F}-E_\mathrm{C}\) is usually only a few hundred meV. Overall, the approximation captures the correct physical trends in both the boundary-dominated and interior-dominated regimes, while ensuring a smooth and continuous crossover between them across the transition region \(E_0 \approx E^\ast\). Importantly, even the lowest-order approximation, \(N=0\), shows good agreement with the exact rates over the full energy range. Since this approximation requires only a single evaluation of the line-shape function at the dominant energy, either at the boundary \(E_0\) or at the interior maximum \(E^\ast\), it is computationally extremely efficient and particularly well suited for TCAD applications.

\subsection{Comparison with Shockley–Read–Hall Theory}

Having reduced the continuum NMP rates to TCAD-compatible expressions, we now relate them to the SRH model, which remains the most widely used description of nonradiative transitions in device simulation. This comparison shows that the NMP rates can be cast into the same operational form as SRH rates, but also exposes a fundamental limitation of the conventional SRH framework: its capture coefficients are purely phenomenological fitting parameters. As a result, SRH theory does not intrinsically contain the temperature and bias dependence of charge trapping, whereas these dependencies arise naturally in the NMP framework from the microscopic defect physics. \\

The leading-order band-edge approximation, given by \Cref{eq:emission_leading,eq:capture_leading}, reduces to a form that is formally identical to the Shockley--Read--Hall (SRH) rate expressions in the regime \(E^\ast < E_{\mathrm{C}}\) and \( E_{\mathrm{F}} < E_\mathrm{C}\). To make this connection explicit, we introduce the local conduction-band electron density in the nondegenerate limit,
\begin{equation}
    n
    \approx
    N_\mathrm{C} \exp[-\beta(E_{\mathrm{C}}-E_{\mathrm{F}})],
\end{equation}
where \(N_\mathrm{C}\) is the effective band weight of the conduction band in the Boltzmann approximation. The capture and emission rates at a single defect can then be written compactly as
\begin{equation}
    k_\mathrm{c}^{\mathrm{CB}}
    =
    c_n^{\mathrm{NMP}}\,
    n,
    \qquad
    k_\mathrm{e}^{\mathrm{CB}}
    =
    e_n^{\mathrm{NMP}}.
\end{equation}
The NMP capture coefficient is given by
\begin{equation}
    c_n^{\mathrm{NMP}}
    =
    \frac{2\pi}{\hbar}
    |\mathcal{W}_{\mathrm{i}}|^2
    \eta_{\mathrm{c}}\!\left(E_{\mathrm{C}}-E_{\mathrm{T}}\right),
\end{equation}
and the NMP emission coefficient is given by
\begin{equation}
    e_n^{\mathrm{NMP}}
    =
    c_n^{\mathrm{NMP}}\, n_1 ,
\end{equation}
where we introduced the reference concentration
\begin{equation}
    n_1 =
    N_\mathrm{C}
    \exp[-\beta(E_{\mathrm{C}}-E_{\mathrm{T}})] .
\end{equation} 
for a compact notation. Thus, in the band-edge-dominated regime, the NMP rates take the same operational form as SRH theory \cite{Schroder2005}. In both descriptions, the capture rate is written as the product of a capture coefficient and the local carrier concentration, whereas the emission rate is independent of the local carrier concentration and is determined by the emission coefficient only. Despite this formal equivalence, the microscopic interpretation of the coefficients is fundamentally different in SRH and NMP theory. \\

Within the NMP framework, the capture and emission coefficients \(c_\mathrm{NMP}^{\mathrm{CB}}\) and \(e_\mathrm{NMP}^{\mathrm{CB}}\) are directly linked to the microscopic properties of the defect and can be obtained from first-principles calculations \cite{Alkauskas2014, Turiansky2021, Waldhoer2021}. These coefficients are therefore not adjustable parameters, but are determined by the underlying defect physics. Importantly, both electron capture and electron emission require the system to overcome a corresponding barrier in the configuration-coordinate diagram (see \Cref{fig:lsf_classic}). These NMP barriers arise naturally from the structural relaxation accompanying the charge transition and enter the transition rates through the NMP line-shape functions, giving rise to the characteristic Arrhenius temperature dependence in the high-temperature limit. At the same time, the NMP barriers are not fixed but depend on the relative alignment of the charge-transition level with respect to the conduction-band edge, which is controlled by the electrostatic potential (see \Cref{fig:charge_transition_level}). Consequently, the NMP transition rates exhibit not only an intrinsic temperature dependence through the thermally activated crossing of the barriers, but also a pronounced bias dependence through the electrostatic modulation of the barriers themselves. Both the experimentally observed temperature and bias dependence of the transition rates \cite{Waldhoer2021} therefore emerge naturally from the microscopic defect physics captured by the NMP framework. \\

In contrast, the SRH model completely neglects the vibrational degrees of freedom of the defect system. It therefore does not describe the structural relaxation following a change in the defect charge state and contains no NMP-type activation barriers. Consequently, the SRH capture and emission coefficients, \(c_\mathrm{SRH}^{\mathrm{CB}}\) and \(e_\mathrm{SRH}^{\mathrm{CB}}\), do not naturally contain the temperature and bias dependence arising from the structural defect relaxation. To reproduce the experimentally observed temperature and bias dependence, one must therefore introduce coefficients that are themselves temperature and bias dependent \cite{Kirton1989}. This effectively shifts the relevant physics into empirical fitting parameters. The limitation of the conventional SRH model becomes particularly apparent at cryogenic temperatures. In its standard form, the capture coefficient is written as
\(c_\mathrm{SRH}^{\mathrm{CB}}=\sigma_n v_{\mathrm{th}}\), where \(\sigma_n\) denotes the capture cross section and \(v_{\mathrm{th}}\) the carrier thermal velocity \cite{Sze2006}. Since \(v_{\mathrm{th}}\) vanishes as \(T \to 0\), the SRH model predicts the complete suppression of both charge capture and emission at cryogenic temperatures. This prediction clearly contradicts experimental observations, which show that charge trapping persists even in the cryogenic regime \cite{Michl2021_1, Michl2021_2}.

\section{Device-Level Applications}
\label{sec:examples}
In the previous chapter, we derived continuum formulations for charge capture and emission processes between localized defect states and electronic bands. In this description, the bands act as charge reservoirs that supply or accept carriers during trapping and detrapping events. Within TCAD simulations, these NMP-based capture and emission rates can be evaluated locally using the spatially resolved band structure and quasi-Fermi levels, thereby enabling time-dependent modeling of charge trapping phenomena under realistic operating conditions. In the following, we demonstrate how the developed framework can be applied to a MOSFET in order to investigate threshold-voltage shifts caused by charge trapping under elevated stress conditions, a degradation phenomenon commonly referred to as bias temperature instability (BTI). \\

Defects located at the semiconductor--oxide interface of a MOSFET typically interact with several charge reservoirs at the same time, such as the conduction and valence bands of the semiconductor and, potentially, the gate electrode. Each reservoir provides an independent channel through which the defect can capture or emit a carrier.  Since these channels are mutually exclusive and compete in parallel, the probability that any capture or emission event occurs within a short time interval is the sum of the individual probabilities. Consequently, the total capture and emission rates are obtained by adding the contributions from all reservoirs,
\begin{equation}
    k^{\mathrm{tot}}_{\mathrm{c}}
    =
    \sum_r k^{(r)}_{\mathrm{c}},
    \qquad
    k^{\mathrm{tot}}_{\mathrm{e}}
    =
    \sum_r k^{(r)}_{\mathrm{e}}.
\end{equation}
where \(r\) labels the individual reservoirs. Each reservoir-specific transition rate can be evaluated either by direct numerical integration (see \Cref{eq:band_emission_ldos,eq:band_capture_ldos}), or by employing the band-edge approximation (see \Cref{eq:emission_leading,eq:capture_leading,eq:clamped_emission_leading,eq:clamped_capture_leading}). \\

Charge trapping is an inherently stochastic process. An individual defect does not capture or emit charge at a deterministic time, but instead undergoes random transitions between its charge states with probabilities determined by the corresponding total capture and emission rates. In large-area devices, however, one is typically interested in the collective behavior of many nearly identical defects within a small volume of the device, rather than the stochastic trajectory of a single defect. Such a local ensemble of $N_{\mathrm{T}}$ defects is conveniently described by the fraction of occupied defects $P_{\mathrm{occu}}(t)=\left\langle P_i(t)\right\rangle$ and the fraction of unoccupied defects $P_{\mathrm{empty}}(t) = 1 - \left\langle P_i(t)\right\rangle$, where $P_i(t)$ denotes the occupancy of the $i$-th defect within the ensemble. Equivalently, these quantities can be interpreted as the probabilities of finding a defect from the ensemble in the occupied or unoccupied state, respectively. Since capture increases the occupied fraction while emission decreases it, the time evolution of the ensemble-averaged occupations reduces to a Pauli master equation \cite{Breuer2007}:
\begin{equation}
    \frac{\mathrm{d} P_{\mathrm{occu}}}{\mathrm{d} t}
    =
    - k_{\mathrm{e}}^{\mathrm{tot}} P_{\mathrm{occu}}
    + k_{\mathrm{c}}^{\mathrm{tot}} P_{\mathrm{empty}},
\end{equation}
\begin{equation}
    \frac{\mathrm{d} P_{\mathrm{empty}}}{\mathrm{d} t}
    =
    + k_{\mathrm{e}}^{\mathrm{tot}} P_{\mathrm{occu}}
    - k_{\mathrm{c}}^{\mathrm{tot}} P_{\mathrm{empty}}.
\end{equation}
Averaging over the ensemble therefore replaces the stochastic switching of individual defects by deterministic master equations for the probabilities of the two charge states. The corresponding average charge of the defect ensemble depends on whether the defects are acceptor-like or donor-like. For acceptor-like defects, the empty state is neutral, while the occupied state is negatively charged. The average defect charge is therefore
\begin{equation}
    Q_{\mathrm{T}}^{\mathrm{acc}}(t)
    =
    - q N_{\mathrm{T}} P_{\mathrm{occu}}(t),
\end{equation}
Conversely, for donor-like defects, the occupied state is neutral, while the empty state is positively charged. In this case, the average defect charge is
\begin{equation}
    Q_{\mathrm{T}}^{\mathrm{don}}(t)
    =
    + q N_{\mathrm{T}} P_{\mathrm{empty}}(t).
\end{equation}
This average charge of the ensemble enters the Poisson equation and results in a time-dependent shift of the device electrostatics, thereby affecting the stability of the device. It is important to note that the Poisson equation and the Pauli master equation form a set of strongly coupled differential equations: the local electrostatic potential influences the capture and emission rates by shifting the band edges and the CTL, while the defects themselves modify the electrostatic potential through their contribution to the space charge. Due to this nonlinear and time-dependent coupling, the resulting system generally cannot be solved analytically and must instead be treated self-consistently using numerical methods. In practical TCAD simulations, the Pauli master equation is commonly integrated in time using implicit Runge--Kutta methods, or Rosenbrock-type methods \cite{Hairer1993, Hairer1996}. These schemes provide robust numerical stability across the wide range of timescales characteristic of charge-trapping phenomena. \\

\begin{figure}[!hbt]
    \centering

    \begin{subfigure}{0.3\textwidth}
        \centering
        \includegraphics[height=\figheight,keepaspectratio]{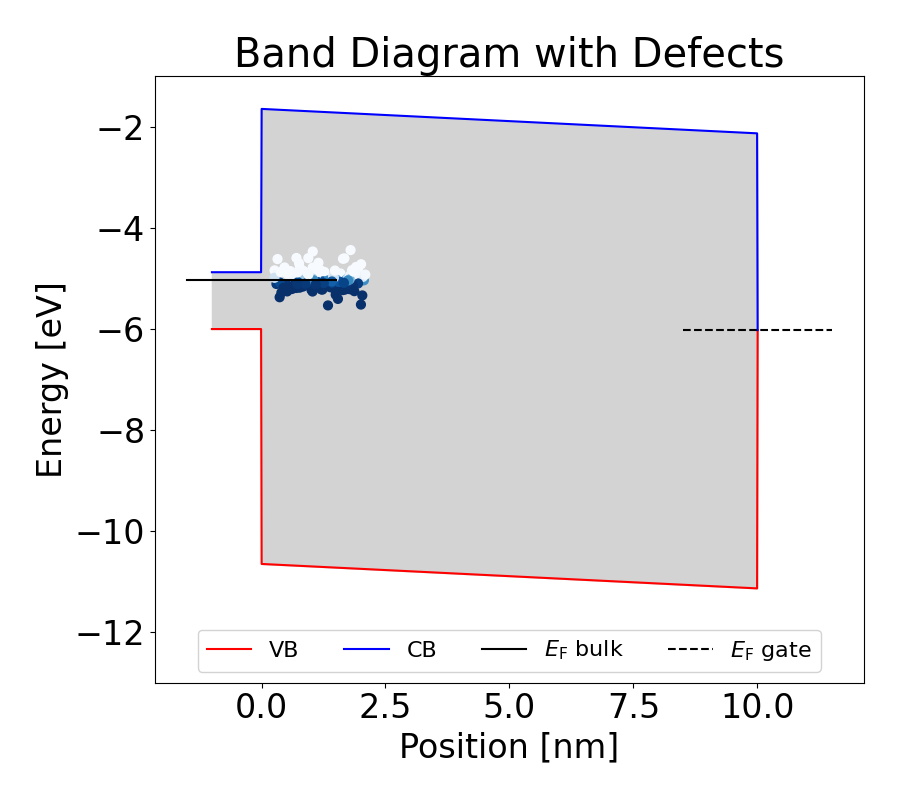}
        \caption{ }
        \label{fig:BTI_experiment_1}
    \end{subfigure}
    \hfill
    \begin{subfigure}{0.3\textwidth}
        \centering
        \includegraphics[height=\figheight,keepaspectratio]{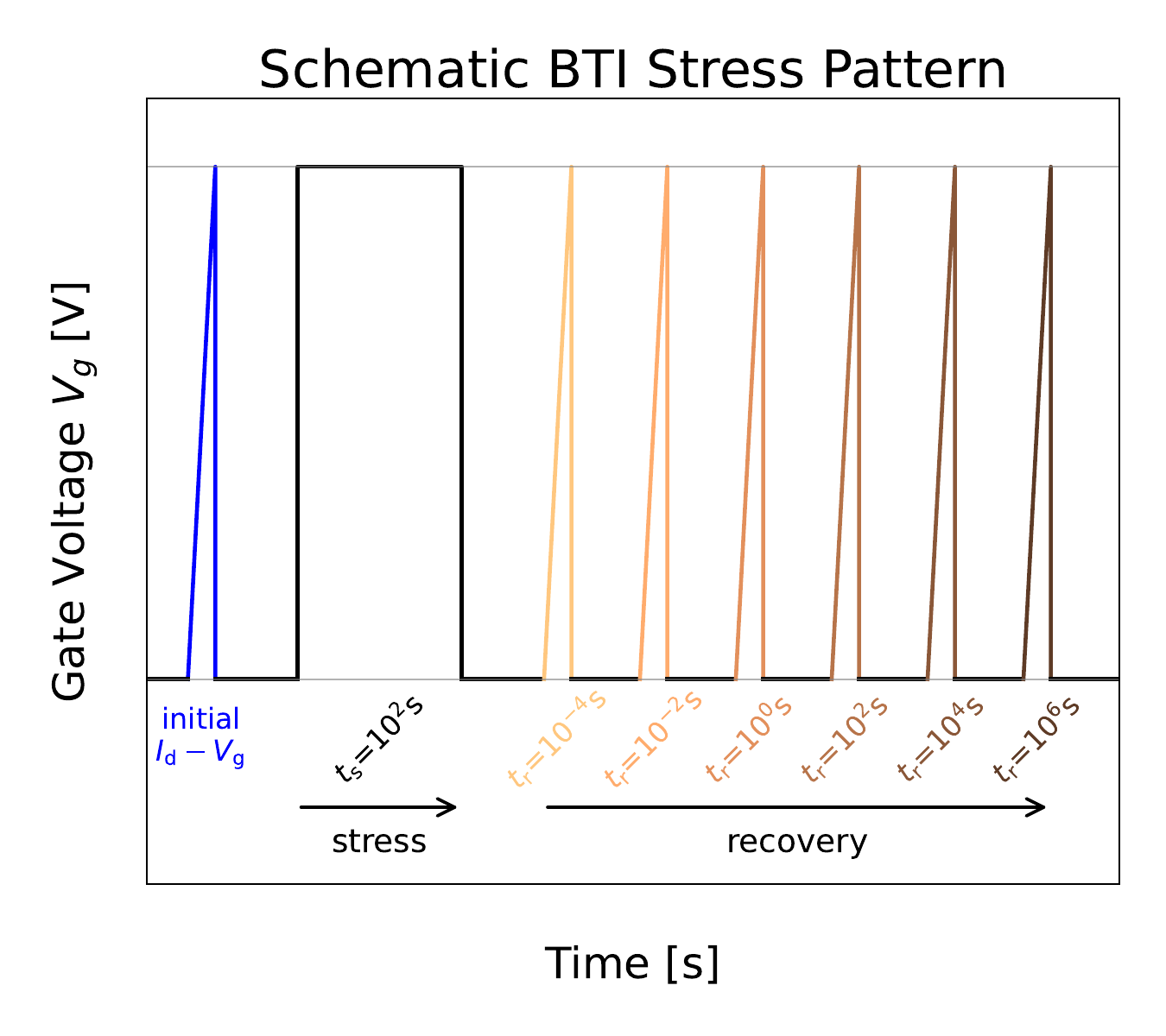}
        \caption{ }
        \label{fig:BTI_experiment_2}
    \end{subfigure}
    \hfill
    \begin{subfigure}{0.3\textwidth}
        \centering
        \includegraphics[height=\figheight,keepaspectratio]{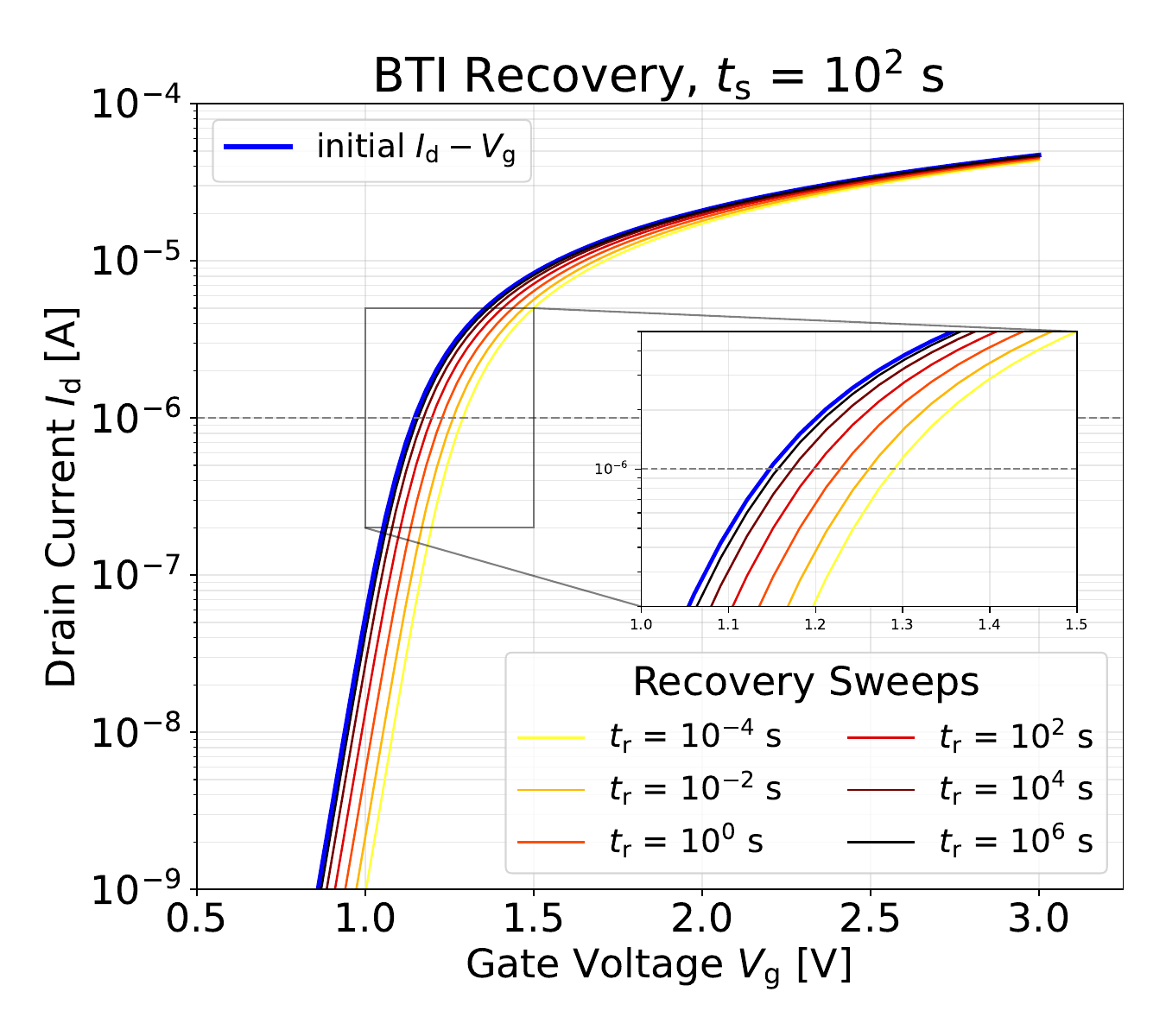}
        \caption{ }
        \label{fig:BTI_experiment_3}
    \end{subfigure}
    \caption{
    Simulation of bias-temperature instability (BTI) in a MOSFET due to acceptor-like NMP defects.
    The defects are sampled uniformly in the first 2 nm of the oxide, while Gaussian distributions are assumed for $E_\mathrm{T}$ and $E_\mathrm{R}$ using the parameters $E_\mathrm{T}=4.9 \pm 0.2\,\mathrm{eV}$ 
    $E_\mathrm{R}=2.25 \pm 0.75\,\mathrm{eV}$ (the CTL is given relative to the oxide valence-band edge).
    (a) Band diagram of the \ce{Si}/\ce{SiO2} MOSFET structure. Oxide defects located near the channel are shown as circles, with blue indicating occupied defects and white indicating empty defects. (b) Schematic BTI stress--recovery pattern. First, an initial $I_\mathrm{d}-V_\mathrm{g}$ sweep is performed to extract the threshold voltage of the unstressed device. The device is then stressed at $V_\mathrm{s}$ for a stress time $\Delta t_\mathrm{s}$. During the subsequent recovery phase at $V_\mathrm{rec}$, several $I_\mathrm{d}-V_\mathrm{g}$ sweeps are performed at logarithmically spaced recovery times to monitor the relaxation of threshold voltage.
    (c) Corresponding simulated $I_\mathrm{d}$--$V_\mathrm{g}$ curves after a stress time of $t_\mathrm{s}=10^2\,\mathrm{s}$, illustrating the stress-induced threshold-voltage shift and its recovery.
    }
    \label{fig:BTI_experiment}
\end{figure}

\Cref{fig:BTI_experiment} illustrates a representative BTI charge-trapping simulation for a \ce{Si}/\ce{SiO2} MOSFET performed with the Compact Physics (Comphy) framework \cite{Rzepa2018,Waldhoer2023}. The band diagram in \Cref{fig:BTI_experiment_1} shows defects in the oxide of the MOSFET, where they can exchange charge with both the channel and the metal gate. When the device is stressed with an elevated gate bias, the channel is driven into strong inversion and the Fermi level $E_\mathrm{F}$ approaches the conduction band of the channel promoting electron capture by oxide defects. This modifies the device electrostatics and shifts the threshold voltage. After the stress is removed, occupied defects may emit charge back to the channel or gate, leading to a recovery of the threshold-voltage shift. This degradation and recovery behavior is commonly characterized using the standard BTI stress--recovery sequence shown in \Cref{fig:BTI_experiment_2}. First, an initial $I_\mathrm{d}$--$V_\mathrm{g}$ curve is recorded to characterize the unstressed device. The device is then subjected to an elevated stress bias $V_\mathrm{s}$ for a duration $t_\mathrm{s}$, allowing the oxide defects to capture charge. After stress, the gate bias is switched to the recovery value $V_\mathrm{rec}$, and $I_\mathrm{d}$--$V_\mathrm{g}$ curves are recorded at logarithmically spaced recovery times $t_\mathrm{rec}$ to monitor the device recovery. The resulting degradation can be quantified from the recorded transfer curves as shown in \Cref{fig:BTI_experiment_3} by extracting the time-dependent threshold voltage shift $\Delta V_\mathrm{th}(t_\mathrm{r})$ between the stressed and unstressed device. For a consistent extraction of $\Delta V_\mathrm{th}(t_\mathrm{r})$, the same criterion is applied to all transfer curves; in this example, a constant-current criterion of $I_\mathrm{d}=1.0 \times 10^{-6}\,\mathrm{A}$ is used.
 \\ 

\begin{figure}[!hbt]
    \centering
    \begin{subfigure}{0.3\textwidth}
        \centering
        \includegraphics[height=\figheight,keepaspectratio]{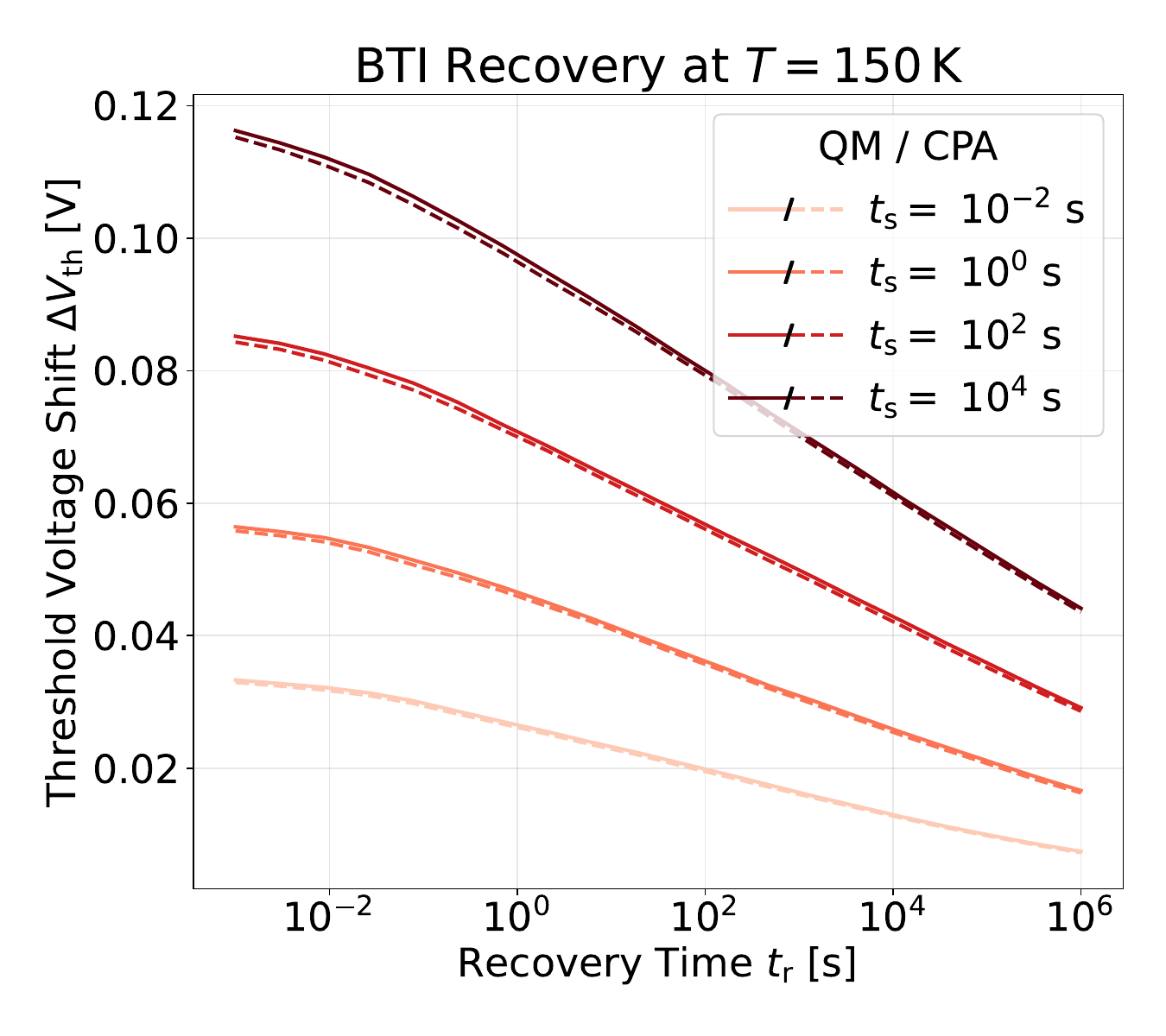}
        \caption{ }
        \label{fig:BTI_recovery_1}
    \end{subfigure}
    \hfill
    \begin{subfigure}{0.3\textwidth}
        \centering
        \includegraphics[height=\figheight,keepaspectratio]{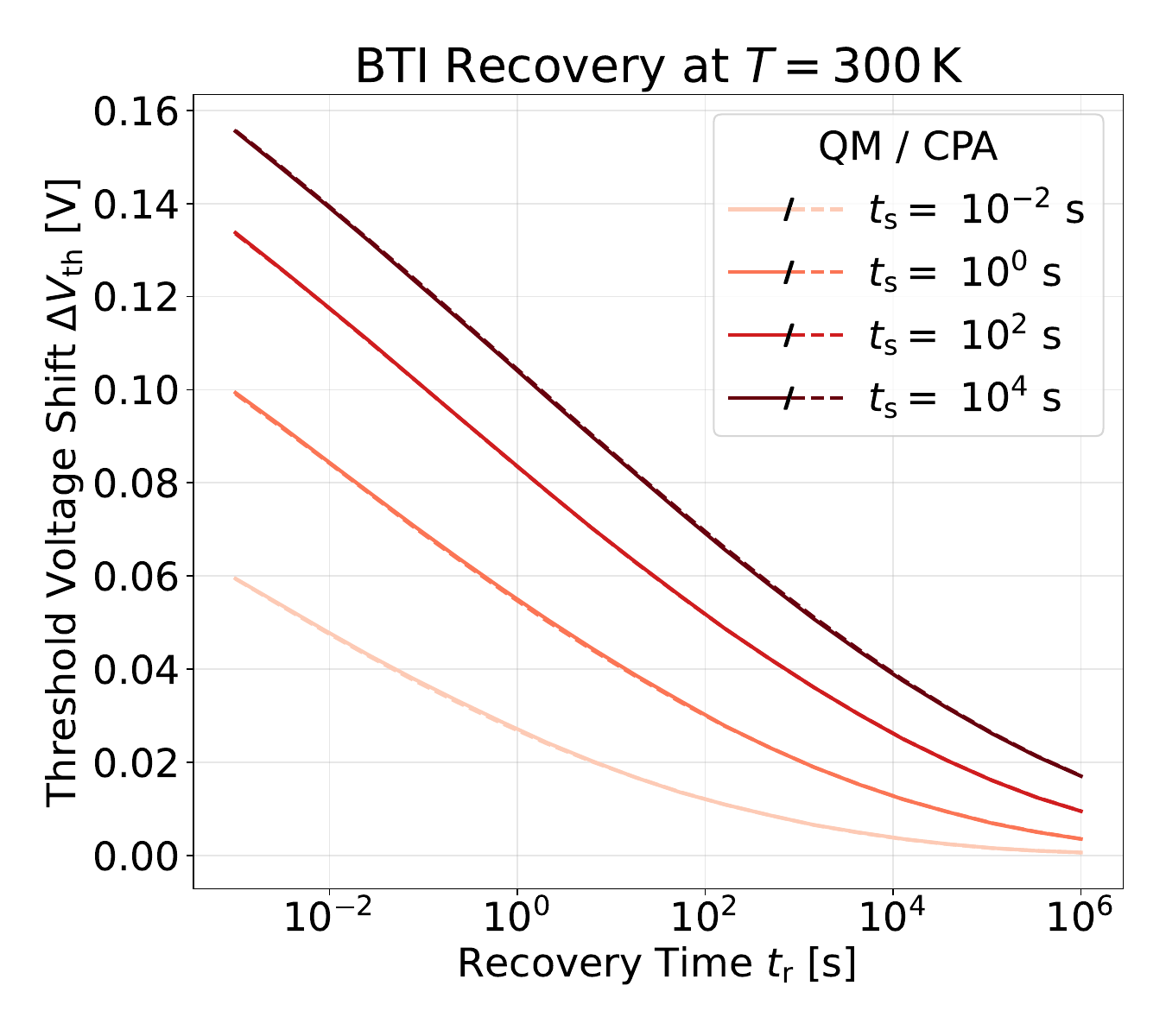}
        \caption{ }
        \label{fig:BTI_recovery_2}
    \end{subfigure}
    \hfill
    \begin{subfigure}{0.3\textwidth}
        \centering
        \includegraphics[height=\figheight,keepaspectratio]{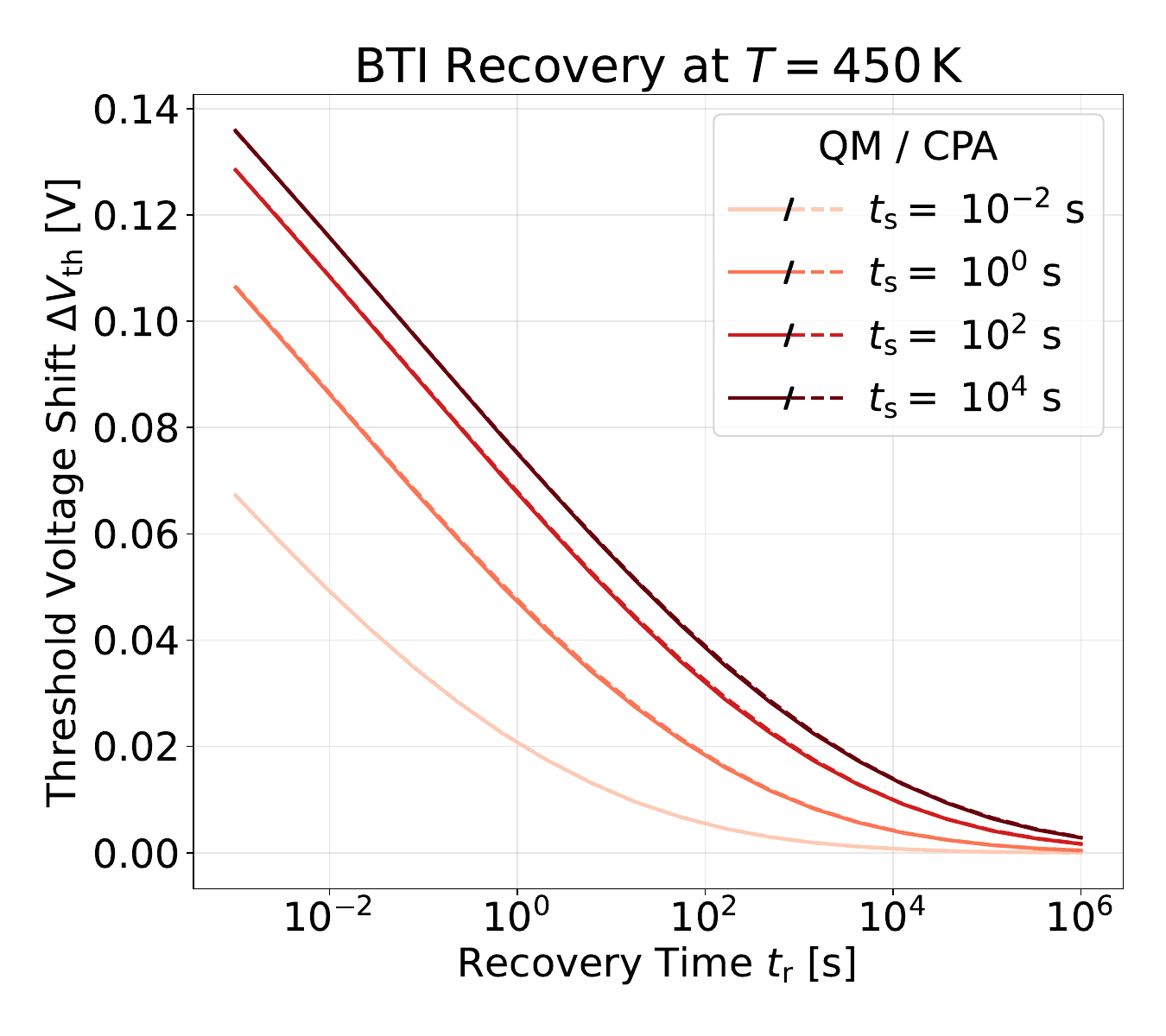}
        \caption{ }
        \label{fig:BTI_recovery_3}
    \end{subfigure}
    
    \vspace{0.1cm}
    
    \begin{subfigure}{0.3\textwidth}
        \centering
        \includegraphics[height=\figheight,keepaspectratio]{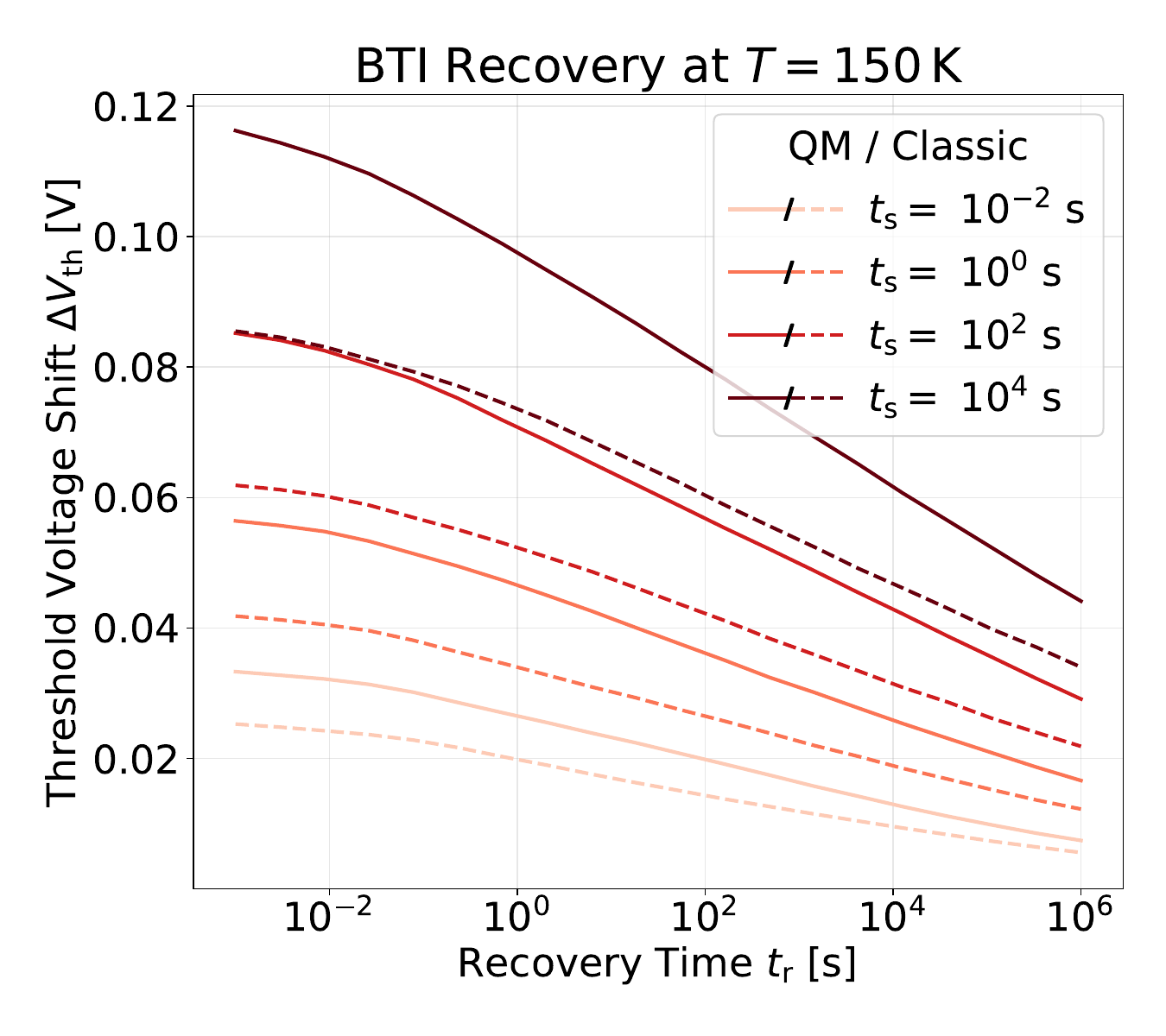}
        \caption{ }
        \label{fig:BTI_recovery_4}
    \end{subfigure}
    \hfill
    \begin{subfigure}{0.3\textwidth}
        \centering
        \includegraphics[height=\figheight,keepaspectratio]{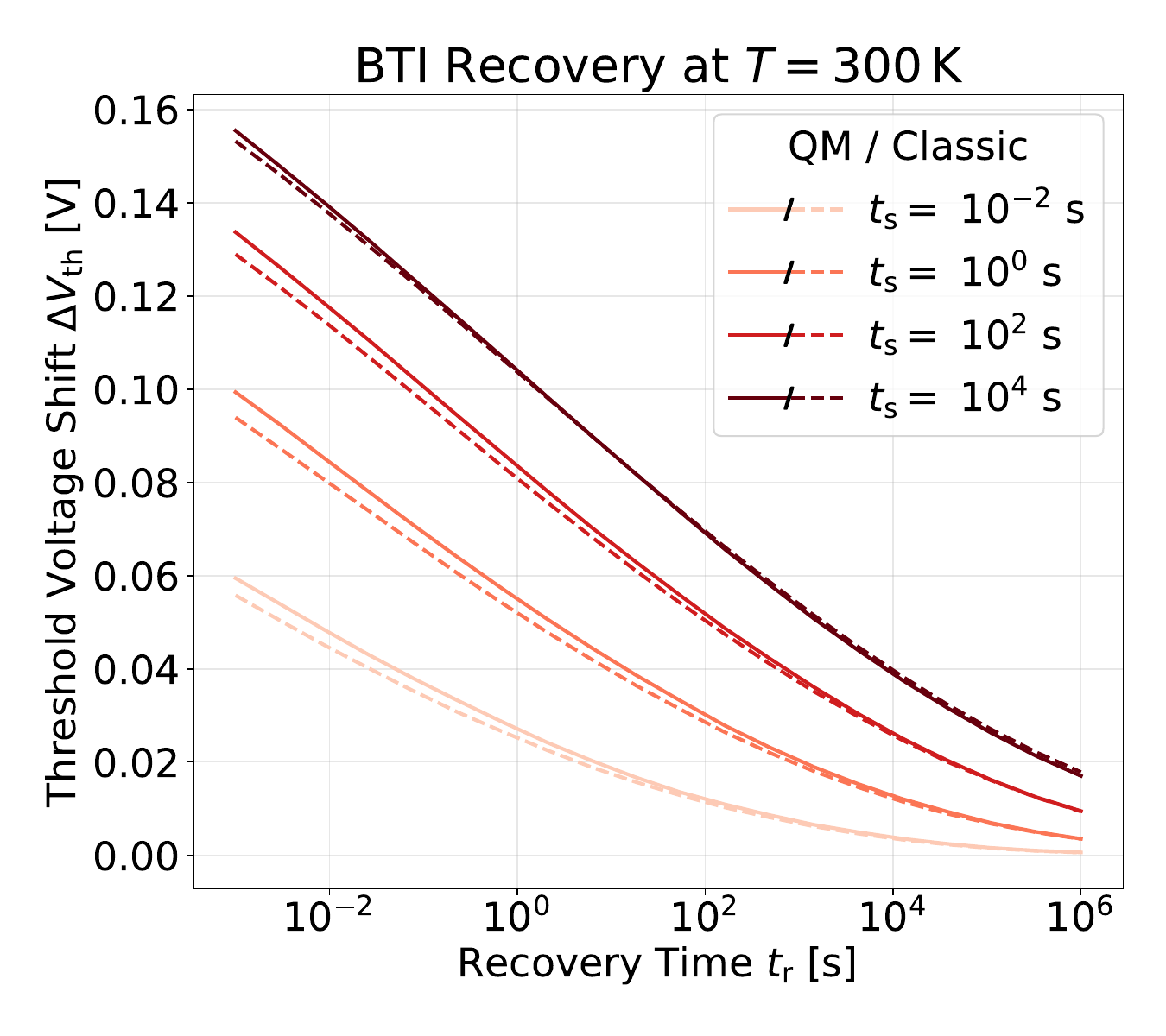}
        \caption{ }
        \label{fig:BTI_recovery_5}
    \end{subfigure}
    \hfill
    \begin{subfigure}{0.3\textwidth}
        \centering
        \includegraphics[height=\figheight,keepaspectratio]{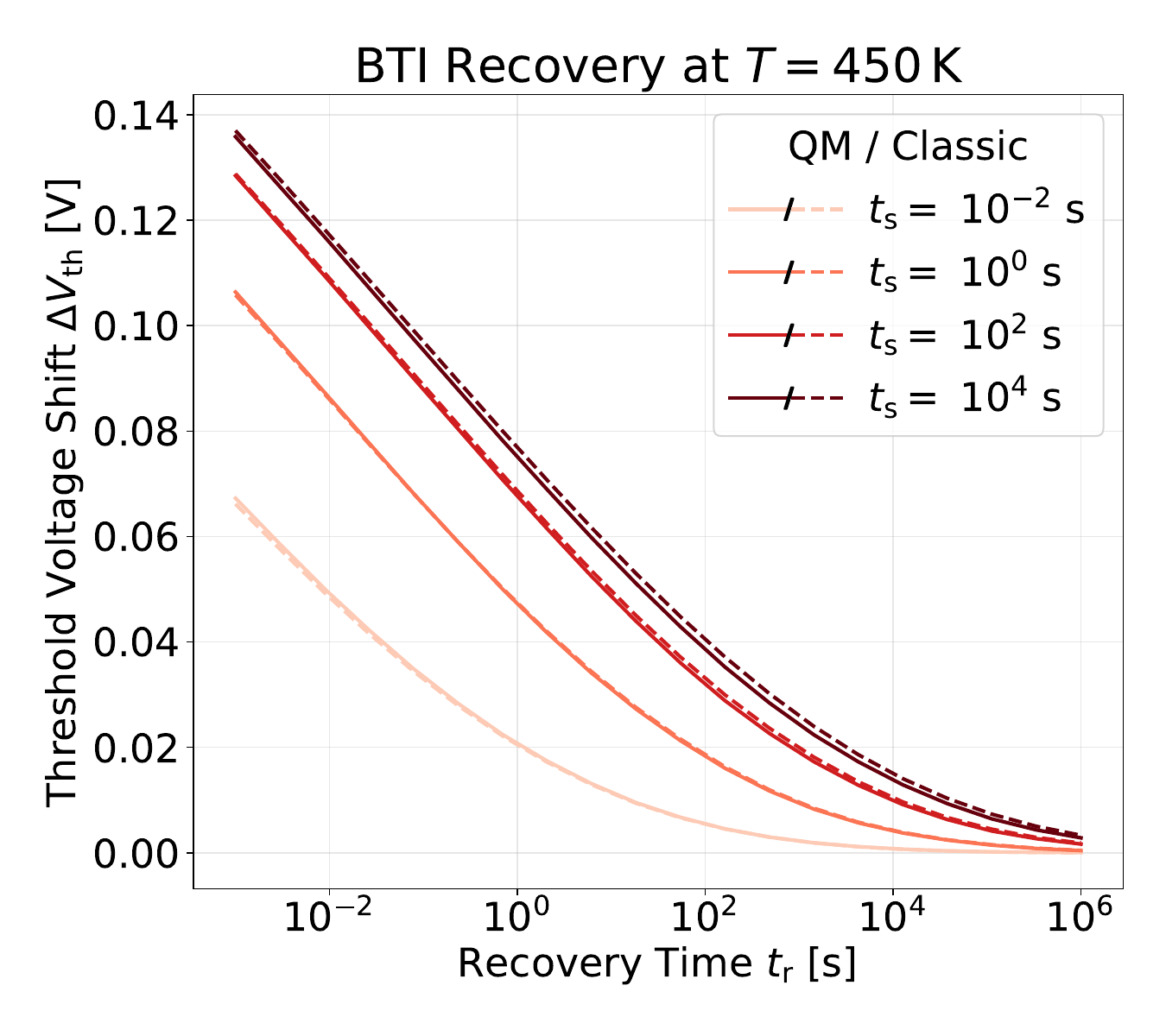}
        \caption{ }
        \label{fig:BTI_recovery_6}
    \end{subfigure}
    \caption{
    Simulated BTI experiments for temperatures of 150\,K, 300\,K, and 450\,K, and stress times of $1.0 \times 10^{-2}\,\mathrm{s}$, $1.0 \times 10^{0}\,\mathrm{s}$, $1.0 \times 10^{2}\,\mathrm{s}$, and $1.0 \times 10^{4}\,\mathrm{s}$.  Panels (a)--(c) compare the full quantum-mechanical rate calculation (QM), shown as solid lines, with the effective crossing-preserving approximation (CPA), shown as dashed lines, demonstrating very good agreement over the full temperature range. 
    Panels (d)--(f) compare the full quantum-mechanical rate calculation (QM), shown as solid lines, with the classical approximation (Classic), shown as dashed lines. The classical approximations shows good agreement with the full quantum-mechanical result at 450\,K and still provides reasonable agreement at 300\,K, but breaks down at 150\,K.
    }
    \label{fig:BTI_recovery}
\end{figure}

To investigate the time and temperature dependence of charge trapping, the BTI experiment is repeated for different temperatures and stress times. \Cref{fig:BTI_recovery} shows the resulting threshold-voltage shift $\Delta V_\mathrm{th}$ for temperatures of $150\,\mathrm{K}$, $300\,\mathrm{K}$, and $450\,\mathrm{K}$, and stress times of $10^{-2}\,\mathrm{s}$, $10^{0}\,\mathrm{s}$, $10^{2}\,\mathrm{s}$, and $10^{4}\,\mathrm{s}$. It compares the simulated recovery curves obtained using different approximations for the transition rates. \Cref{fig:BTI_recovery_1,fig:BTI_recovery_2,fig:BTI_recovery_3} show that the effective crossing-preserving approximation (CPA)  closely reproduces the full quantum-mechanical rate calculation at all three temperatures. In contrast, \Cref{fig:BTI_recovery_4,fig:BTI_recovery_5,fig:BTI_recovery_6} reveal a clear limitation of the classical approximation: while it shows good agreement with the full quantum-mechanical result at 450\,K and still provides reasonable agreement at 300\,K, it breaks down at 150\,K and fails to accurately capture the device degradation. This underscores the importance of quantum effects for describing BTI degradation at low temperatures, where a purely classical treatment is no longer sufficient. This behavior is consistent with earlier theoretical and experimental studies of charge trapping at cryogenic temperatures~\cite{Michl2021_1,Michl2021_2}, which demonstrated that the experimentally observed charge-trapping dynamics at cryogenic temperatures cannot be explained by classical NMP theory. Instead, a quantum-mechanical description of the transition rates, accounting for nuclear tunneling, is required to accurately reproduce the measured degradation and recovery kinetics.

\clearpage
\newpage

\section{Conclusions}

In this work, we developed a systematic framework for modeling nonradiative charge transfer processes in semiconductor devices within the nonradiative multiphonon (NMP) formalism. Starting from the quantum-mechanical description of coupled electron–nuclei dynamics, we derived a hierarchy of increasingly efficient rate models and established their connection to practical TCAD simulations. \\

Specifically, to overcome the main limitation of the full quantum-mechanical NMP model, namely its high computational cost, we introduced the crossing-preserving approximation (CPA). The CPA maps general unequal-curvature systems onto an effective equal-curvature representation while preserving the dominant crossing configuration that governs the transition dynamics. This construction yields a fully analytic, numerically stable, and computationally efficient approximation to the quantum-mechanical line-shape function. Comparisons with numerically exact calculations demonstrated that the CPA accurately reproduces quantum-mechanical transition rates over a broad parameter range, including regimes where the well established classical approximation fails due to tunneling effects. At the same time, the efficiency of the CPA makes quantum-mechanical NMP simulations of large defect populations feasible in practical TCAD environments. \\

Furthermore, we extended the description to defects that interact with electronic bands rather than only with discrete electronic states. For this case, the continuum transition rates are expressed as energy integrals over the band LDOS weighted by the corresponding carrier occupation probabilities, quantities readily available in semiconductor-device simulation frameworks. For near-equilibrium conditions, where the carrier occupation is described by Fermi--Dirac statistics, we then developed a band-edge approximation that replaces the full energy integral by a single-point evaluation at the dominant electronic energy. This greatly reduces the computational cost while retaining the relevant microscopic defect physics, thereby enabling efficient and physically consistent modeling of charge trapping phenomena in semiconductor devices. \\

Another major objective of this work was to provide a comprehensive guide for implementing NMP-based models. To this end, the manuscript presents a complete derivation of the formalism, discusses the physical interpretation of the model parameters, compares different evaluation strategies, and explicitly identifies the regimes in which common approximations remain valid. In particular, we clarified the limitations of the widely used classical approximation and derived criteria for the onset of quantum-mechanical tunneling effects. The resulting framework therefore serves not only as a source of efficient rate expressions but also as a practical reference for researchers and engineers implementing NMP models in simulation tools. \\

\clearpage

\section*{Author contributions}
The authors contributed equally to all aspects of the article. 

\section*{Competing interests}
The authors declare no competing interests. 

\section*{Data availability}
The data that support the findings of this study are available from the corresponding author upon reasonable request.

\section*{Supplementary Information}
\subsection*{Supplementary Information SI-1: Detailed-Balance Condition}
\label{sec:detailed_balance_derivation}

In this section, we derive the detailed-balance relation satisfied by the nonradiative multiphonon (NMP) transition rates. We first present the derivation within the full quantum-mechanical formalism, where the transition rates are expressed in terms of vibronic line-shape functions. This derivation is completely general and does not rely on the harmonic approximation. We then specialize the result to harmonic potential-energy surfaces, obtaining an explicit expression for the free-energy difference and its high-temperature limit. Next, we derive the corresponding detailed-balance relation directly within the classical NMP formalism. Finally, we compare the quantum and classical expressions, demonstrating that the latter is recovered as the high-temperature limit of the former, thereby establishing the consistency between the two formulations.


\subsubsection*{A. Quantum-mechanical detailed balance}

We begin by considering the quantum-mechanical expression for the NMP line-shape function describing the forward transition from the initial electronic state $i$ to the final state $f$. The line-shape function is obtained by summing over all vibronic transitions weighted by the thermal occupation of the initial vibronic states,
\begin{equation}
\eta_{if}
=
\sum_m w_m^{(i)}
\sum_n
\left|
\langle \chi_{i,m}|Q|\chi_{f,n}\rangle
\right|^2
\delta(E_{i,m}-E_{f,n}),
\end{equation}
where
\begin{equation}
w_m^{(i)}
=
\frac{\exp(-\beta E_{i,m})}{Z_i},
\qquad
Z_i
=
\sum_m
\exp(-\beta E_{i,m}).
\end{equation}

The corresponding reverse line-shape function is obtained by interchanging the initial and final vibronic manifolds,
\begin{equation}
\eta_{fi}
=
\sum_n w_n^{(f)}
\sum_m
\left|
\langle \chi_{f,n}|Q|\chi_{i,m}\rangle
\right|^2
\delta(E_{f,n}-E_{i,m}),
\end{equation}
with
\begin{equation}
w_n^{(f)}
=
\frac{\exp(-\beta E_{f,n})}{Z_f},
\qquad
Z_f
=
\sum_n
\exp(-\beta E_{f,n}).
\end{equation}

The transition matrix elements satisfy
\begin{equation}
\left|
\langle \chi_{f,n}|Q|\chi_{i,m}\rangle
\right|^2
=
\left|
\langle \chi_{i,m}|Q|\chi_{f,n}\rangle
\right|^2,
\end{equation}
as a consequence of the Hermiticity of the configuration coordinate operator. Furthermore, the Dirac delta function is symmetric,
\begin{equation}
\delta(E_{f,n}-E_{i,m})
=
\delta(E_{i,m}-E_{f,n}).
\end{equation}
Therefore, the forward and reverse line-shape functions contain identical vibronic transition matrix elements, differing only in the thermal occupation factors associated with the initial states.

Since the delta function enforces energy conservation,
\begin{equation}
E_{i,m}=E_{f,n},
\end{equation}
the ratio of the Boltzmann weights becomes
\begin{equation}
\frac{w_m^{(i)}}{w_n^{(f)}}
=
\frac{Z_f}{Z_i}
\frac{\exp(-\beta E_{i,m})}
{\exp(-\beta E_{f,n})}
=
\frac{Z_f}{Z_i},
\end{equation}
which immediately gives
\begin{equation}
\frac{\eta_{if}}{\eta_{fi}}
=
\frac{Z_f}{Z_i}.
\end{equation}
Introducing the Helmholtz free energy,
\begin{equation}
F_a
=
-\frac{1}{\beta}\ln Z_a,
\qquad
a\in\{i,f\},
\end{equation}
we obtain the exact detailed-balance relation
\begin{equation}
\frac{k_{if}}{k_{fi}}
=
\frac{\eta_{if}}{\eta_{fi}}
=
\exp(-\beta\Delta F),
\end{equation}
where
\begin{equation}
\Delta F=F_f-F_i.
\end{equation}

This relation is exact and holds for arbitrary vibronic spectra. No assumptions regarding the shape of the potential-energy surfaces have been made up to this point. To obtain an explicit expression for the free-energy difference, we now specialize to harmonic potential-energy surfaces. In this case, the vibronic energies are given by
\begin{equation}
E_{a,n}
=
E_a
+
\hbar\Omega_a
\left(
n+\frac12
\right),
\qquad
a\in\{i,f\},
\end{equation}
and the corresponding partition function can be evaluated analytically,
\begin{equation}
Z_a
=
\sum_{n=0}^{\infty}
\exp\!\left[
-\beta
\left(
E_a
+
\hbar\Omega_a
\left(
n+\frac12
\right)
\right)
\right]
=
\frac{\exp(-\beta E_a)}
{2\sinh(\beta\hbar\Omega_a/2)}.
\end{equation}
The free-energy difference therefore becomes
\begin{equation}
\Delta F
=
\Delta E
+
\frac{1}{\beta}
\ln
\left[
\frac{
\sinh(\beta\hbar\Omega_f/2)}
{\sinh(\beta\hbar\Omega_i/2)}
\right],
\end{equation}
where
\begin{equation}
\Delta E
=
E_f
-
E_i.
\end{equation}
Substituting this expression into the general detailed-balance relation yields
\begin{equation}
\boxed{
\frac{k_{if}}{k_{fi}}
=
\exp(-\beta\Delta E)
\,
\frac{
\sinh(\beta\hbar\Omega_i/2)}
{\sinh(\beta\hbar\Omega_f/2)}
}.
\end{equation}
This expression explicitly shows that, in the quantum-mechanical description, detailed balance depends not only on the electronic energy difference but also on the quantum vibrational free-energy contribution arising from the different phonon frequencies.


\subsubsection*{B. Classical detailed balance}

We now derive the detailed-balance relation independently within the classical NMP formalism. This provides an explicit verification that the classical theory satisfies the same thermodynamic constraints.
The classical forward line-shape function is
\begin{equation}
\eta_{if}^{\rm cl}
=
\int dQ\,
\frac{\exp[-\beta V_i(Q)]}
{Z_i^{\rm cl}}
Q^2
\delta\!\left[V_i(Q)-V_f(Q)\right],
\end{equation}
where
\begin{equation}
Z_i^{\rm cl}
=
\int dQ\,
\exp[-\beta V_i(Q)].
\end{equation}
The corresponding reverse line-shape function is obtained by interchanging the initial and final potential-energy surfaces,
\begin{equation}
\eta_{fi}^{\rm cl}
=
\int dQ\,
\frac{\exp[-\beta V_f(Q)]}
{Z_f^{\rm cl}}
Q^2
\delta\!\left[V_f(Q)-V_i(Q)\right],
\end{equation}
with
\begin{equation}
Z_f^{\rm cl}
=
\int dQ\,
\exp[-\beta V_f(Q)].
\end{equation}
Since the delta function is symmetric,
\begin{equation}
\delta\!\left[V_f(Q)-V_i(Q)\right]
=
\delta\!\left[V_i(Q)-V_f(Q)\right],
\end{equation}
and on its support one has
\begin{equation}
V_i(Q)=V_f(Q),
\end{equation}
the Boltzmann factors satisfy
\begin{equation}
\frac{
\exp[-\beta V_i(Q)]/Z_i^{\rm cl}
}{
\exp[-\beta V_f(Q)]/Z_f^{\rm cl}
}
=
\frac{Z_f^{\rm cl}}{Z_i^{\rm cl}},
\end{equation}
which immediately yields
\begin{equation}
\frac{\eta_{if}^{\rm cl}}
{\eta_{fi}^{\rm cl}}
=
\frac{Z_f^{\rm cl}}
{Z_i^{\rm cl}}.
\end{equation}
For harmonic potentials,
\begin{equation}
Z_a^{\rm cl}
=
e^{-\beta E_a}
\sqrt{\frac{2\pi}{\beta\Omega_a^2}},
\end{equation}
which leads to
\begin{equation}
\boxed{
\frac{k_{if}^{\rm cl}}
{k_{fi}^{\rm cl}}
=
\frac{\Omega_i}{\Omega_f}
\exp(-\beta\Delta E).
}
\end{equation}


\subsubsection*{C. Comparison between the quantum and classical results}

The quantum and classical derivations follow the same underlying physical principle: detailed balance originates from the symmetry of the transition probability together with the thermal occupation of the initial states. The distinction between the two formulations arises entirely from the different forms of the partition functions describing quantum and classical vibrational statistics.
The quantum-mechanical expression,
\begin{equation}
\frac{k_{if}}{k_{fi}}
=
\exp(-\beta\Delta E)
\,
\frac{
\sinh(\beta\hbar\Omega_i/2)}
{\sinh(\beta\hbar\Omega_f/2)},
\end{equation}
contains the complete quantum vibrational contribution through the harmonic partition functions. In contrast, the classical treatment yields
\begin{equation}
\frac{k_{if}^{\rm cl}}
{k_{fi}^{\rm cl}}
=
\frac{\Omega_i}{\Omega_f}
\exp(-\beta\Delta E).
\end{equation}
Expanding the quantum expression in the limit $\beta\hbar\Omega\ll1$ gives
\begin{equation}
\sinh(\beta\hbar\Omega/2)
\simeq
\frac{\beta\hbar\Omega}{2},
\end{equation}
from which the quantum result reduces exactly to
\begin{equation}
\frac{k_{if}}{k_{fi}}
\longrightarrow
\frac{\Omega_i}{\Omega_f}
\exp(-\beta\Delta E).
\end{equation}
Thus, the independently derived classical detailed-balance relation is recovered as the high-temperature limit of the full quantum-mechanical theory. This agreement provides an important consistency check on both formulations of the NMP transition-rate theory and confirms that the quantum formalism correctly reproduces the expected classical behavior in the appropriate thermodynamic limit.

\subsection*{Supplementary Information SI-2: Recursive Calculation of Franck-Condon Matrix Elements}

The direct numerical evaluation of the Franck--Condon overlaps
\begin{equation}
A_{mn}
=
\langle \chi_{i,m} | \chi_{f,n} \rangle
\end{equation}
from the explicit harmonic-oscillator wave functions rapidly becomes numerically inefficient for high vibrational quantum numbers. A direct implementation requires repeated evaluation of Hermite polynomials and highly oscillatory Gaussian integrals, which becomes increasingly susceptible to overflow, underflow, and cancellation errors for large $m$ and $n$. To avoid these issues, the overlap matrix is most efficiently constructed using recursive relations \cite{Drallos1986,Schmidt2010}.

The recursion is initialized with the ground-state overlap
\begin{equation}
A_{00}
=
\left(
\frac{2\sqrt{\Omega_i\Omega_f}}{\Omega_i+\Omega_f}
\right)^{1/2}
\exp\left[
-\frac{\Omega_i\Omega_f \Delta Q^2}
{2(\Omega_i+\Omega_f)}
\right].
\end{equation}

The first column and first row are then generated from
\begin{equation}
A_{m+2,0}
=
\sqrt{\frac{m+1}{m+2}}
\frac{\Omega_i-\Omega_f}{\Omega_i+\Omega_f}
A_{m,0}
+
\sqrt{\frac{2}{m+2}}
\frac{\sqrt{\Omega_i}\Omega_f}{\Omega_i+\Omega_f}
\Delta Q\,A_{m+1,0},
\end{equation}
and
\begin{equation}
A_{0,n+2}
=
-\sqrt{\frac{n+1}{n+2}}
\frac{\Omega_i-\Omega_f}{\Omega_i+\Omega_f}
A_{0,n}
-
\sqrt{\frac{2}{n+2}}
\frac{\Omega_i\sqrt{\Omega_f}}{\Omega_i+\Omega_f}
\Delta Q\,A_{0,n+1}.
\end{equation}

All remaining matrix elements are obtained from the two-dimensional recursion
\begin{align}
A_{m+1,n+1}
={}&
K_1\sqrt{\frac{1}{n+1}}A_{m+1,n}
+
K_2\sqrt{\frac{n}{n+1}}A_{m+1,n-1}
+
K_3\sqrt{\frac{1}{m+1}}A_{m,n+1}
\nonumber\\
&+
K_4\sqrt{\frac{1}{(m+1)(n+1)}}A_{m,n}
+
K_5\sqrt{\frac{n}{(m+1)(n+1)}}A_{m,n-1}
\nonumber\\
&+
K_6\sqrt{\frac{m}{m+1}}A_{m-1,n+1}
+
K_7\sqrt{\frac{m}{(m+1)(n+1)}}A_{m-1,n}
\nonumber\\
&+
K_8\sqrt{\frac{mn}{(m+1)(n+1)}}A_{m-1,n-1},
\end{align}
with
\begin{align}
K_1 &= -\frac{\Omega_i\sqrt{\Omega_f}\Delta Q}
{\sqrt{2}(\Omega_i+\Omega_f)},&
K_2 &= -\frac{\Omega_i-\Omega_f}{\Omega_i+\Omega_f},&
K_3 &= \frac{\sqrt{\Omega_i}\Omega_f\Delta Q}
{\sqrt{2}(\Omega_i+\Omega_f)},&
K_4 &= \frac{2\sqrt{\Omega_i\Omega_f}}{\Omega_i+\Omega_f},
\nonumber\\
K_5 &= -K_3,&
K_6 &= -K_2,&
K_7 &= -K_1,&
K_8 &= 1.
\end{align}

Terms containing negative indices are omitted. The recursion generates the complete Franck--Condon overlap matrix from the single element $A_{00}$ without explicitly evaluating the vibrational wave functions. The resulting algorithm scales approximately as $\mathcal{O}(N M)$ for a truncation at vibrational quantum numbers $(N, M)$ and only requires local arithmetic operations on previously computed matrix elements. \\

For large relaxation energy, the dominant overlaps are distributed over many phonon sidebands, whereas for weak lattice relaxation the matrix becomes concentrated close to the diagonal. Nevertheless, accurate evaluation of the line-shape function generally requires a large number of vibrational states ($N, M > 100$) to capture the exponentially decaying tails of the Franck--Condon distribution and to ensure convergence of the thermally weighted sums.

\subsection*{Supplementary Information SI-3: Estimate for the Breakdown of the Classical Approximation}

\begin{figure}[!hbt]
    \centering
    \begin{subfigure}{0.33\textwidth}
        \centering
        \includegraphics[width=\textwidth]{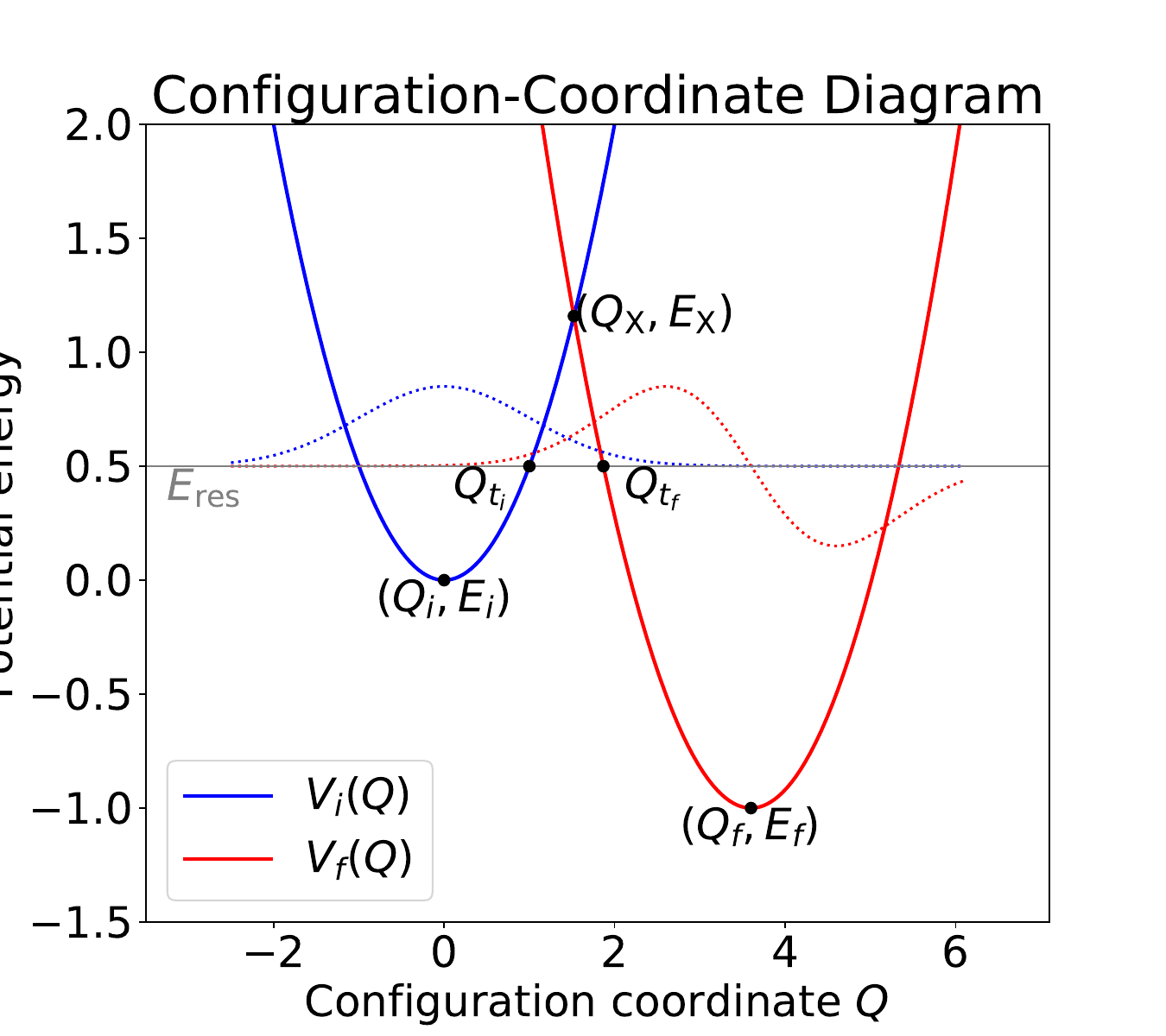}
    \end{subfigure}
    \caption{
    Configuration-coordinate diagram illustrating the WKB estimate for the breakdown of the classical approximation. The initial and final harmonic potential-energy surfaces \(V_i(Q)\) and \(V_f(Q)\) are shown together with their minima \((Q_i,E_i)\) and \((Q_f,E_f)\), the diabatic crossing point \((Q_\mathrm X,E_\mathrm X)\), and the resonant energy \(E_\mathrm{res}\). The turning points \(Q_{t_i}\) and \(Q_{t_f}\) define the classically forbidden region of the resonant vibrational states. The plotted wavefunctions \(\chi_{i,0}(Q)\) and \(\chi_{f,n^\ast}(Q)\) illustrate the opposing evanescent tails whose overlap dominates the low-temperature transition rate.
    }
    \label{fig:wkb_tunneling_geometry}
\end{figure}

In this section, we derive a simple estimate for the temperature below which the classical approximation to the NMP transition rate is expected to break down. The purpose of this estimate is not to define a sharp crossover temperature, but rather to identify the characteristic scale below which the purely classical picture is no longer sufficient and quantum-mechanical overlap of vibrational wavefunctions must be taken into account. We keep the notation independent of any specific choice of coordinate origin and the harmonic potentials are thus given by
\begin{equation}
V_a(Q)
=
E_a+\frac12\Omega_a^2(Q-Q_a)^2 \qquad \mathrm{with} \qquad a \in \{i, f\}
\end{equation}
Let \(Q_\mathrm X\) denote the dominant diabatic crossing point, and define the corresponding crossing energy by
$E_\mathrm X
=
V_i(Q_\mathrm X)
=
V_f(Q_\mathrm X).
$
In the classical limit, the transition rate follows an Arrhenius law governed by the activation barrier for reaching the crossing point,
\begin{equation}
k_{if}^\mathrm{cl}
\propto
\exp\!\left(
-\frac{\Delta E_\mathrm X}{k_\mathrm B T}\right)
\qquad \mathrm{with} \qquad
\Delta E_\mathrm X
=
E_\mathrm X - E_i
.
\label{eq:SI_classical_Arrhenius}
\end{equation}
At low temperature, the classical picture ceases to be adequate because the initial vibrational population is dominated by the ground state. The quantum-mechanical line-shape function is then governed by transitions from \(\chi_{i,0}\) into final vibrational states \(\chi_{f,n}\) whose energies approximately satisfy energy conservation. We denote this common resonant energy by
\begin{equation}
E_\mathrm{res}
\equiv
E_{i,0}
\approx
E_{f,n^\ast},
\label{eq:SI_Eres_def}
\end{equation}
where \(n^\ast\) is the resonant final vibrational state. The relevant low-temperature matrix element is therefore
\begin{equation}
\langle \chi_{i,0}|Q|\chi_{f,n^\ast}\rangle
=
\int \mathrm dQ\,
\chi_{i,0}(Q)\,Q\,\chi_{f,n^\ast}(Q).
\end{equation}
In the following, we assume that the the crossing point between the two potential energy surfaces lies between the classical turning points, i.e. $Q_{t_i} < Q_\mathrm{X} < Q_{t_f}$. As illustrated in \Cref{fig:wkb_tunneling_geometry}, this ordering ensures that the evanescent tails of the initial and final vibrational wavefunctions decay in opposite directions in the crossing region. As a result, the product \(\chi_{i,0}(Q)\chi_{f,n^\ast}(Q)\) exhibits an interior maximum $Q^\ast$ between the two turning points \(Q_{t_i}\) and \(Q_{t_f}\).

To locate this maximum, we approximate the exponential part of the overlap integral using the WKB approximation. In the classically forbidden region, the vibrational wavefunctions can be approximated by the WKB form
\begin{equation}
\chi(Q)
\sim
\frac{1}{\sqrt{\kappa(Q)}}
\exp\!\left[
-\int^Q \mathrm dQ'\,\kappa(Q')
\right],
\qquad
\kappa(Q)
=
\frac{1}{\hbar}
\sqrt{2\bigl(V(Q)-E_\mathrm{res}\bigr)} .
\end{equation}
For \(Q_{t_i}<Q<Q_{t_f}\), the initial-state wavefunction decays from \(Q_{t_i}\) toward increasing \(Q\), while the final-state wavefunction decays from \(Q_{t_f}\) toward decreasing \(Q\). Thus,
\begin{equation}
\chi_{i,0}(Q)\sim e^{-S_i(Q)/\hbar},
\qquad
\chi_{f,n^\ast}(Q)\sim e^{-S_f(Q)/\hbar},
\end{equation}
with
\begin{equation}
S_i(Q)
=
\int_{Q_{t_i}}^{Q}
\mathrm dQ'\,
\sqrt{2\bigl(V_i(Q')-E_\mathrm{res}\bigr)}
\end{equation}
and
\begin{equation}
S_f(Q)
=
\int_{Q}^{Q_{t_f}}
\mathrm dQ'\,
\sqrt{2\bigl(V_f(Q')-E_\mathrm{res}\bigr)} .
\end{equation}
The product of the two evanescent tails is therefore exponentially suppressed as
\begin{equation}
\chi_{i,0}(Q)\chi_{f,n^\ast}(Q)
\sim
\exp\!\left[
-\frac{1}{\hbar}
\left(
S_i(Q)+S_f(Q)
\right)
\right].
\end{equation}
The overlap integral is therefore dominated by the coordinate \(Q^\ast\) that minimizes the total action
\begin{equation}
S_\mathrm{tot}(Q)=S_i(Q)+S_f(Q).
\end{equation}
Minimization gives
\begin{equation}
V_i(Q^\ast)-E_\mathrm{res}
=
V_f(Q^\ast)-E_\mathrm{res},
\end{equation}
or simply
\begin{equation}
V_i(Q^\ast)=V_f(Q^\ast).
\end{equation}
Thus the dominant contribution arises from the vicinity of the diabatic crossing point,
\begin{equation}
Q^\ast \approx Q_\mathrm X .
\end{equation}
Therefore, to leading exponential accuracy,
\begin{equation}
\left|
\langle \chi_{i,0}|Q|\chi_{f,n^\ast}\rangle
\right|^2
\propto
\exp\!\left[
-\frac{2}{\hbar}
\left(
S_i(Q_\mathrm X)+S_f(Q_\mathrm X)
\right)
\right].
\label{eq:SI_overlap_WKB}
\end{equation}
The breakdown of the classical approximation can now be estimated by comparing the thermal Arrhenius suppression in \Cref{eq:SI_classical_Arrhenius} with the quantum-mechanical tunneling suppression in \Cref{eq:SI_overlap_WKB}. Equating the exponents gives
\begin{equation}
k_\mathrm B T_\mathrm{quant}
\sim
\frac{\hbar\,\Delta E_\mathrm X}
{2\left[
S_i(Q_\mathrm X)+S_f(Q_\mathrm X)
\right]} .
\label{eq:SI_Tquant_exact_action_general}
\end{equation}
We now evaluate the actions exactly for harmonic potential-energy surfaces. For a general harmonic potential
\begin{equation}
V_a(Q)
=
E_a+\frac12\Omega_a^2(Q-Q_a)^2 ,
\end{equation}
the exact under-barrier action from the classical turning point to \(Q_\mathrm X\) is
\begin{equation}
S_a(Q_\mathrm X;E_\mathrm{res})
=
\int_{Q_{t,a}}^{Q_\mathrm X}
\mathrm dQ\,
\sqrt{2\left[V_a(Q)-E_\mathrm{res}\right]} .
\end{equation}
Introducing
\begin{equation}
x=|Q-Q_a|,
\qquad
x_t=\frac{\sqrt{2(E_\mathrm{res}-E_a)}}{\Omega_a},
\qquad
x_\mathrm X=|Q_\mathrm X-Q_a|,
\end{equation}
the action becomes
\begin{equation}
S_a(Q_\mathrm X;E_\mathrm{res})
=
\Omega_a
\int_{x_t}^{x_\mathrm X}
\mathrm dx\,
\sqrt{x^2-x_t^2}.
\end{equation}
Using
\begin{equation}
\int \mathrm dx\,\sqrt{x^2-x_t^2}
=
\frac12
\left[
x\sqrt{x^2-x_t^2}
-
x_t^2
\operatorname{arcosh}
\!\left(\frac{x}{x_t}\right)
\right],
\end{equation}
one obtains
\begin{equation}
S_a(Q_\mathrm X;E_\mathrm{res})
=
\frac{\Omega_a}{2}
\left[
x_\mathrm X\sqrt{x_\mathrm X^2-x_t^2}
-
x_t^2
\operatorname{arcosh}
\!\left(\frac{x_\mathrm X}{x_t}\right)
\right].
\end{equation}
Since
\begin{equation}
x_\mathrm X^2
=
\frac{2\left(E_\mathrm X-E_a\right)}{\Omega_a^2},
\qquad
x_\mathrm X^2-x_t^2
=
\frac{2\left(E_\mathrm X-E_\mathrm{res}\right)}{\Omega_a^2},
\end{equation}
and
\begin{equation}
\frac{x_\mathrm X}{x_t}
=
\sqrt{
\frac{E_\mathrm X-E_a}
{E_\mathrm{res}-E_a}
},
\end{equation}
the exact harmonic action can be written as
\begin{equation}
S_a(Q_\mathrm X;E_\mathrm{res})
=
\frac{1}{\Omega_a}
\left[
\sqrt{
\bigl(E_\mathrm X-E_\mathrm{res}\bigr)
\bigl(E_\mathrm X-E_a\bigr)
}
-
\bigl(E_\mathrm{res}-E_a\bigr)
\operatorname{arcosh}
\!\left(
\sqrt{
\frac{E_\mathrm X-E_a}
{E_\mathrm{res}-E_a}
}
\right)
\right].
\label{eq:SI_exact_harmonic_action}
\end{equation}
To obtain a simpler analytic estimate, we now simplify the exact
harmonic actions in \Cref{eq:SI_exact_harmonic_action}. The purpose of
this step is not to retain all subleading prefactors and zero-point
corrections, but only to extract the leading exponential scale that
controls the crossover between thermal activation and nuclear tunneling. We specialize to the near-resonant case in which the energy offset
between the two minima is small compared to the crossing energy,
\(\Delta E=E_f-E_i\simeq0\). Thus \(E_f\simeq E_i\), and the two
actions have the same energy dependence, differing only through the
curvatures \(\Omega_i\) and \(\Omega_f\). It is then convenient to measure
all energies relative to the initial minimum and to define
\begin{equation}
\Delta E_\mathrm X = E_\mathrm X-E_i,
\qquad
\Delta E_\mathrm{res}=E_\mathrm{res}-E_i .
\end{equation}
Under these assumptions, the action for either surface may be written as
\begin{equation}
S_a^\mathrm X
=
\frac{1}{\Omega_a}
\left[
\sqrt{
\Delta E_\mathrm X
\left(
\Delta E_\mathrm X-\Delta E_\mathrm{res}
\right)}
-
\Delta E_\mathrm{res}
\operatorname{arcosh}
\left(
\sqrt{
\frac{\Delta E_\mathrm X}{\Delta E_\mathrm{res}}
}
\right)
\right],
\qquad
a\in\{i,f\}.
\label{eq:SI_action_small_Eres_start}
\end{equation}
We further assume that the crossing energy measured relative to the
initial minimum is much larger than the resonant energy measured relative
to the same reference. Defining
\begin{equation}
\varepsilon
\equiv
\frac{\Delta E_\mathrm{res}}{\Delta E_\mathrm X},
\qquad
\varepsilon \ll 1 ,
\end{equation}
we expand the action in powers of \(\varepsilon\). The algebraic
square-root term gives
\begin{equation}
\sqrt{
\Delta E_\mathrm X
\left(
\Delta E_\mathrm X-\Delta E_\mathrm{res}
\right)}
=
\Delta E_\mathrm X\sqrt{1-\varepsilon}
=
\Delta E_\mathrm X
\left[
1-\frac{\varepsilon}{2}
+\mathcal O(\varepsilon^2)
\right].
\end{equation}
The inverse hyperbolic cosine reduces to
\begin{equation}
\operatorname{arcosh}
\left(
\sqrt{
\frac{\Delta E_\mathrm X}{\Delta E_\mathrm{res}}
}
\right)
=
\ln
\left(
2\sqrt{
\frac{\Delta E_\mathrm X}{\Delta E_\mathrm{res}}
}
\right)
+
\mathcal O(\varepsilon).
\end{equation}
Substitution into \Cref{eq:SI_action_small_Eres_start} therefore yields
\begin{equation}
S_a^\mathrm X
=
\frac{\Delta E_\mathrm X}{\Omega_a}
-
\frac{\Delta E_\mathrm{res}}{\Omega_a}
\left[
\frac12
+
\ln
\left(
2\sqrt{
\frac{\Delta E_\mathrm X}{\Delta E_\mathrm{res}}
}
\right)
\right]
+
\mathcal O
\left(
\frac{\Delta E_\mathrm{res}^2}
{\Omega_a \Delta E_\mathrm X}
\right).
\label{eq:SI_action_expanded}
\end{equation}
The logarithmic correction in Eq.~\eqref{eq:SI_action_expanded} is negligible in the asymptotic limit \(\varepsilon=\Delta E_\mathrm{res}/\Delta E_\mathrm X\to0 \), since its relative size scales as \(\varepsilon\ln(1/\varepsilon)\). Thus, when \(\Delta E_\mathrm{res}\ll\Delta E_\mathrm X \) sufficiently strongly, the leading exponential scale is 
\begin{equation}
S_a^\mathrm X
\approx
\frac{\Delta E_\mathrm X}{\Omega_a}.
\label{eq:SI_action_leading}
\end{equation}
Consequently, the total tunneling action entering the low-temperature
overlap is
\begin{equation}
S_i^\mathrm X+S_f^\mathrm X
\approx
\Delta E_\mathrm X
\left(
\frac{1}{\Omega_i}
+
\frac{1}{\Omega_f}
\right).
\label{eq:SI_total_action_leading}
\end{equation}
Inserting this leading action into
\Cref{eq:SI_Tquant_exact_action_general} gives
\begin{equation}
k_\mathrm B T_\mathrm{quant}
\sim
\frac{\hbar \Delta E_\mathrm X}
{2\Delta E_\mathrm X
\left(
\Omega_i^{-1}+\Omega_f^{-1}
\right)}
=
\frac{\hbar}
{2\left(
\Omega_i^{-1}+\Omega_f^{-1}
\right)} .
\end{equation}
The crossing energy cancels because both the classical Arrhenius exponent
and the leading WKB tunneling exponent scale linearly with
\(\Delta E_\mathrm X\) in the deep-forbidden-region limit. We therefore
obtain
\begin{equation}
T_\mathrm{quant}
\sim
\frac{\hbar}{2k_\mathrm B}
\frac{\Omega_i\Omega_f}{\Omega_i+\Omega_f}.
\label{eq:SI_Tquant_frequency}
\end{equation}
Finally, using the definitions of the relaxation energies,
\begin{equation}
E_\mathrm R^i=\frac12\Omega_i^2\Delta Q^2,
\qquad
E_\mathrm R^f=\frac12\Omega_f^2\Delta Q^2,
\end{equation}
the estimate can be expressed directly in terms of the NMP parameters as
\begin{equation}
T_\mathrm{quant}
\sim
\frac{\hbar}{\sqrt{2}\,k_\mathrm B\,\Delta Q}
\frac{
\sqrt{E_\mathrm R^iE_\mathrm R^f}
}{
\sqrt{E_\mathrm R^i}+\sqrt{E_\mathrm R^f}
}.
\label{eq:Tquant_NMP}
\end{equation}
For equal curvatures,
\begin{equation}
E_\mathrm R^i=E_\mathrm R^f\equiv E_\mathrm R,
\end{equation}
this reduces to
\begin{equation}
T_\mathrm{quant}
\sim
\frac{\hbar}{2\sqrt{2}\,k_\mathrm B}
\frac{\sqrt{E_\mathrm R}}{\Delta Q}.
\label{eq:Tquant_NMP_equal}
\end{equation}
This final expression should be interpreted as a leading exponential estimate rather than as a sharp crossover temperature. Its validity is limited to the near-resonant case, in which the energy offset between the two minima is small, \(\Delta E=E_f-E_i\simeq0 \), so that the initial and final potential-energy surfaces have nearly aligned minima in energy. It also assumes a deep-forbidden-region limit, in which the crossing energy measured from the initial minimum is much larger than the resonant vibrational energy, \(\Delta E_\mathrm X\gg\Delta E_\mathrm{res}\). In physical terms, this means that the diabatic crossing lies high above the inital ground state, so that the zero point energy can be safely neglected in the WKB action.

\subsection*{Supplementary Information SI-4: Equal-Curvature Line-Shape Function}

In this section we derive the analytic equal-curvature line-shape function used in the main text. The derivation proceeds in three steps. First, we start from the unequal-curvature case and relate the coordinate-weighted matrix elements
\(\langle \chi_{i,m}|Q|\chi_{f,n}\rangle\)
to the Franck--Condon overlaps
\(\langle \chi_{i,m}|\chi_{f,n}\rangle\). Second, we take the equal-curvature limit, where the line-shape function factorizes as \(\Delta Q_\mathrm{X}^2\) times the conventional Franck--Condon line-shape function. Third, we take the continuum limit of this line-shape function, replacing the discrete Dirac comb by a smooth envelope that can be written in terms of a modified Bessel function. \\

We first keep the general unequal-curvature formulation in order to clarify which step is special to the equal-curvature limit. In the one-dimensional configuration-coordinate model used in the main text, the one-dimensional vibrational Hamiltonians are
\begin{equation}
H_i
=
\frac{P^2}{2}
+
V_i(Q),
\qquad
\text{with}
\qquad
V_i(Q)=\frac12\Omega_i^2 Q^2 ,
\label{eq:si_Hi_vi}
\end{equation}
and
\begin{equation}
H_f
=
\frac{P^2}{2}
+
V_f(Q),
\qquad
\text{with}
\qquad
V_f(Q)=\Delta E+\frac12\Omega_f^2(Q-\Delta Q)^2 .
\label{eq:si_Hf_vf}
\end{equation}
where \(P\) is the momentum conjugate to the configuration coordinate \(Q\). \\

The final-state potential can be expanded as
\begin{equation}
V_f(Q)
=
\Delta E
+
\frac12\Omega_f^2(Q-\Delta Q)^2
=
\Delta E
+
\frac12\Omega_f^2\Delta Q^2
-
\Omega_f^2\Delta Q\,Q
+
\frac12\Omega_f^2Q^2 .
\end{equation}
Therefore, the final-state vibrational Hamiltonian can be written in terms of the initial-state Hamiltonian as
\begin{equation}
H_f
=
H_i
+
\Delta E
+
\frac12\Omega_f^2\Delta Q^2
-
\Omega_f^2\Delta Q\,Q
+
\frac12(\Omega_f^2-\Omega_i^2)Q^2 .
\label{eq:si_Hf_unequal}
\end{equation}
We define the Franck--Condon overlaps
\begin{equation}
A_{mn}=\langle \chi_{i,m}|\chi_{f,n}\rangle,
\end{equation}
and the coordinate-weighted overlaps
\begin{equation}
M_{mn}=\langle \chi_{i,m}|Q|\chi_{f,n}\rangle,
\qquad
B_{mn}=\langle \chi_{i,m}|Q^2|\chi_{f,n}\rangle .
\end{equation}
Taking matrix elements of Eq.~\eqref{eq:si_Hf_unequal} between
\(\langle\chi_{i,m}|\) and \(|\chi_{f,n}\rangle\), and using
\begin{equation}
H_i|\chi_{i,m}\rangle=E_{i,m}|\chi_{i,m}\rangle,
\qquad
H_f|\chi_{f,n}\rangle=E_{f,n}|\chi_{f,n}\rangle,
\end{equation}
gives
\begin{equation}
E_{f,n}A_{mn}
=
\left(
E_{i,m}
+
\Delta E
+
\frac12\Omega_f^2\Delta Q^2
\right)A_{mn}
-
\Omega_f^2\Delta Q\,M_{mn}
+
\frac12(\Omega_f^2-\Omega_i^2)B_{mn}.
\end{equation}
Solving for the coordinate-weighted overlap yields
\begin{equation}
M_{mn}
=
\frac{
\left(
E_{i,m}
+
\Delta E
+
\frac12\Omega_f^2\Delta Q^2
-
E_{f,n}
\right)A_{mn}
+
\frac12(\Omega_f^2-\Omega_i^2)B_{mn}
}{
\Omega_f^2\Delta Q
}.
\label{eq:si_Mmn_unequal}
\end{equation}
The quantum-mechanical line-shape function can therefore be written as
\begin{equation}
\eta_{if}
=
\sum_{m=0}^{\infty}w_m^{(i)}
\sum_{n=0}^{\infty}
|M_{mn}|^2
\delta(E_{i,m}-E_{f,n}) .
\label{eq:si_eta_general_Mmn}
\end{equation}
On the support of the delta function, \(E_{i,m}=E_{f,n}\), so Eq.~\eqref{eq:si_Mmn_unequal} becomes
\begin{equation}
M_{mn}
=
\left[
\frac{
\Delta E+\frac12\Omega_f^2\Delta Q^2
}{
\Omega_f^2\Delta Q
}
+
\frac{\Omega_f^2-\Omega_i^2}{2\Omega_f^2\Delta Q}
\frac{B_{mn}}{A_{mn}}
\right]A_{mn}.
\label{eq:si_Mmn_onshell_unequal}
\end{equation}
It is useful to denote the transition-dependent coordinate prefactor by
\begin{equation}
\mathcal Q_{mn}
=
\frac{
\Delta E+\frac12\Omega_f^2\Delta Q^2
}{
\Omega_f^2\Delta Q
}
+
\frac{\Omega_f^2-\Omega_i^2}{2\Omega_f^2\Delta Q}
\frac{B_{mn}}{A_{mn}} .
\label{eq:si_Qmn_unequal}
\end{equation}
Then
\begin{equation}
M_{mn}=\mathcal Q_{mn}A_{mn},
\end{equation}
and the line-shape function may be written as
\begin{equation}
\eta_{if}
=
\sum_{m=0}^{\infty}w_m^{(i)}
\sum_{n=0}^{\infty}
|\mathcal Q_{mn}|^2
|A_{mn}|^2
\delta(E_{i,m}-E_{f,n}) .
\label{eq:si_eta_unequal_Qmn}
\end{equation}
Equation~\eqref{eq:si_Qmn_unequal} shows that, in the general unequal-curvature case, \(\mathcal Q_{mn}\) depends on the particular vibronic transition \(m\to n\). This transition dependence enters through the ratio \(B_{mn}/A_{mn}\), which is weighted by the curvature mismatch \(\Omega_f^2-\Omega_i^2\). Consequently, each Franck--Condon contribution \(|A_{mn}|^2\) carries its own transition-dependent prefactor \(|\mathcal Q_{mn}|^2\). We now specialize to the equal-curvature case,
\begin{equation}
\Omega_i=\Omega_f\equiv\Omega,
\qquad
E_\mathrm{R}^i=E_\mathrm{R}^f\equiv E_\mathrm{R}
=
\frac12\Omega^2\Delta Q^2 .
\end{equation}
In this limit the \(Q^2\)-weighted term in Eq.~\eqref{eq:si_Qmn_unequal} vanishes identically. The coordinate prefactor is no longer transition dependent and becomes
\begin{equation}
\mathcal Q_{mn}
=
\frac{\Delta E+E_\mathrm{R}}{\Omega^2\Delta Q}
=
\Delta Q\frac{\Delta E+E_\mathrm{R}}{2E_\mathrm{R}}
\equiv
\Delta Q_\mathrm{X} .
\label{eq:si_Qmn_equal_QX}
\end{equation}
This is the classical crossing coordinate of the two equal-curvature parabolas, measured relative to the initial minimum. Therefore
\begin{equation}
M_{mn}=\Delta Q_\mathrm{X}A_{mn},
\end{equation}
and the coordinate-weighted line-shape function factorizes exactly,
\begin{equation}
\eta_{if}
=
\Delta Q_\mathrm{X}^2\,\eta_\mathrm{FC},
\label{eq:si_eta_factorized_equal}
\end{equation}
where
\begin{equation}
\eta_\mathrm{FC}
=
\sum_{m=0}^{\infty}w_m^{(i)}
\sum_{n=0}^{\infty}
|A_{mn}|^2
\delta(E_{i,m}-E_{f,n})
\end{equation}
is the ordinary Franck--Condon line-shape function. For equal curvatures,
\begin{equation}
E_{i,m}
=
\hbar\Omega\left(m+\frac12\right),
\qquad
E_{f,n}
=
\Delta E
+
\hbar\Omega\left(n+\frac12\right).
\end{equation}
Thus the energy-conservation condition \(E_{i,m}=E_{f,n}\) implies
\begin{equation}
\Delta E=(m-n)\hbar\Omega .
\label{eq:si_energy_conservation_equal}
\end{equation}
The thermally weighted Franck--Condon spectrum can therefore be written in the Poisson-product form
\begin{equation}
\eta_\mathrm{FC}(\Delta E)
=
e^{-(a+b)}
\sum_{m=0}^{\infty}
\sum_{n=0}^{\infty}
\frac{a^n}{n!}
\frac{b^m}{m!}
\delta\!\left[\Delta E-(m-n)\hbar\Omega\right],
\label{eq:si_eta_FC_poisson}
\end{equation}
with
\begin{equation}
a=S(1+\bar n),
\qquad
b=S\bar n,
\qquad
S=\frac{E_\mathrm{R}}{\hbar\Omega},
\qquad
\bar n=\frac{1}{e^{\beta\hbar\Omega}-1} .
\end{equation}
Introducing the signed dimensionless energy offset
\begin{equation}
p=\frac{\Delta E}{\hbar\Omega},
\end{equation}
we use
\begin{equation}
\delta\!\left[\Delta E-(m-n)\hbar\Omega\right]
=
\frac{1}{\hbar\Omega}
\delta\!\left[p-(m-n)\right] .
\end{equation}
The delta function enforces \(p=m-n\). We now take the continuum limit of the discrete vibronic spectrum by promoting one of the vibrational sums to an integral. Since \(m,n\ge0\), this step must be carried out separately for \(p\ge0\) and \(p<0\). This branch split is essential: for positive \(p\), the initial vibrational quantum number must exceed the final one by \(p\), whereas for negative \(p\), the final vibrational quantum number must exceed the initial one by \(|p|\). For \(p\ge0\), the constraint \(p=m-n\) gives \(m=n+p\). Taking the continuum envelope of the discrete spectrum gives
\begin{align}
\eta_\mathrm{FC}^{(+)}(\Delta E)
&\approx
\frac{e^{-(a+b)}}{\hbar\Omega}
\int_0^\infty \mathrm dn\,
\frac{a^n}{\Gamma(n+1)}
\frac{b^{n+p}}{\Gamma(n+p+1)} \\
&=
\frac{e^{-(a+b)}}{\hbar\Omega}
b^p
\sum_{n=0}^{\infty}
\frac{(ab)^n}{n!\Gamma(n+p+1)} .
\end{align}
Using the series representation of the modified Bessel function of the first kind,
\begin{equation}
I_p(2\sqrt{ab})
=
\sum_{n=0}^{\infty}
\frac{(ab)^{n+p/2}}{n!\Gamma(n+p+1)},
\qquad p\ge0,
\end{equation}
we obtain
\begin{equation}
\eta_\mathrm{FC}^{(+)}(\Delta E)
\approx
\frac{1}{\hbar\Omega}
\exp[-(a+b)]
\left(\frac{b}{a}\right)^{p/2}
I_p(2\sqrt{ab}),
\qquad p\ge0 .
\label{eq:si_eta_FC_positive_p}
\end{equation}
For \(p<0\), we write \(q=-p>0\). The constraint \(p=m-n\) then gives \(n=m+q\). The corresponding continuum envelope is
\begin{align}
\eta_\mathrm{FC}^{(-)}(\Delta E)
&\approx
\frac{e^{-(a+b)}}{\hbar\Omega}
\sum_{m=0}^{\infty}
\frac{b^m}{m!}
\frac{a^{m+q}}{\Gamma(m+q+1)} \\
&=
\frac{e^{-(a+b)}}{\hbar\Omega}
a^q
\sum_{m=0}^{\infty}
\frac{(ab)^m}{m!\Gamma(m+q+1)} .
\end{align}
Using the series representation of the modified Bessel function of the first kind,
\begin{equation}
I_q(2\sqrt{ab})
=
\sum_{m=0}^{\infty}
\frac{(ab)^{m+q/2}}{m!\Gamma(m+q+1)},
\qquad q>0,
\end{equation}
we obtain
\begin{equation}
\eta_\mathrm{FC}^{(-)}(\Delta E)
\approx
\frac{1}{\hbar\Omega}
\exp[-(a+b)]
\left(\frac{a}{b}\right)^{q/2}
I_q(2\sqrt{ab}),
\qquad q=-p>0 .
\label{eq:si_eta_FC_negative_p}
\end{equation}
Since \(q=|p|\), we can combine both branches which results in a compact expression for the continuum limit of the Franck-Condon lineshape function
\begin{equation}
\eta_\mathrm{FC}(\Delta E)
\approx
\frac{1}{\hbar\Omega}
\exp[-(a+b)]
\left(\frac{b}{a}\right)^{p/2}
I_{|p|}(2\sqrt{ab}) .
\label{eq:si_eta_FC_bessel}
\end{equation}
Combining Eq.~\eqref{eq:si_eta_factorized_equal} with Eq.~\eqref{eq:si_eta_FC_bessel} gives the equal-curvature line-shape envelope for the NMP model
\begin{equation}
\eta_{if}
\approx
\frac{\Delta Q_\mathrm{X}^2}{\hbar\Omega}
\exp\!\left[-S(1+2\bar n)\right]
\left(\frac{\bar n}{1+\bar n}\right)^{p/2}
I_{|p|}\!\left(2S\sqrt{\bar n(1+\bar n)}\right),
\label{eq:si_eta_equal_bessel}
\end{equation}
with
\begin{equation}
\Omega=\frac{\sqrt{2E_\mathrm{R}}}{\Delta Q},
\qquad
S=\frac{E_\mathrm{R}}{\hbar\Omega},
\qquad
p=\frac{\Delta E}{\hbar\Omega},
\qquad
\bar n=\frac{1}{e^{\beta\hbar\Omega}-1} .
\label{eq:si_equal_parameters}
\end{equation}
Equivalently, since
\begin{equation}
\frac{\bar n}{1+\bar n}
=
e^{-\beta\hbar\Omega},
\end{equation}
Eq.~\eqref{eq:si_eta_equal_bessel} may be written as
\begin{equation}
\eta_{if}
\approx
\frac{\Delta Q_\mathrm{X}^2}{\hbar\Omega}
\exp\!\left[
-\frac{\Delta E}{2k_\mathrm{B}T}
-
S(1+2\bar n)
\right]
I_{|\Delta E|/(\hbar\Omega)}\!\left(
2S\sqrt{\bar n(1+\bar n)}
\right).
\label{eq:si_eta_equal_bessel_alt}
\end{equation}
Equation~\eqref{eq:si_eta_equal_bessel} is the single-Bessel-function representation of the line-shape-function envelope in the equal-curvature limit. The equivalent form in Eq.~\eqref{eq:si_eta_equal_bessel_alt} makes the detailed-balance-related Boltzmann factor explicit. In the exact discrete theory, energy conservation restricts \(p\) to integer values and the spectrum is a Dirac comb. The continuum approximation promotes \(p\) to a continuous variable and yields the smooth envelope used for the analytic rate expression.
\subsection*{Supplementary Information SI-5: Intrinsic Maxima of the Classical Line-Shape Function}
\label{si:classical_lsf_maximum}
In the classical limit, the line-shape function is given by
\begin{equation}
\eta^{\mathrm{cl}}
\propto
\Delta Q_\mathrm{X}^2
\exp\!\left(-\beta\Delta E_\mathrm{X}\right),
\end{equation}
where \(\Delta E_\mathrm{X}\) is the classical barrier measured from the minimum of the initial potential-energy surface and \(\Delta Q_\mathrm{X}\) is the corresponding crossing coordinate. Consequently, the classical line-shape function is governed by the competition between the exponentially increasing Boltzmann factor, which favors small barriers \(\Delta E_\mathrm{X}\), and the coordinate prefactor \(\Delta Q_\mathrm{X}^2\), which vanishes when the crossing point coincides with the minimum of the initial potential-energy surface. In this section, we derive the intrinsic maxima of the classical line-shape function resulting from this interplay. Using the notation introduced in the main text, the diabatic potential-energy surfaces can be rewritten in the following form
\begin{equation}
V_i(Q)
=
E_\mathrm{R}^i
\left(\frac{Q}{\Delta Q}\right)^2,
\qquad
V_f(Q)
=
\Delta E
+
E_\mathrm{R}^f
\left(\frac{Q-\Delta Q}{\Delta Q}\right)^2 .
\end{equation}
Introducing the dimensionless crossing coordinate
\begin{equation}
x
=
\frac{\Delta Q_\mathrm{X}}{\Delta Q},
\end{equation}
the crossing condition
\begin{equation}
V_i(\Delta Q_\mathrm{X})
=
V_f(\Delta Q_\mathrm{X})
\end{equation}
gives
\begin{equation}
\Delta E(x)
=
(E_\mathrm{R}^i-E_\mathrm{R}^f)x^2
+
2E_\mathrm{R}^f x
-
E_\mathrm{R}^f,
\label{eq:si_deltaE_x}
\end{equation}
while the corresponding crossing energy measured from the minimum of the initial potential-energy surface is
\begin{equation}
\Delta E_\mathrm{X}(x)
=
E_\mathrm{R}^i x^2 .
\end{equation}
According to the classical high-temperature expression derived in the main text, the line-shape function is proportional to (up to slowly varying prefactors)
\begin{equation}
\eta^{\mathrm{cl}}(x)
\propto
x^2
\exp\!\left(-\beta E_\mathrm{R}^i x^2\right).
\end{equation}
Taking the logarithm yields
\begin{equation}
\ln \eta^{\mathrm{cl}}
=
2\ln|x|
-
\beta E_\mathrm{R}^i x^2
+
\mathrm{const.},
\end{equation}
such that
\begin{equation}
\frac{\mathrm d}{\mathrm dx}
\ln\eta^{\mathrm{cl}}
=
\frac{2}{x}
-
2\beta E_\mathrm{R}^i x.
\end{equation}
The stationary points therefore satisfy
\begin{equation}
x_\pm
=
\pm
\frac{1}{\sqrt{\beta E_\mathrm{R}^i}}.
\end{equation}
Since
\begin{equation}
\frac{\mathrm d^2}{\mathrm dx^2}
\ln\eta^{\mathrm{cl}}
=
-\frac{2}{x^2}
-
2\beta E_\mathrm{R}^i
<
0,
\end{equation}
both stationary points correspond to maxima. Substituting these values into \Cref{eq:si_deltaE_x} gives the corresponding energy offsets
\begin{equation}
\Delta E_\pm
=
-E_\mathrm{R}^f
+
\frac{E_\mathrm{R}^i-E_\mathrm{R}^f}
{\beta E_\mathrm{R}^i}
\pm
\frac{2E_\mathrm{R}^f}
{\sqrt{\beta E_\mathrm{R}^i}}.
\end{equation}
For equal curvatures,
\begin{equation}
E_\mathrm{R}^i
=
E_\mathrm{R}^f
\equiv
E_\mathrm{R},
\end{equation}
this expression simplifies to
\begin{equation}
\Delta E_\pm
=
-E_\mathrm{R}
\pm
2\sqrt{\frac{E_\mathrm{R}}{\beta}}
=
-E_\mathrm{R}
\pm
2\sqrt{E_\mathrm{R}k_\mathrm{B}T}.
\end{equation}
In general, the classical line-shape function therefore possesses two maxima, corresponding to the two possible crossing branches (\(\Delta Q_\mathrm{X}>0\) and \(\Delta Q_\mathrm{X}<0\)). Both maxima are located close to
$\Delta E \approx -E_\mathrm{R}^f$
which marks the point at which the classical barrier vanishes, i.e., where the crossing point coincides with the minimum of the initial potential-energy surface. Exactly at this point, however, the coordinate prefactor \(\Delta Q_\mathrm{X}^2\) vanishes, causing the classical line-shape function itself to become zero. Consequently, the maxima are shifted slightly away from the point $\Delta E \approx -E_\mathrm{R}^f$. The full quantum-mechanical line-shape function exhibits a similar behavior, with its maximum typically occurring close to the $\Delta E \approx -E_\mathrm{R}^f$, although its precise position is further modified by quantum effects such as zero-point motion and tunneling.

\subsection*{Supplementary Information SI-6: Derivation of the Electronic Capture and Emission Rates from Time-Dependent Perturbation Theory}

In this section, we derive the electronic capture and emission rates used in
the main text using time-dependent perturbation theory in the density-matrix
formalism. The Hamiltonian is decomposed into an exactly solvable part and a weak
perturbation,
\begin{equation}
H = H_0 + H_{\mathrm{int}},
\end{equation}
where the unperturbed Hamiltonian is diagonal in the diabatic electronic basis,
\begin{equation}
H_0 =
|\phi_i\rangle\langle\phi_i|
\left(
\frac{P^2}{2}+V_i(Q)
\right)
+
|\phi_f\rangle\langle\phi_f|
\left(
\frac{P^2}{2}+V_f(Q)
\right).
\end{equation}
The perturbation responsible for electronic transitions is given by
\begin{equation}
H_{\mathrm{int}}
=
W_{if}\,Q
\left(
|\phi_i\rangle\langle\phi_f|
+
|\phi_f\rangle\langle\phi_i|
\right).
\end{equation}
The eigenstates of $H_0$ are the vibronic product states
\begin{equation}
|\Psi_{a,m}\rangle
=
|\phi_a\rangle
\otimes
|\chi_{a,m}\rangle,
\qquad
a\in\{i,f\},
\end{equation}

We first consider the forward transition
$i\rightarrow f$. The system is assumed to be initially prepared in the
vibronic manifold associated with the electronic state
$|\phi_i\rangle$, while the vibrational subsystem has already relaxed to
thermal equilibrium on the corresponding potential-energy surface. The
initial density operator is therefore
\begin{equation}
\rho(0)
=
|\phi_i\rangle\langle\phi_i|
\otimes
\rho_i^{\mathrm{eq}},
\end{equation}
where
\begin{equation}
\rho_i^{\mathrm{eq}}
=
\sum_m
w_m^{(i)}
|\chi_{i,m}\rangle
\langle\chi_{i,m}|,
\qquad
w_m^{(i)}
=
\frac{\exp(-\beta E_{i,m})}{Z_i},
\end{equation}
and
\begin{equation}
Z_i
=
\sum_m
\exp(-\beta E_{i,m}).
\end{equation}
Passing to the interaction picture,
\begin{equation}
\tilde{\rho}(t)
=
e^{iH_0t/\hbar}
\rho(t)
e^{-iH_0t/\hbar},
\qquad
\tilde{H}_{\mathrm{int}}(t)
=
e^{iH_0t/\hbar}
H_{\mathrm{int}}
e^{-iH_0t/\hbar},
\end{equation}
the density operator obeys
\begin{equation}
\frac{d\tilde{\rho}(t)}{dt}
=
-\frac{i}{\hbar}
\left[
\tilde{H}_{\mathrm{int}}(t),
\tilde{\rho}(t)
\right].
\end{equation}
Integrating once and substituting back yields the second-order Dyson
expansion,
\begin{equation}
\tilde{\rho}(t)
=
\rho(0)
-
\frac{i}{\hbar}
\int_0^t
dt_1
\left[
\tilde{H}_{\mathrm{int}}(t_1),
\rho(0)
\right]
-
\frac{1}{\hbar^2}
\int_0^t
dt_1
\int_0^{t_1}
dt_2
\left[
\tilde{H}_{\mathrm{int}}(t_1),
\left[
\tilde{H}_{\mathrm{int}}(t_2),
\rho(0)
\right]
\right],
\end{equation}
which is accurate up to second order in the weak coupling
$H_{\mathrm{int}}$.
The probability of finding the electronic subsystem in the final state
$|\phi_f\rangle$ at time $t$ is
The probability of finding the electronic subsystem in the final charge
state $|\phi_f\rangle$ at time $t$ is obtained by projecting onto the
electronic state $|\phi_f\rangle$ while remaining insensitive to the
vibrational state. This is achieved using the projector
\begin{equation}
P_f(t)
=
\mathrm{Tr}
\left\{
\Pi_f
\tilde{\rho}(t)
\right\},
\qquad
\Pi_f
=
|\phi_f\rangle\langle\phi_f|
\otimes
I_{\mathrm{vib}},
\end{equation}
where $I_{\mathrm{vib}}$ denotes the identity operator on the
vibrational Hilbert space. Consequently, the trace is taken over all
vibrational states, reflecting the fact that experiments resolve only
the electronic charge state and not the accompanying vibrational
transition.
Since the initial state contains no population in
$|\phi_f\rangle$, the first non-vanishing contribution is second order in
$H_{\mathrm{int}}$. Inserting the complete vibronic basis gives
\begin{equation}
P_f(t)
=
\sum_m
w_m^{(i)}
\sum_n
\frac{
\left|
\langle
\Psi_{f,n}
|
H_{\mathrm{int}}
|
\Psi_{i,m}
\rangle
\right|^2
}{\hbar^2}
\left|
\int_0^t
dt'
e^{i(E_{f,n}-E_{i,m})t'/\hbar}
\right|^2.
\end{equation}
The time integral evaluates to
\begin{equation}
\left|
\int_0^t
dt'
e^{i(E_{f,n}-E_{i,m})t'/\hbar}
\right|^2
=
\frac{
4
\sin^2
\!\left[
(E_{f,n}-E_{i,m})t/(2\hbar)
\right]
}{
(E_{f,n}-E_{i,m})^2/\hbar^2
}.
\end{equation}
In the long-time limit,
\begin{equation}
\lim_{t\rightarrow\infty}
\frac{1}{t}
\left|
\int_0^t
dt'
e^{i(E_{f,n}-E_{i,m})t'/\hbar}
\right|^2
=
2\pi\hbar\,
\delta(E_{f,n}-E_{i,m}),
\end{equation}
such that the transition probability grows linearly with time,
\begin{equation}
P_f(t)
\simeq
k_{if}t.
\end{equation}
The transition rate is therefore
\begin{equation}
k_{if}
=
\frac{2\pi}{\hbar}
\sum_m
w_m^{(i)}
\sum_n
\left|
\langle
\Psi_{f,n}
|
H_{\mathrm{int}}
|
\Psi_{i,m}
\rangle
\right|^2
\delta(E_{i,m}-E_{f,n}).
\end{equation}
Using the explicit form of the perturbation,
\begin{equation}
\langle
\Psi_{f,n}
|
H_{\mathrm{int}}
|
\Psi_{i,m}
\rangle
=
W_{if}
\,
\langle
\chi_{f,n}
|
Q
|
\chi_{i,m}
\rangle,
\end{equation}
we obtain the final form of the transition rate
\begin{equation}
k_{if}
=
\frac{2\pi}{\hbar}
|W_{if}|^2
\eta_{if}, \qquad \eta_{if}
=
\sum_m
w_m^{(i)}
\sum_n
\left|
\langle
\chi_{i,m}
|
Q
|
\chi_{f,n}
\rangle
\right|^2
\delta(E_{i,m}-E_{f,n}).
\end{equation}

The reverse transition rate is obtained analogously by preparing the
system initially in the vibronic manifold associated with the electronic
state $|\phi_f\rangle$,
\begin{equation}
\rho(0)
=
|\phi_f\rangle
\langle\phi_f|
\otimes
\rho_f^{\mathrm{eq}},
\end{equation}
with
\begin{equation}
\rho_f^{\mathrm{eq}}
=
\sum_n
w_n^{(f)}
|\chi_{f,n}\rangle
\langle
\chi_{f,n}
|,
\qquad
w_n^{(f)}
=
\frac{\exp(-\beta E_{f,n})}{Z_f}.
\end{equation}
Repeating the same calculation yields
\begin{equation}
k_{fi}
=
\frac{2\pi}{\hbar}
|W_{if}|^2
\eta_{fi}, \qquad \eta_{fi}
=
\sum_n
w_n^{(f)}
\sum_m
\left|
\langle
\chi_{f,n}
|
Q
|
\chi_{i,m}
\rangle
\right|^2
\delta(E_{f,n}-E_{i,m})
\end{equation}
Thus, the observable electronic transition rates are obtained as conditional
thermal averages of the textbook Fermi golden-rule transition rates, where
the thermal average is taken over the equilibrium vibrational ensemble
associated with the initially occupied electronic charge state.

\newpage

\printbibliography

\end{document}